\documentclass[12pt]{article}
\usepackage{packages}

\title{The economic and health impacts of contact tracing and quarantine programs}

\author{Darija Barak, Edoardo Gallo, Alastair Langtry}

\begin{document}

\maketitle
\thispagestyle{empty}
\blfootnote{\noindent Authors affiliation: University of Cambridge, Faculty of Economics, Sidgwick Avenue, Cambridge, UK. 
}
\blfootnote{\noindent Corresponding author: edo[at]econ.cam.ac.uk (E. Gallo)}

\begin{abstract}
    \noindent Contact tracing and quarantine programs have been one of the leading Non-Pharmaceutical Interventions against COVID-19. Some governments have relied on mandatory programs, whereas others embrace a voluntary approach. However, there is limited evidence on the relative effectiveness of these different approaches. In an interactive online experiment conducted on 731 subjects representative of the adult US population in terms of sex and region of residence, we find there is a clear ranking. A fully mandatory program is better than an optional one, and an optional system is better than no intervention at all. The ranking is driven by reductions in infections, while economic activity stays unchanged. We also find that political conservatives have higher infections and levels of economic activity, and they are less likely to participate in the contact tracing program.\\
    
\end{abstract}

\clearpage
\pagenumbering{arabic}
\section*{Title}
Full title: The economic and health impacts of contact tracing and quarantine programs \\
Short title: The economic and health impacts of contact tracing

\section*{Authors}
Darija Barak, Edoardo Gallo$^{*}$, Alastair Langtry \\
\noindent $^{*}$ Corresponding author (edo@cam.ac.uk) \\

\section*{Affiliations}
University of Cambridge, Faculty of Economics, Sidgwick Avenue, Cambridge, UK.

\section*{Abstract}
Contact tracing and quarantine programs have been one of the leading Non-Pharmaceutical Interventions against COVID-19. Some governments have relied on mandatory programs, whereas others embrace a voluntary approach. However, there is limited evidence on the relative effectiveness of these different approaches. In an interactive online experiment conducted on 731 subjects representative of the adult US population in terms of sex and region of residence, we find there is a clear ranking. A fully mandatory program is better than an optional one, and an optional system is better than no intervention at all. The ranking is driven by reductions in infections, while economic activity stays unchanged. We also find that political conservatives have higher infections and levels of economic activity, and they are less likely to participate in the contact tracing program.

\section*{Teaser}
An optional contact tracing program reduces infections, but a mandatory one is more effective. Neither have a significant impact on economic activity.

\section*{MAIN TEXT}

\section{Introduction}

Contact tracing and quarantine programs have been one of the most important Non-Pharmaceutical Interventions governments have used to contain the COVID-19 pandemic \cite{WHO25June2021, johnshopkins2021}. However, governments worldwide have taken a variety of different approaches. For instance, China, Singapore, and Taiwan made these programs compulsory from the start of the pandemic \cite{H_2020, KC_2020, iwasaki2020does}. In contrast, the United States implemented these tools by leaving the choice of whether to participate to citizens, while Brazil largely did not use them at all \cite{O_2020}. Intermediate paths are also common -- in the United Kingdom, quarantining when asked to is a legal requirement, but contact tracing was left largely up to individual citizens and businesses \cite{blasimme2021digital, UKselfisolate}.

The unprecedented scale and urgency of the COVID-19 pandemic meant that there was no time to test which approach to setting up contact tracing and quarantine programs worked better to achieve the optimal trade-off between containment of the disease and minimizing damage to the economy. Studies based on surveys and mobility data have documented attitudes to Non-Pharmaceutical Interventions and the resulting changes in health behaviors typically in specific national context \cite{kraemer2020effect, gadarian2021partisanship, pei2020differential}. However, these studies cannot identify the causal impact of the policy, and they are unable to examine the counterfactual of the impact of a different policy in the same context. Moreover, surveys rely on self-reported data which may not necessarily map onto actual behavior. Field experiments have been extensively used to test the effectiveness of health interventions \cite{duflo2020field, NBERw27496}, but the restrictions imposed by the pandemic posed significant logistical challenges to their implementation to test the effectiveness of tracing and quarantine policies, with the exception of small-scale studies in specific settings \cite{wells2021optimal}. At the moment, there is a lack of causal evidence on the effectiveness of different contact tracing and quarantine programs to prevent the spread of COVID-19, and how they impact the economy. 

We conduct a large-scale interactive online experiment to investigate the causal impacts of the existence and/or mandatory nature of contact tracing and quarantine programs. Our sample of 731 subjects is representative of the adult US population in terms of region of residence and sex. In the environment of our experiment, participants choose a level of economic activity. Higher activity is beneficial, but, at the same time, increases the chance of getting infected with COVID-19. The option to sign up to the contact tracing system allows participants to get an alert when they have been exposed to the disease. In case they have been alerted, participants can opt to quarantine so that they are out of circulation for the duration of their illness to prevent the diffusion of the disease.

The core part of our experimental design is four treatments that systematically make tracing and/or quarantine mandatory. This allows us to compare outcomes in a society where both tracing and quarantine are optional to one where they are both mandatory, as well as what happens when just one of the two is mandatory. We also examine two additional benchmark treatments. The first one is a setting where neither program is available as doing nothing is a natural benchmark against which to compare the effectiveness of policy interventions. The second one is a setting where the level of individual economic activity is fixed to the maximum, but participants have the option to opt into tracing and/or quarantine.

The first finding is that making both tracing and quarantine mandatory works better than a system where both of them are optional. A fully mandatory system decreases the rate of infections by taking infected participants out of circulation, and it has no discernible impact on the overall level of activity. Making either only tracing or only quarantine mandatory is also inferior to a fully mandatory system.

The second finding is that providing optional tracing and quarantine programs is significantly better than doing nothing. They reduce the prevalence of COVID-19 without any corresponding reduction in economic activity. Leaving tracing optional and making quarantine mandatory does not provide additional benefits compared to a system with both being optional. A program with mandatory tracing and optional quarantine provides at most mixed benefits compared to when both of them are optional, and these benefits are not robust in the long-term. 

The third finding is that, in the US context, political ideology plays an important role in individuals' response to interventions, and, consequently, to their effectiveness. This is consistent with other empirical studies which document partisan difference in behavioral responses to Non-Pharmaceutical Interventions to stem COVID-19 in the US context \cite{allcott2020polarization, clinton2021partisan, gadarian2021partisanship}. In our experiment, political conservatives choose higher levels of economic activity, are less likely to participate in the contact tracing program, and are more likely to become infected. The benefits from higher activity more than outweigh the costs of potential infection, so they are individually better-off. This is because conservatives can partially free-ride on liberals, who push down overall rates of infection in a group with their lower choices of activity and greater sign-up to the tracing scheme. 

A consequence of these differences in behavioral responses is that the effectiveness of a policy depends on the ideological make-up of a society. We use the experimental data to calibrate behavior of a representative conservative/liberal participant in each treatment, and then simulate the evolution of decisions and outcomes for ideologically homogeneous groups. A liberal group experiences higher welfare than a conservative one in every treatment with the exception of a fully mandatory system where they fare similarly. In other words, the best policy for a conservative society may be exactly what many conservatives vocally oppose -- a mandate that makes tracing and quarantine interventions compulsory.

\section{Experimental Design} 

\begin{figure}[h]
\centering
\includegraphics[width=0.9\textwidth]{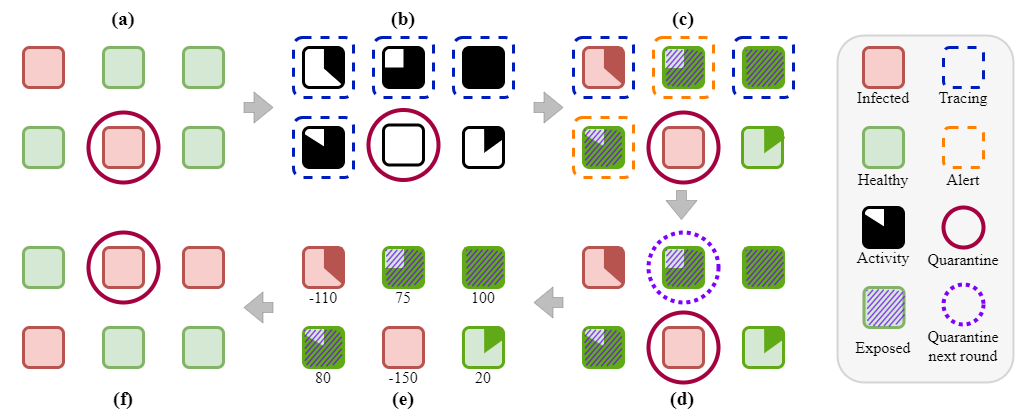}
\caption{Flow of a typical round, $t$, in treatment OO. \textbf{(a)} Initial conditions for round $t$. Shows which agents are healthy/infected and in quarantine. \textbf{(b)} Agents choose activity and tracing for round $t$. Agents in quarantine have activity defaulted to zero, but can still choose tracing. \textbf{(c)} COVID-19 spreads, and some agents become exposed. The contact tracing scheme also sends alerts. \textbf{(d)} Agents who received an alert in panel (c) choose whether to quarantine in round $t+1$. \textbf{(e)} Agents realize payoffs. \textbf{(f)} Transition to round $t+1$. Agents who were infected [resp. were exposed] during round $t$ recover [resp. become infected]. Agents leave/enter quarantine as appropriate. \textbf{Other Treatments:} as described for treatment OO with the following changes. Treatment NI
items in right-hand column of the legend are not present. Treatment NA: unless in quarantine, activity is fixed to $100$ in panel (b). Treatments OM and MM: all agents must sign up to tracing in panel (b). Treatments MO and MM: all agents who receive an alert in panel (c) must go into quarantine in round $t+1$.}
\label{fig1}
\end{figure}

\subsection{Game} The game is based on an augmented Susceptible-Infected-Susceptible (SIS) model \cite{kermack1927contribution, allen1994some} in discrete time. Agents play the game in a fixed group of 12-15. Fig.\ref{fig1} explains how a typical round (i.e. time period) works, and the transition between rounds. The schema refers to the version of the game where agents choose their level of activity, participation in tracing, and whether or not to quarantine. The end of this section explains how we restrict these choices in our other treatments. 

Depending on their decisions in the previous round, Fig.\ref{fig1}(a) shows agents' health status can be either healthy or infected, and they can be either in quarantine or not. Note that agents do not learn of their health status until the end of the round in order to capture the pre-symptomatic period of COVID-19 \cite{he2020temporal, furukawa2020evidence}. In Fig.\ref{fig1}(b) agents choose a level of economic activity -- an integer between 0 and 100 -- and whether or not to participate in the contact tracing scheme. An agent in quarantine has her activity set to 0 by design, but can still choose whether to participate in the contact tracing scheme.

Fig.\ref{fig1}(c) shows the spread of COVID-19. Each infected agent not in quarantine goes on to expose $R_0 A_t^2$ healthy agents. $R_0$ is the infectiousness of COVID-19 and $A_t$ is the mean of the level of economic activity decisions in Fig.\ref{fig1}(b). This captures the fact that lower average activity reduces contacts and so leaves fewer opportunities for COVID-19 to spread. Further, the probability an individual healthy agent gets exposed is proportional to their own economic activity. Fig.\ref{fig1}(c) also shows that one third of the agents who are participating in the contact tracing scheme \emph{and} were exposed receive an alert from the scheme. In our stylized environment, the scheme never sends false positive alerts.

\begin{figure}[t]
\centering
\includegraphics[width=0.9\textwidth]{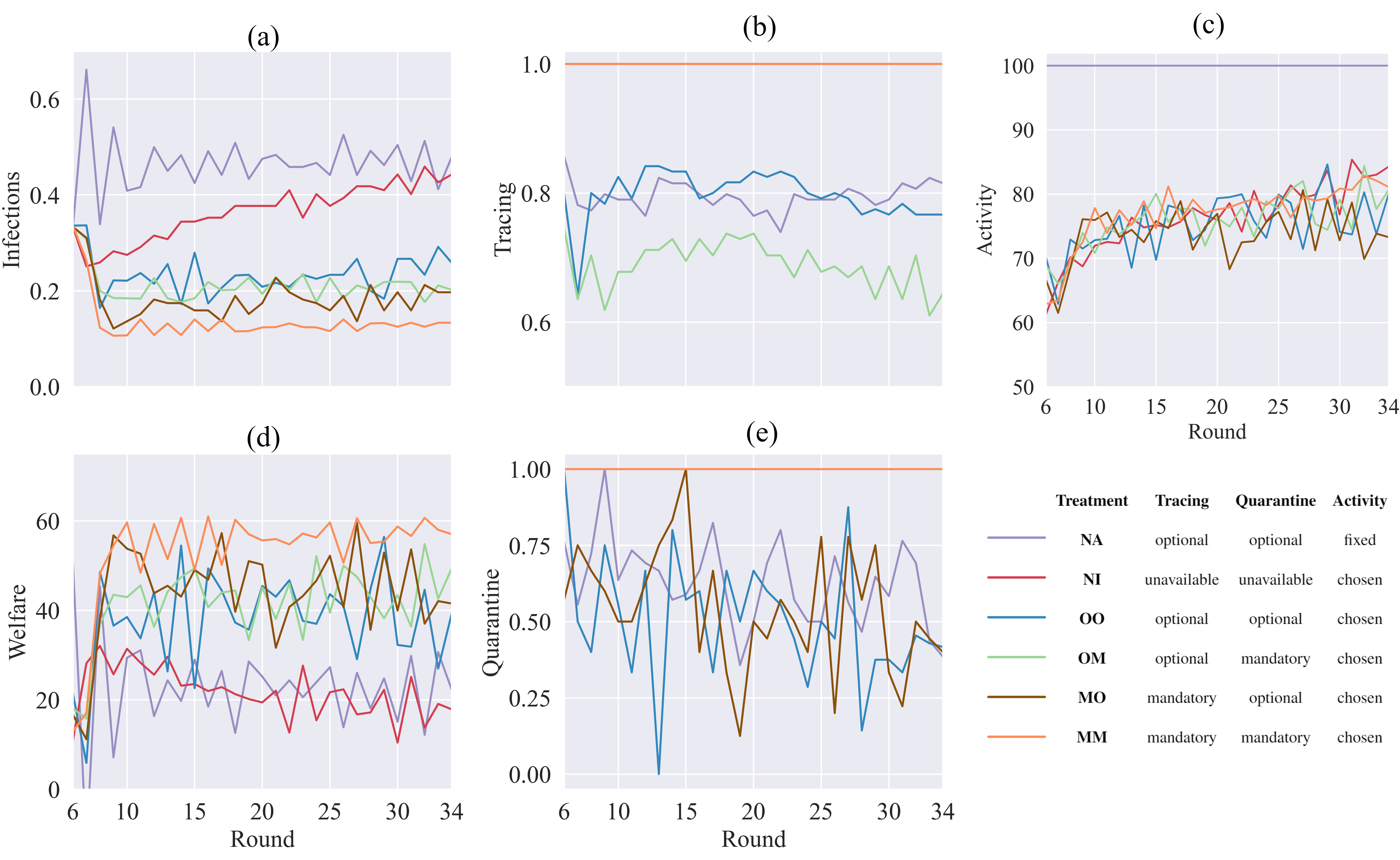}
\caption{Treatment-level means of outcomes and decisions over time. COVID-19 is not introduced until Round 6, so Rounds 1-5 are omitted. \textbf{(a)} Fraction of subjects infected. \textbf{(b)} Fraction of subjects participating in the contact tracing scheme. \textbf{(c)} Average activity decision for subjects not in quarantine. \textbf{(d)} Average individual welfare. \textbf{(e)} Fraction of subjects choosing to go into quarantine after receiving an alert.}
\label{fig2}
\end{figure}

Agents who received an alert then choose whether to go into quarantine in the next round (Fig.\ref{fig1}(d)). Agents realize payoffs in Fig.\ref{fig1}(e). In our experiment, agents get 1 point per unit of activity they do, and lose 150 points if infected. Tracing is free. Quarantine has no costs beyond the effect of setting activity to zero. Finally, Fig.\ref{fig1}(f) shows the transition to the next round. Infected agents recover, exposed agents become infected, and agents who chose to do so go into quarantine. In line with the SIS framework, an agent cannot be infected in two consecutive rounds.

For the first 5 rounds of our game there is no COVID-19, so subjects are always healthy. During these rounds, subjects only choose a level of economic activity. This gives them time to learn about the interface. After the end of Round 5, 4-5 subjects are randomly chosen to be infected with COVID-19. The experiment then proceeds for a further 30 rounds, at which point it terminates with 50\% probability after each round.

\subsection{Treatments} The purpose of our study is to understand how the presence and/or mandatory nature of tracing and quarantine measures affect the diffusion of COVID-19 and impact economic activity. The design, therefore, examines how making an individual's choice of joining tracing and/or going into quarantine mandatory affects the evolution of COVID-19 and economic activity. This leads to a total of 6 treatments.

Our baseline is the \emph{No Intervention} (\emph{NI}) scenario with neither tracing nor quarantine. The first treatment is a society with \emph{Optional tracing \& Optional quarantine} (\emph{OO}) that is the game described above. Our next treatments systematically make tracing and/or quarantine mandatory rather than a choice available to participants: \emph{Mandatory tracing \& Optional quarantine} (\emph{MO}), \emph{Optional tracing \& Mandatory quarantine} (\emph{OM}), and \emph{Mandatory tracing \& Mandatory quarantine} (\emph{MM}). Finally, we have one \emph{No Activity choice} (\emph{NA}) treatment that is identical to \emph{OO} but with activity level fixed at 100 unless a participant is in quarantine.   

\vspace{15mm}
\section{Results}

\begin{table*}[h]
\centering
\caption{Main regression results for individual subject decisions and outcomes.}
\label{tab:regressions}
\resizebox{\textwidth}{!}{%
\begin{tabular}{@{}lrcrcrcrcrc@{}}
\toprule
\multicolumn{1}{r}{\textbf{Dependent Variable:}} & \multicolumn{2}{c}{Activity} & \multicolumn{2}{c}{Tracing} & \multicolumn{2}{c}{Quarantine} & \multicolumn{2}{c}{Infection} & \multicolumn{2}{c}{Welfare} \\ \midrule
\multicolumn{1}{r}{Model$^{a}$:} & \multicolumn{2}{c}{M1} & \multicolumn{2}{c}{M2} & \multicolumn{2}{c}{M3} & \multicolumn{2}{c}{M4} & \multicolumn{2}{c}{M5} \\
\multicolumn{1}{r}{Data from treatments$^{b}$:} & \multicolumn{2}{c}{all but NA} & \multicolumn{2}{c}{NA, OO, OM} & \multicolumn{2}{c}{NA, OO, MO} & \multicolumn{2}{c}{all} & \multicolumn{2}{c}{all} \\ \midrule
\multicolumn{11}{l}{\textbf{Independent Variables$^{c}$:}} \\
Treat.NA & -\hspace{3mm} &  & -0.0576*\hspace{2mm} & (0.0331) & 0.0398\hspace{3mm} & (0.0732) & 0.2280*** & (0.0178) & -8.177*** & (1.886) \\
Treat.NI & -2.382\hspace{3mm} & (2.341) & -\hspace{3mm} &  & -\hspace{3mm} &  & 0.0463\hspace{3mm} & (0.0365) & -7.407*\hspace{2mm} & (3.909) \\
Treat.OM & 1.986\hspace{3mm} & (2.041) & -0.1370*** & (0.0486) & -\hspace{3mm} &  & 0.0017\hspace{3mm} & (0.0218) & 2.560\hspace{3mm} & (2.852) \\
Treat.MO & 0.618\hspace{3mm} & (2.185) & -\hspace{3mm} &  & -0.0056\hspace{3mm} & (0.0710) & -0.0348\hspace{3mm} & (0.0221) & 6.836*** & (2.636) \\
Treat.MM & 0.182\hspace{3mm} & (1.818) & -\hspace{3mm} &  & -\hspace{3mm} &  & -0.0643*** & (0.0175) & 10.110*** & (2.112) \\
Round & 0.263*** & (0.065) & -0.0006\hspace{3mm} & (0.0015) & -0.0100**\hspace{1mm} & (0.0041) & 0.0001\hspace{3mm} & (0.0010) & 0.281**\hspace{1mm} & (0.113) \\
Round \# Treat.NA & -\hspace{3mm} &  & 0.0007\hspace{3mm} & (0.0016) & 0.0033\hspace{3mm} & (0.0045) & 0.0005\hspace{3mm} & (0.0010) & -0.369*** & (0.120) \\
Round \# Treat.NI & 0.303**\hspace{1mm} & (0.128) & -\hspace{3mm} &  & -\hspace{3mm} &  & 0.0061*** & (0.0015) & -0.630*** & (0.158) \\
Round \# Treat.OM & 0.041\hspace{3mm} & (0.081) & -0.0009\hspace{3mm} & (0.0022) & -\hspace{3mm} &  & -0.0011\hspace{3mm} & (0.0013) & 0.175\hspace{3mm} & (0.172) \\
Round \# Treat.MO & -0.129\hspace{3mm} & (0.082) & -\hspace{3mm} &  & 0.0032\hspace{3mm} & (0.0053) & -0.0009\hspace{3mm} & (0.0012) & -0.016\hspace{3mm} & (0.166) \\
Round \# Treat.MM & 0.116\hspace{3mm} & (0.080) & -\hspace{3mm} &  & -\hspace{3mm} &  & -0.0021**\hspace{1mm} & (0.0010) & 0.403*** & (0.123) \\
Ideology & 1.696*** & (0.415) & -0.0611*** & (0.0190) & 0.0020\hspace{3mm} & (0.0235) & 0.0066**\hspace{1mm} & (0.0029) & 1.546*** & (0.490) \\
Prosocial values & -3.934*** & (1.127) & 0.0588*\hspace{2mm} & (0.0355) & 0.1580*** & (0.0480) & -0.0177*** & (0.0061) & -4.071*** & (1.383) \\
Risk score & 0.093*** & (0.028) & -0.0006\hspace{3mm} & (0.0009) & -0.0017\hspace{3mm} & (0.0014) & 0.0004**\hspace{1mm} & (0.0002) & 0.079**\hspace{1mm} & (0.036) \\
Own activity @ t-1 & 0.468*** & (0.022) & -\hspace{3mm} &  & -\hspace{3mm} &  & -\hspace{3mm} &  & -\hspace{3mm} &  \\
Infected @ t-1 & 6.646*** & (1.042) & -\hspace{3mm} &  & -\hspace{3mm} &  & -\hspace{3mm} &  & -\hspace{3mm} &  \\
Alert @ t-1 & 4.434**\hspace{1mm} & (2.112) & -\hspace{3mm} &  & -\hspace{3mm} &  & -\hspace{3mm} &  & -\hspace{3mm} &  \\
Global activity @ t-1 & -0.356*** & (0.053) & -\hspace{3mm} &  & -\hspace{3mm} &  & -\hspace{3mm} &  & -\hspace{3mm} &  \\
Number of infected @ t & -1.178*\hspace{2mm} & (0.705) & -\hspace{3mm} &  & -\hspace{3mm} &  & -\hspace{3mm} &  & -\hspace{3mm} &  \\
Constant & 68.32*** & (4.995) & 1.1540*** & (0.2000) & 0.4630*\hspace{2mm} & (0.2520) & 0.2450*** & (0.0377) & 39.11*** & (6.722) \\
\midrule
Observations: & \multicolumn{2}{c}{17,246} & \multicolumn{2}{c}{10,353} & \multicolumn{2}{c}{859} & \multicolumn{2}{c}{21,199} & \multicolumn{2}{c}{21,199} \\
R-squared: & \multicolumn{2}{c}{0.292} & \multicolumn{2}{c}{0.103} & \multicolumn{2}{c}{0.109} & \multicolumn{2}{c}{0.070} & \multicolumn{2}{c}{0.028} \\
\midrule
\multicolumn{11}{l}{\multirow{6}{*}{\parbox{25cm}{Notes: 
Standard errors (reported in parentheses) are clustered at the group level. *** $p<0.01$, ** $p<0.05$, * $p<0.1$. Treatment OO is the baseline in all models.
\textbf{(a)} Activity and Welfare are Ordinary Least-Squares regressions, and the others are Linear Probability Models. All models use a two-stage least squares instrumental variable approach for the ideology variable. 
\textbf{(b)} Activity, tracing, and quarantine regressions only use data from treatments where subjects freely made the respective decision. The quarantine regression contains fewer observations because subjects only make this decision after receiving an alert. 
\textbf{(c)} All regressions include the following demographics controls: age, gender, race, years of schooling, marital and employment statuses, self-reported household income, region, and a proxy for how well subjects understood the instructions.}}} \\
\multicolumn{11}{l}{} \\ 
\multicolumn{11}{l}{} \\
\multicolumn{11}{l}{} \\
\multicolumn{11}{l}{} \\
\multicolumn{11}{l}{} \\ \bottomrule
\end{tabular}
}
\end{table*}

\subsection{Intervention effectiveness} Fig.\ref{fig2} shows the decisions and outcomes for each treatment over time. While only summary statistics, these line plots show all of the main results and are consistent with the regression analysis in Table \ref{tab:regressions}. 

The first finding is that making both tracing and quarantine mandatory works better than a system where both of them are optional. Fig.\ref{fig2}(d) shows that welfare is higher in treatment MM than in OO. Both non-parametric (MW, $p=0.0002$) and parametric (Table \ref{fig2} M5 $p=0.001$) analyses show that the difference is highly significant. There are two drivers of the superior performance of a fully mandatory program. First, Fig. \ref{fig2}(a) shows that MM has a lower rate of infections compared to OO (M4 $p=0.0002$, MW $p=0.0008$) because making tracing and quarantine mandatory takes infected subjects ``out of circulation'' and prevents them from spreading the disease. Second, Fig. \ref{fig2}(c) shows that mandatory tracing and quarantine have no discernible negative impact on the level of economic activity (M1 $p=0.92$, MW $p=0.73$) so there is no opportunity cost from making the programs mandatory. Additionally, welfare is rising at a \emph{faster} rate over time in treatment MM (M5 $p=0.001$). This is because a high rate of quarantining is \emph{maintained} in treatment MM, unlike in OO (Table \ref{tab:regressions}, M3).

The second finding is that having both tracing and quarantine optional is better than having no intervention at all. Fig.\ref{fig2}(d) shows that welfare is lower in treatment NI than in OO. Non-parametric analysis shows a highly significant difference (MW, $p=0.004$). Parametric analysis requires more careful interpretation. In the first round where COVID-19 is present (Round 6 of the experiment), the difference is only marginally significant (Table \ref{fig2} M5 $p=0.06$). However, treatment OO shows a strong positive time trend (M5 $p=0.01$) while treatment NI shows a strong \emph{negative} time trend (t-test $p=0.002$). This means that welfare is diverging over time. For example, after only 1 round (i.e. Round 7 of the experiment) the difference in welfare is significant at the 5\% level (t-test $p=0.04$), and at the 1\% level after 5 rounds (t-test $p=0.007$). By halfway through (Round 20 of the experiment), the point estimate for the difference in welfare \emph{doubles} (t-test $p=0.04$) compared to when COVID-19 was first introduced. Fig. \ref{fig2}(b) and \ref{fig2}(e) show that welfare improves because individuals decide to opt into the tracing and quarantine schemes, and this has no negative impact on economic activity. On average, $80\%$ of participants join the tracing scheme in OO, and, conditional on receiving an alert, $52\%$ of subjects decide to go into quarantine. Fig. \ref{fig2}(c) shows that the optional tracing and quarantine system has no downside in terms of reduced economic activity (M1 $p=0.31$, MW $p=0.62$).

Leaving tracing optional and making quarantine mandatory is clearly inferior to having both mandatory, and it does not provide additional benefits compared to a system with both being optional. As Fig. \ref{fig2}(d) shows, welfare is significantly lower in OM compared to MM (M5, $p=0.002$, MW, $p=0.0004$). The reason is that fewer participants in OM participate in the tracing scheme compared to, trivially, MM (MW $p=0.0001$), but also compared to OO where both tracing and quarantine are optional (M2 $p=0.005$, MW $p=0.05$). Conditional on receiving an alert, more subjects (trivially) go into quarantine in OM compared to OO (MW $p=0.0001$), but this effect does not dominate the inferior participation in the tracing scheme resulting in no difference in welfare between OO and OM (M5 $p=0.37$, MW $p=0.34$).
 
A system with mandatory tracing and optional quarantine is also inferior to a fully mandatory system, and provides at most mixed benefits compared to when both of them are optional. As Fig. \ref{fig2}(d) shows, welfare is significantly lower in MO compared to MM (MW, $p=0.0004$). Although the difference is not significant in the parametric analysis (M5, t-test $p=0.10$), there is a strong positive time trend for treatment MM that is absent for treatment MO. For example, by Round 20, the difference in combined point estimates is statistically significant (t-test $p<0.0001$). Making tracing mandatory in MO does not improve welfare compared to OO (MW $p=0.08$). Even if the difference in welfare between MO and OO is significant in the parametric analysis (M5 $p=0.009$), in the long-term this treatment effect may not be robust because standard errors are increasing over the rounds (see SI for further discussion).

Treatment NA provides a benchmark where subjects cannot reduce their economic activity. It helps isolate the relative importance of social distancing (i.e. reductions in economic activity), and the tracing and quarantine programs. Treatments NA and NI are indistinguishable in terms of welfare (Fig.\ref{fig2}(d), M5 $p=0.82$, MW $p=0.14$) and both lower than treatment OO (NA vs OO M5 $p<0.0001$, MW $p=0.002$; NI vs OO M5 $p=0.06$, MW $p=0.004$). This suggests that the effect of tracing and quarantine is similar in magnitude to that of voluntary social distancing.

\subsection{Behavioral implications of intervention}
Beyond implications for overall welfare, the existence of the tracing and quarantine programs also induces noteworthy changes in individual behavior. These behaviors are driven by individuals acting in their self-interest, and have harmful social consequences. 

First, some subjects use the contact tracing scheme to ``game the system''. When an individual is infected in Round $T$, they know they are healthy in Round $T+1$ because they cannot have contracted COVID-19 in the previous round. A consequence is that high activity in $T$ provides additional benefits and has no downside for the individual who cannot get infected, but it is very harmful for society because the infected individual is more likely to infect others. We observe that subjects who do not go into quarantine choose \emph{higher} economic activity in the round after receiving an alert when they know they are infected for sure (M1 $p=0.04$). In other words, they game the system because they sign up to the tracing scheme to learn when they are infected so that they can go about their activities with no risk of getting infected, rather than for its purpose of knowing when they should quarantine so that they do not put others at risk.

Second, mandatory quarantine in the OM treatment dissuades subjects from participating in the contact tracing scheme. Subjects in the OM treatment are less likely to sign up to tracing compared to subjects in the OO treatment (M2 $p=0.005$). This is in agreement with the prediction of a self-interested agent framework, as mandatory quarantine makes participation in the contact tracing scheme in OM tantamount to choosing to quarantine, and therefore giving up on any benefit of activity once they receive an alert.

\subsection{Individual characteristics} Social values and risk preferences are associated with decisions in a highly intuitive manner. \emph{Prosocial} subjects do less economic activity (Table \ref{tab:regressions}, M1 $p=0.0005$) and are much more likely to go into quarantine when given the opportunity (M3 $p=0.001$). Risk-seeking preferences are associated with choosing higher economic activity (M1 $p=0.001$) and are not associated with quarantine decisions. This should not be surprising. Activity yields certain benefits and entails risky costs, while quarantine yields a certain outcome.

We include a battery of self-reported demographic information in all regressions in Table \ref{tab:regressions}. These demographics are not significantly associated with any of the decisions or outcomes, and so are omitted from Table \ref{tab:regressions}. Quantitatively, the most important individual characteristic is political ideology.

\subsection{Ideology} 
\begin{figure}[t]
\centering
\begin{subfigure}[]{0.5\textwidth}
  \includegraphics[width=\linewidth]{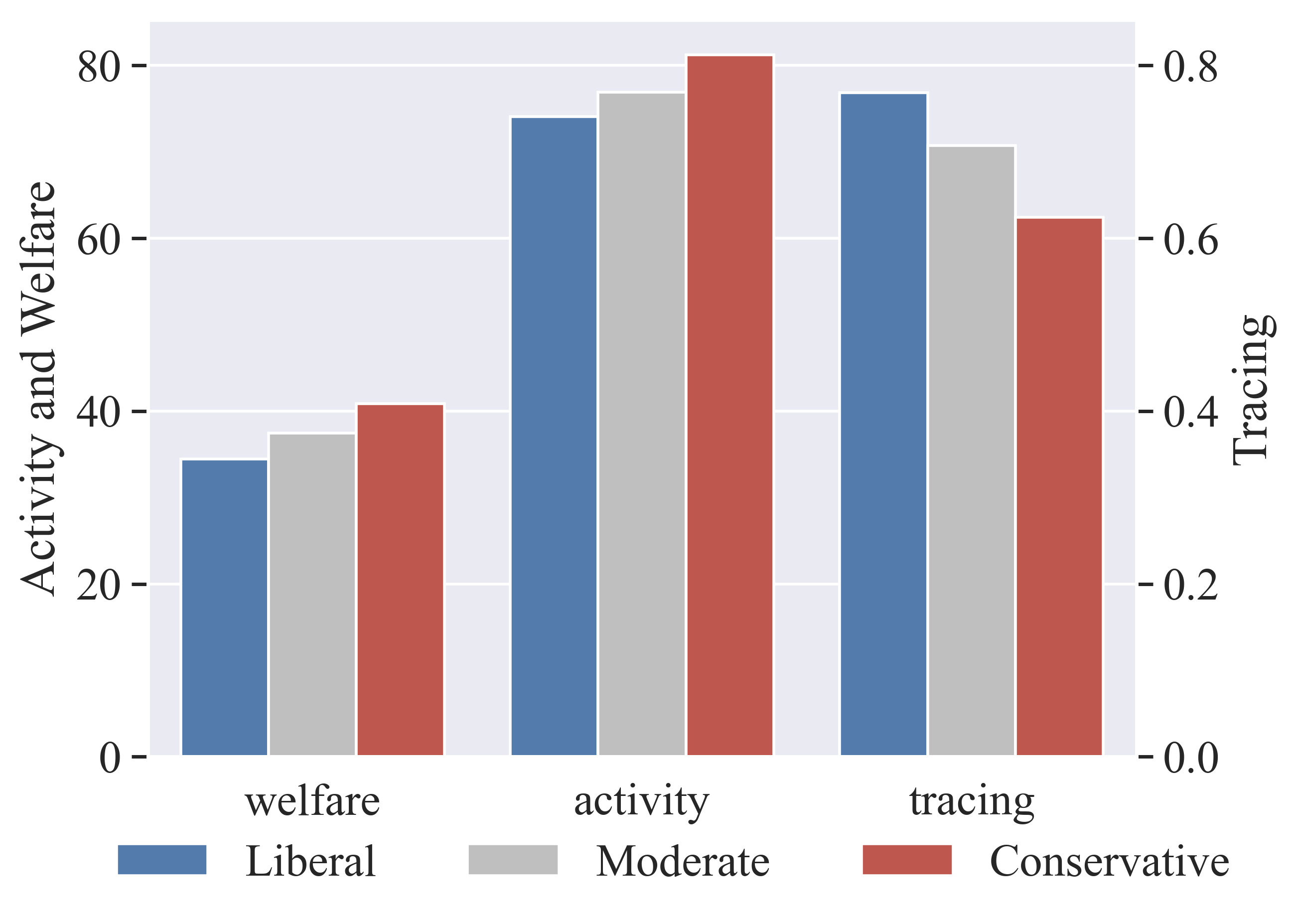}
  \caption{}
\end{subfigure}%
\begin{subfigure}[]{0.5\textwidth}
  \includegraphics[width=\linewidth]{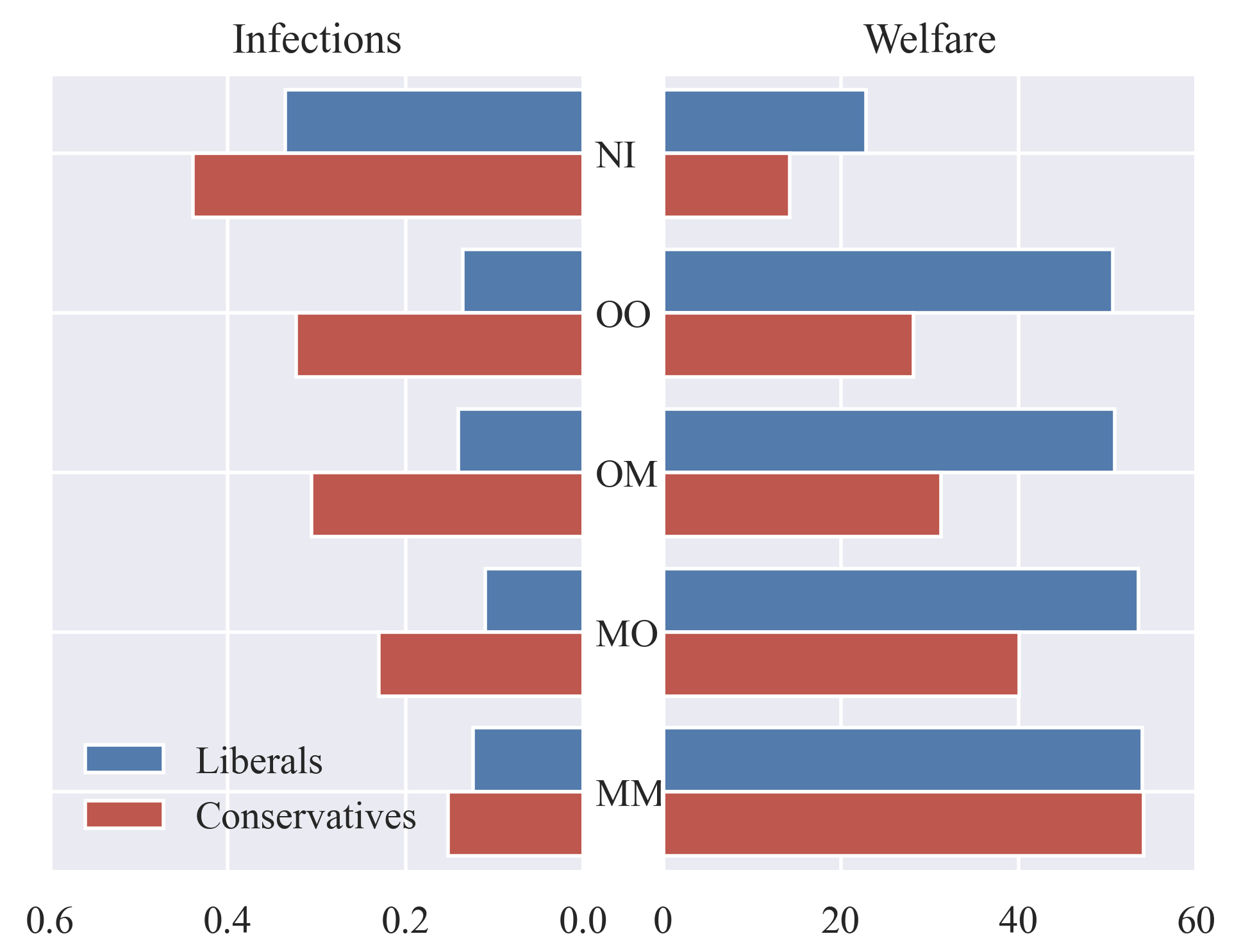}
  \caption{}
\end{subfigure}

\caption{\textbf{(a)} Summary statistics for activity and welfare (left-hand axis), and tracing (right-hand axis) from experimental data, split by self-reported political ideology. \textbf{(b)} Simulated average outcomes (infections and welfare) for ideologically homogeneous groups.}
\label{fig3}
\end{figure}

Fig.\ref{fig3}(a) show that political ideology is strongly associated with subjects' decisions and outcomes. Self-reported conservative participants sign up to tracing $62\%$ of the time, while liberals have a $77\%$ average sign-up rate. When it comes to activity, conservatives select an average of $81$ compared to $74$ for liberals. These choices mean that conservatives are on average better off than liberals in the experiment -- they earn an average of $6$ points per round more than liberals -- because their benefits from increased activity more than offset their higher risk of infection.


These results are consistent with existing evidence showing that ideology can affect the extent to which people engage in social distancing \cite{allcott2020polarization, gollwitzer2020partisan,fines} and tracing \cite{zhang2020americans}. However, ideology is clearly not randomly assigned, so we cannot attach a causal interpretation to the differences observed in our experiment because there may be an unobserved variable influencing both ideology and decisions. We use an Instrumental Variable (IV) approach (see Methods) to show that ideology is a \emph{causal} determinant of behavior. Conservative ideology causes subjects to choose higher economic activity (M1 $p<0.0001$), and makes them less likely to sign up to tracing (M2 $p=0.001$). There is no evidence that it affects subjects' propensity to go into quarantine conditional on receiving an alert (M3 $p=0.93$). The consequence of these behavioral differences is that conservatives are more likely to get infected (M4 $p=0.02$), but experience higher overall welfare (M5 $p=0.002$). This is because the losses from more frequent infection are more than offset by the gains from higher economic activity.

A naive interpretation of these results is that conservatives are better at coping in a world with COVID-19. This may, however, simply happen because they free-ride on liberals' willingness to join tracing and lower their economic activity. What would be the effectiveness of different policies in an ideologically homogeneous group? We carry out a set of simulations to explore this question. In particular, we use the experimental data to calibrate the behavior of a representative conservative/liberal participant in each treatment, and then simulate the evolution of decisions and outcomes for an ideologically homogeneous group (see Methods for details).

Fig.\ref{fig3}(b) reports the average simulated probability of being infected and welfare for ideologically homogeneous groups in each treatment. A liberal group experiences a higher welfare than a conservative group in all treatments (MW power $=1.0$) with the exception of MM (MW power $=0.09$). The reason is that contact tracing uptake is significantly higher in liberal groups compared to conservative ones whenever it is optional (MW power $=1.0$). As a result, liberal groups are able to achieve and sustain lower levels of infection, leading to overall higher welfare. Throughout the pandemic, even at its' early stages, opposition to mandatory tracing and quarantine was much stronger among the Conservatives in the US \cite{UTX2020}. Our simulations suggest that, paradoxically, mandatory programs are exactly what is needed in a world populated by conservatives to avoid a lower welfare outcome. 

\section{Policy Implications}

A widespread narrative to oppose tracing and quarantine programs is that they damage the economy \cite{FT2021}. Our results suggest that fears of significant economic collateral damage from tracing and quarantine programs are unfounded. Independent of the specific set-up of the contact and tracing program, economic activity is unchanged compared to a no intervention benchmark in our experimental environment. Infection rates are, however, significantly higher when there is no intervention compared to setting up any type of contact and tracing system, which is consistent with a large body of evidence showing that these programs reduce infections \cite{kerr2021controlling, aleta2020modelling, wymant2021epidemiological, R_2021}. There is, therefore, a significant upside and a limited downside to setting up a tracing and quarantine system.

The details of the program design matter for its effectiveness. Making both tracing and quarantine mandatory is more effective than a fully optional or a hybrid program where only either tracing or quarantine is mandatory. In contexts where a fully mandatory program is not politically palatable, our results suggest that the benefits of a hybrid program are at best very limited so there is little gain from pushing a part of the program to be mandatory. Optimizing the design of the program is clearly an area where further interactive online experiments are needed. For instance, it would be interesting to explore how varying the accuracy of the tracing program affects our results.

In the US context, we examine how political ideology affects the \emph{relative} efficacy of different policy decisions. Simulations based on our experimental results suggest that the benefits of optional tracing and quarantine programs may be greater in liberal communities. Further, they suggest that the additional benefits of enforcing participation in the programs is substantial in conservative communities, but modest in liberal ones. Survey evidence points towards greater resistance to mandatory participation in some conservative communities \cite{mcclain2020challenges}, but policymakers should weight this against the likely greater benefits. While political ideology stands out compared to other sociodemographic factors as a driver of decisions in the US, this may not be the case in other countries. A promising avenue for future experimental work is to investigate which sociodemographic factors affect the effectiveness of tracing and quarantine programs in other countries to design policies that are optimized for a specific cultural context.

Our study also highlights the important role interactive web-based experiments can play in providing timely initial information to policymakers to compare the effectiveness of different policy designs. Experiments are the gold standard to test the causal impact of policy interventions, but lab and field experiments were severely constrained during the COVID-19 pandemic \cite{HM_2020}. Web-based experiments, however, were unaffected -- they can be deployed quickly, scaled up at minimal cost, and easily reach a diverse sample as well as specific populations of particular interest. They can play a complementary role to fieldwork by providing a testing environment to narrow down the options for pilot tests in the field. 

\section{Methods}

Full details on methods -- theoretical framework, data collection, data analysis, and simulations, -- along with a detailed description of the dataset are in the SI.\\

\noindent \textbf{Ethical approval.} This research received ethical approval for the use of human subjects from the Faculty of Economics Ethical Committee at the University of Cambridge (approval no. UCAM-FoE-21-01). Informed consent was obtained from all subjects before participation.\\

\noindent \textbf{Pre-registration.} The design of this study was pre-registered on Aspredicted. A private copy of pre-registration form is available upon request.\\

\noindent \textbf{Recruitment and sessions.}
We recruited a US-resident standing panel through Amazon Mechanical Turk that is representative of the US adult population in terms of sex and region of residence. Recruitment involved a short survey collecting basic demographic information and a simple qualification task to check potential subjects' understanding of basic probability. Recruitment survey was hosted on Qualtrics. We randomly invited members of the panel to a total of 22 sessions conducted between June 14th and July 22nd, 2021, with a total of 731 subjects across 61 groups. Each session had between one and four independent groups of 12-14 subjects playing simultaneously. The experiment lasted on average 32.8 minutes (st.d. 5.4 minutes), and subjects earned an average of \$13.54 USD per hour (st.d. \$4.4). Subjects remained completely anonymous throughout the recruitment survey and the experiment, and repeated participation was not allowed.\\

\noindent \textbf{Software.} The experiment was coded in oTree (v3.3.7) \cite{CSW_2016} with a server hosted on Heroku. \\

\noindent \textbf{Infinite populations to finite-size groups.} Our implementation of the SIS model for discrete populations relies on the following procedure. When determining the number of infected agents, we calculate the expected number of infections and round the resulting number to the nearest integer. We then randomly determine who gets infected. Similarly, when determining the number of alerts sent by the contact tracing scheme, we round the expectation to the nearest integer, and then randomly send alerts. See SI for details.\\

\noindent \textbf{Statistical analysis.} Non-parametric analysis uses a two-sided Mann-Whitney U-test (MW) \cite{MW_1947}. Parametric analysis uses Ordinary Least Squares (OLS) regression where the dependent variable is approximately continuous, and a Linear Probability Model (LPM) where the dependent variable is binary \cite{G_1964, A_1984}. Additionally, we use an Instrumental Variables (IV) approach to investigate the effect of ideology. We use survey responses to four questions concerning partisan issues in current US politics that are unrelated to COVID-19 or medicine/science more generally to create an index (see SI for full text of these questions).\\

\noindent \textbf{Preferences elicitation.} 
We elicit social values using the Social Values Orientation scale~\cite{MAH_2011} and risk preferences using the `Bomb' risk elicitation task~\cite{CF_2013}. Both tasks are incentivized.\\

\noindent \textbf{Simulations.} We generate 10 groups of 12 agents, and simulate their behavior (activity level, tracing and quarantine choices where relevant) across all 6 treatments. Our simulations exclude demographic characteristics and other variables that are insignificant in models M1-M3 in Table \ref{tab:regressions} at 5\%. The scores are drawn uniformly at random, using the distribution from our dataset. Similarly, other significant characteristics (e.g. risk score) are drawn randomly from distributions of our dataset. For each ideology, we repeat this exercise 1,000 times. In our analysis, we report the estimates of power of hypothesis tests such that the null hypothesis of no difference between outcomes for the two ideological groups is false at 5\% significance level. Estimated power is the share of times (out of 1,000) the p-value of the relevant test is less than $5\%$. The test is a two-sided Mann-Whitney U-test throughout. In line with standard practices, we consider 80\% power to be sufficiently strong evidence of a rejection of the null hypothesis \cite{C_1992}.

\clearpage
\bibliographystyle{ieeetr}
\bibliography{bibliography}

\clearpage

\section*{Data and code availability statement} 

The code for the experiment and the analysis, together with the dataset and further robustness checks are available upon request. 

\section*{Acknowledgements}
We are grateful to Matthew Elliott for helpful comments, and to Ilia Shumailov for technical assistance. 

\section*{Author Contributions}
Experiment design: DB, EG, AL \\
Data collection: DB, AL \\
Analysis: DB, EG, AL \\
Writing paper: DB, EG, AL

\section*{Competing Interests Statement}
All authors declare they have no competing interests.

\section*{Funding}
Cambridge Endowment for Research in Finance: EG \\ 
Economic and Social Research Council award reference ES/P000738/1: AL \\ 
Economic and Social Research Council award reference ES/J500033/1: DB \\
Morgan-Jones Fund (Trinity College, Cambridge): DB 

\end{document}


\maketitle

\renewcommand{\thefigure}{S\arabic{figure}}
\renewcommand{\thetable}{S\arabic{table}}
\setcounter{figure}{0}

This Appendix contains additional information. \Cref{sec:model} sets out the theoretical framework, specifies treatments and parameterization used in our experiment, and presents solutions to the model using simulations. \Cref{sec:methodology:data_collection} describes the methodology of data collection, including subject recruitment, and the workflow and implementation of the actual experiment. Next, \Cref{sec:dataset} presents our dataset, while \Cref{sec:methodology:data_analysis} contains a summary of the tools we use to analyze experimental data. Results are presented in \Cref{sec:analysis}. In \Cref{sec:simulations}, we present the methodology and results of a set of simulations calibrated on experimental data. Finally, sample instructions for the experiment are in \Cref{sec:instructions}.


\section{Theory} \label{sec:model}
We begin with the canonical Susceptible-Infected-Susceptible (SIS) model of epidemics in discrete time \cite{kermack1927contribution, allen1994some}. In this simplest version of the model, the population is fixed (there are no births or deaths), infection lasts for one round and confers no immunity. It is characterized by the single equation $I_t /N = (1 - I_t / N) \ R$ and the initial condition $I_0$, where $I_t$ is the number of agents \emph{infected} at time $t$, $N$ is the total number of agents, and $R$ is the ``infectiousness'' (or basic reproduction rate, or $R_0$) of the disease.\footnote{Most expositions assume that the number of agents is large (and so working with fractions of the population is more convenient). This allows a single agent's impact on aggregate variables, and any issues rounding to an integer, to be ignored. For clarity of exposition, we assume a large number of agents when setting out the theoretical model. We show simulations which do account for rounding to whole numbers for the parameterized model.}

We then augment this SIS model to include \emph{economic activity}. Let $a_{jt} \in [0,100]$ be the economic activity of agent $j$ at time $t$, and let $A_t \in [0,100]$ be the mean activity at time $t$. We assume that it affects the epidemic dynamics in the following way:
\begin{align}\label{eq:disease dynamics}
    \frac{I_{t+1} }{N} = \left(1 - \frac{I_t}{N}\right) R \left(\frac{A_t}{100}\right)^{\gamma}
\end{align}

Intuitively, a reduction in economic activity reduces the number of agents infected in the next round by reducing the number of contacts, and so the opportunity for the disease to spread. The parameter $\gamma$ governs the shape of this relationship. For $\gamma > 1$, the relationship is convex. This means there are ``increasing returns'' to activity in terms of disease spread. A given increase in economic activity has a larger impact on the spread of the disease when activity is high to begin with. Conversely, $\gamma < 1$ implies ``decreasing returns'' (and the logic is reversed).

The probability that an individual agent becomes infected varies linearly with her own economic activity. Therefore: 
\begin{align}\label{eq:individual infection}
Pr(I_{jt} = 1 | I_{jt-1} = 0) = \frac{I_t}{N - I_t} \frac{a_{jt}}{A_{t}}
\end{align}
where $I_{jt} \in \{ 0,1 \}$ indicates whether agent $j$ is infected at time $t$. Each agent's marginal benefit of economic activity is $\pi > 0$. They also bear a cost $\kappa > 0$ when they are infected. Agents discount the future at a rate $\beta \in [0,1]$. Therefore, agent $j$ has utility:
\begin{align}\label{eq:utility}
    u_j = \sum_{t=1}^{\infty} \beta^t (\pi a_{jt} - \kappa I_{jt})
\end{align}

Further, we introduce a \emph{contact tracing} scheme, and a \emph{quarantine} program. The contact tracing scheme is free for agents to use, and provides information to an agent as to whether they are infected in the next round. If an agent $j$ participates in the contact tracing scheme, she receives a signal $C_{jt} \in \{ 0,1 \}$ such that:
\begin{align}\label{eq:tracing info}
    Pr(C_{jt} = 1 | I_{jt+1} = 1) = \phi \ , \ Pr(C_{jt} = 1 | I_{jt+1} = 0) = 0
\end{align}
where $\phi$ is the \emph{efficacy} of the contact tracing scheme (the proportion of infections that are picked up by the scheme). Note that we assume the scheme generates no false positives.\footnote{Consequently, $Pr(I_{jt+1} = 1 | C_{jt} = 1) = 1$ and, by Bayes Theorem, $Pr(I_{jt+1} = 1 | C_{jt} = 0) = \frac{(1 - \phi) Pr(I_{jt+1} = 1)}{ 1 - \phi Pr(I_{jt+1} = 1)}$. This is derived by the law of total conditional probability.} 

An agent can only enter the quarantine program for round $t+1$ if she receives a signal $C_t = 1$ (which in turn requires that she participated in the contact tracing scheme at time $t$). If agent $j$ enters the quarantine program for round $t+1$, then $a_{jt+1}$ is set to zero. Due to the operation of the model, if agent $j$ enters quarantine for round $t+1$ then she: (i) must be infected during $t+1$, (ii) cannot be infected during round $t+2$, (iii) cannot cause any other agent to be infected in round $t+2$. Both the contact tracing scheme and the quarantine program can be optional (individual agents can choose for themselves), mandatory, or unavailable. We abstract from enforcement issues, and assume that when these are mandatory all agents do in fact participate.\footnote{Of course complete participation is unlikely to occur in reality, but is the natural benchmark when examining the effects of mandatory contact tracing schemes and quarantine programs.} \\

\noindent \textbf{Treatments.} The experiment focuses on the impacts of the contact tracing scheme and the quarantine program. In our four main treatments, both are present (we can think of them as being provided by a government). Each of the contact tracing scheme and the quarantine program can be optional or mandatory, and these four treatments cover all possible combinations.

We have two further treatments which act as benchmarks to help uncover the impact of contact tracing and quarantine. In the fifth treatment, agents still choose their economic activity, but the contact tracing scheme and the quarantine program are unavailable. This provides a no intervention benchmark against which we can compare treatments where contact tracing and quarantine are present. The final treatment shuts down the endogenous activity channel, and instead fixes activity at 100 (except when an agent is in quarantine, when activity is set to zero).\footnote{Note that average activity in the group can still fluctuate, but only due to quarantine.} This helps to isolate the effect of the contact tracing scheme and quarantine program from the effect of endogenous activity choices.

\Cref{tab:treatments list} sets out these six treatments for reference. For the remainder of the paper, we will refer to each treatment by the codes shown in the first column.\\

\begin{table}[ht]
\centering
\caption{List of treatments}\label{tab:treatments list}
\begin{tabular}{ccccc}
\toprule
 &
  \begin{tabular}[c]{@{}c@{}}Contact Tracing \\ scheme\end{tabular} &
  \begin{tabular}[c]{@{}c@{}}Quarantine\\ program\end{tabular} &
  \begin{tabular}[c]{@{}c@{}}Activity\end{tabular} &
  \textbf{} \\
  \midrule
\textbf{NI} & unavailable & unavailable & chosen & \\
\textbf{OO} & optional & optional & chosen & \\
\textbf{MM} & mandatory & mandatory & chosen & \\
\textbf{MO} & mandatory & optional & chosen & \\
\textbf{OM} & optional & mandatory & chosen & \\
\textbf{NA} & optional & optional & fixed & \\ 
\bottomrule
\end{tabular}
\end{table}

\noindent \textbf{Parameterization.} We now set out the parameter values we use in the experiment. We set $\gamma$, the parameter governing the relationship between economic activity and infections in the next round, to $2$, and basic reproduction rate, $R$, to $3$. We set the benefit of each unit of economic activity, $\pi$, equal to $1$, and the cost of being infected, $\kappa$, equal to $150$. We also set $\beta$ to $1$, and the efficacy of the contact tracing scheme, $\phi$, to $1/3$. Finally, we set $N$ between $12$ and $15$ (the exact value varies across groups in the experiment) and $I_0$ equal to $4$ (or 5 when $N$ is 14 or 15).\footnote{This ensures that the fraction of initially infected is approximately the same.}

An additional feature of our parameterization is that we only introduce initial infections and the tracing and quarantine programs after 5 rounds (rather than before the first round as is standard in SIS-style models). This is an experimental design consideration and has no impact on the theoretical model. For the first 5 rounds, the theory predicts that all agents choose activity equal to 100, and nothing else happens.

\begin{table}[ht]
\centering
\caption{Parameter values}
\label{tab:parameter values}
\begin{tabular}{@{}cll@{}}
\toprule
Parameter & Value & Comment \\ \midrule
$\gamma$ & $2$ & convexity parameter \\
$R$ & $3$ & basic reproduction rate of COVID-19 \\
$\pi$ & $1$ & marginal benefit of economic activity \\
$\kappa$ & $150$ & cost of infection \\
$\beta$ & $1$ & discount factor \\
$\phi$ & $1/3$ & efficacy of contact tracing system \\
$N$ & $\in \{12,13,14,15\}$ & size of the group \\
$I_0$ & $\in \{4,5\}$ & number of initially infected agents \\ \bottomrule
\end{tabular}%
\end{table}

\begin{figure}[!ht]
\centering
\begin{subfigure}{.5\textwidth}
  \includegraphics[width=\linewidth]{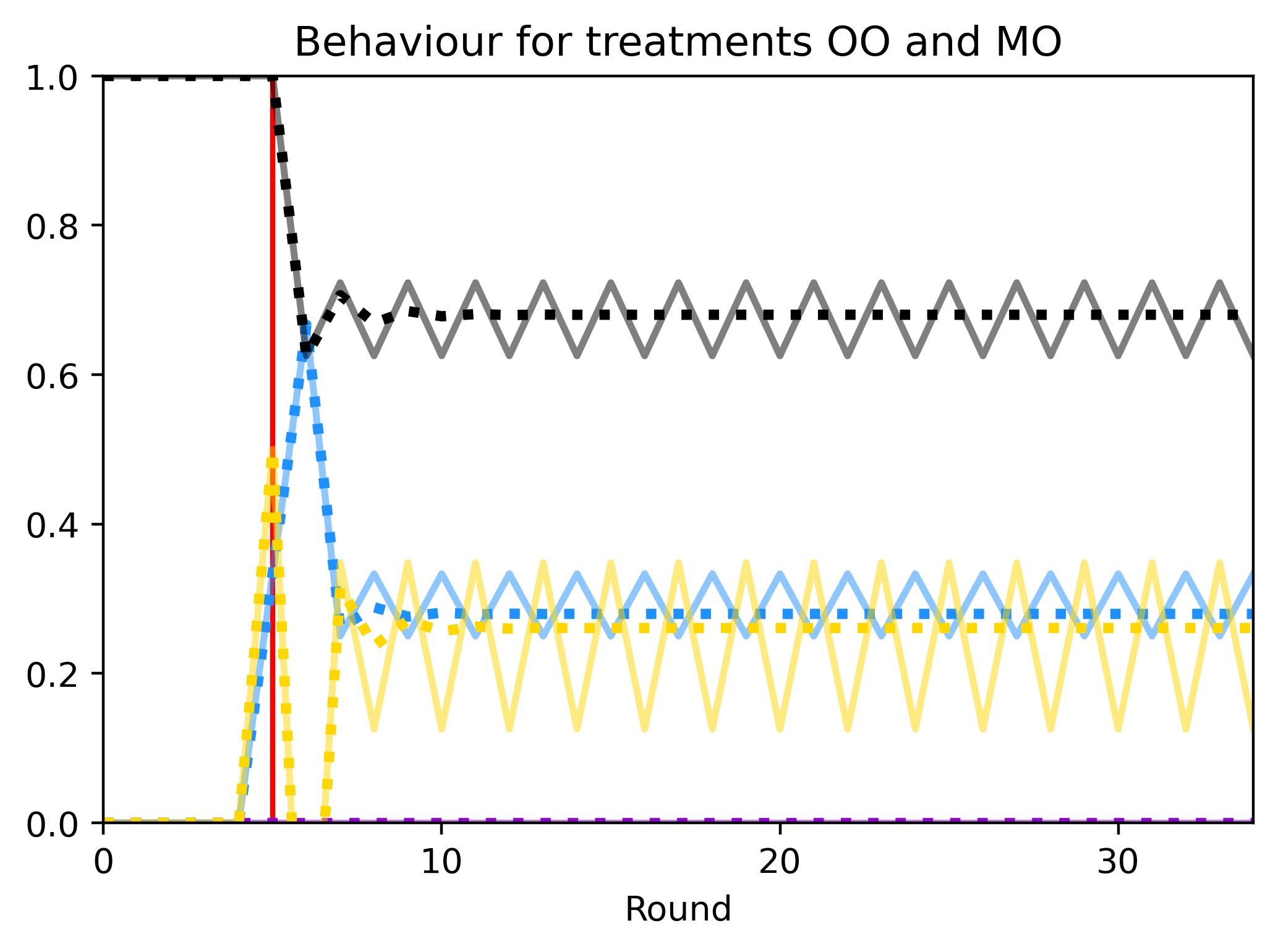}
\end{subfigure}%
\begin{subfigure}{.5\textwidth}
  \includegraphics[width=\linewidth]{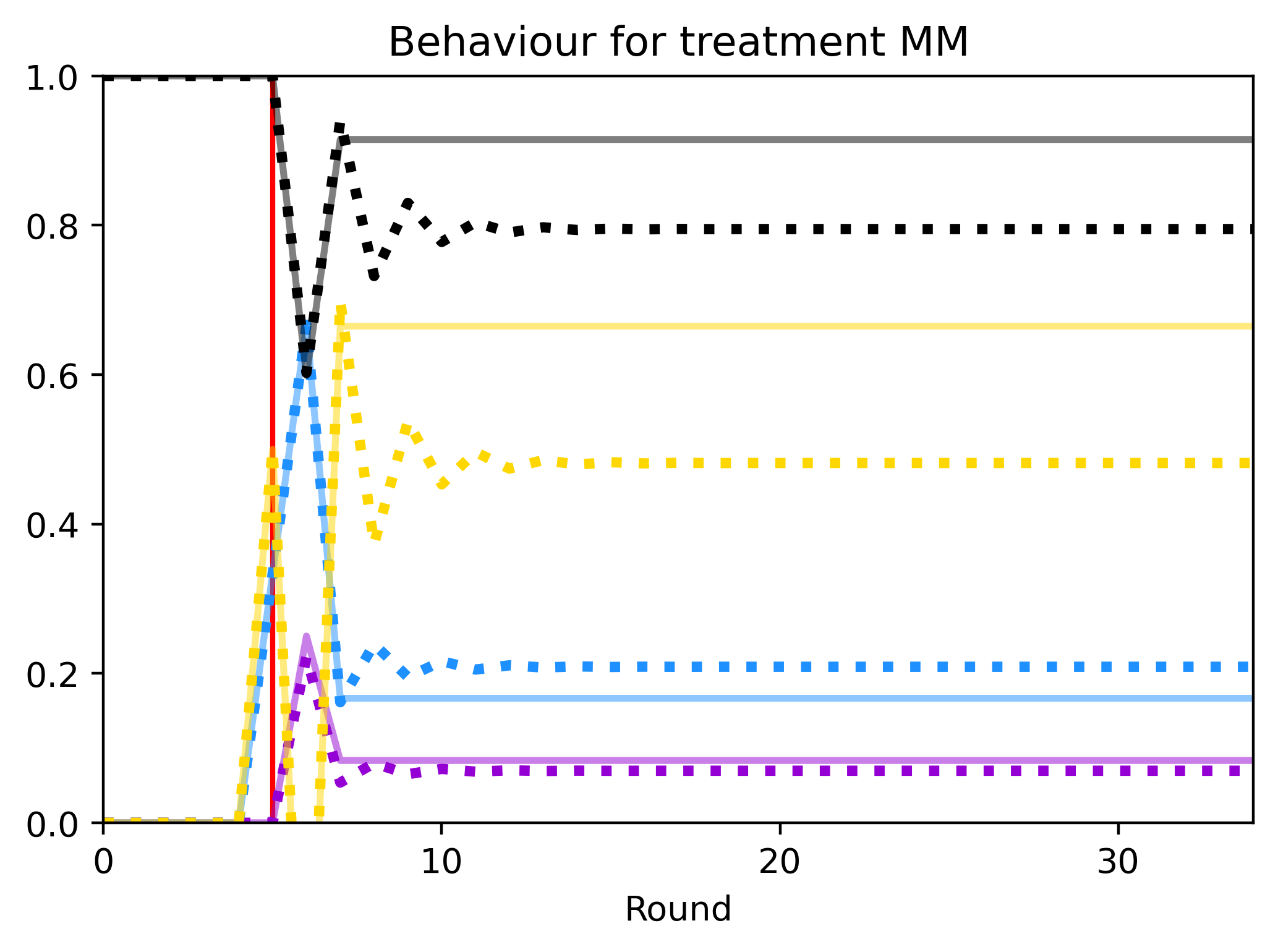}
\end{subfigure}

\hfill
\vspace{5mm}
\centering
\begin{subfigure}{.5\textwidth}
  \includegraphics[width=\linewidth]{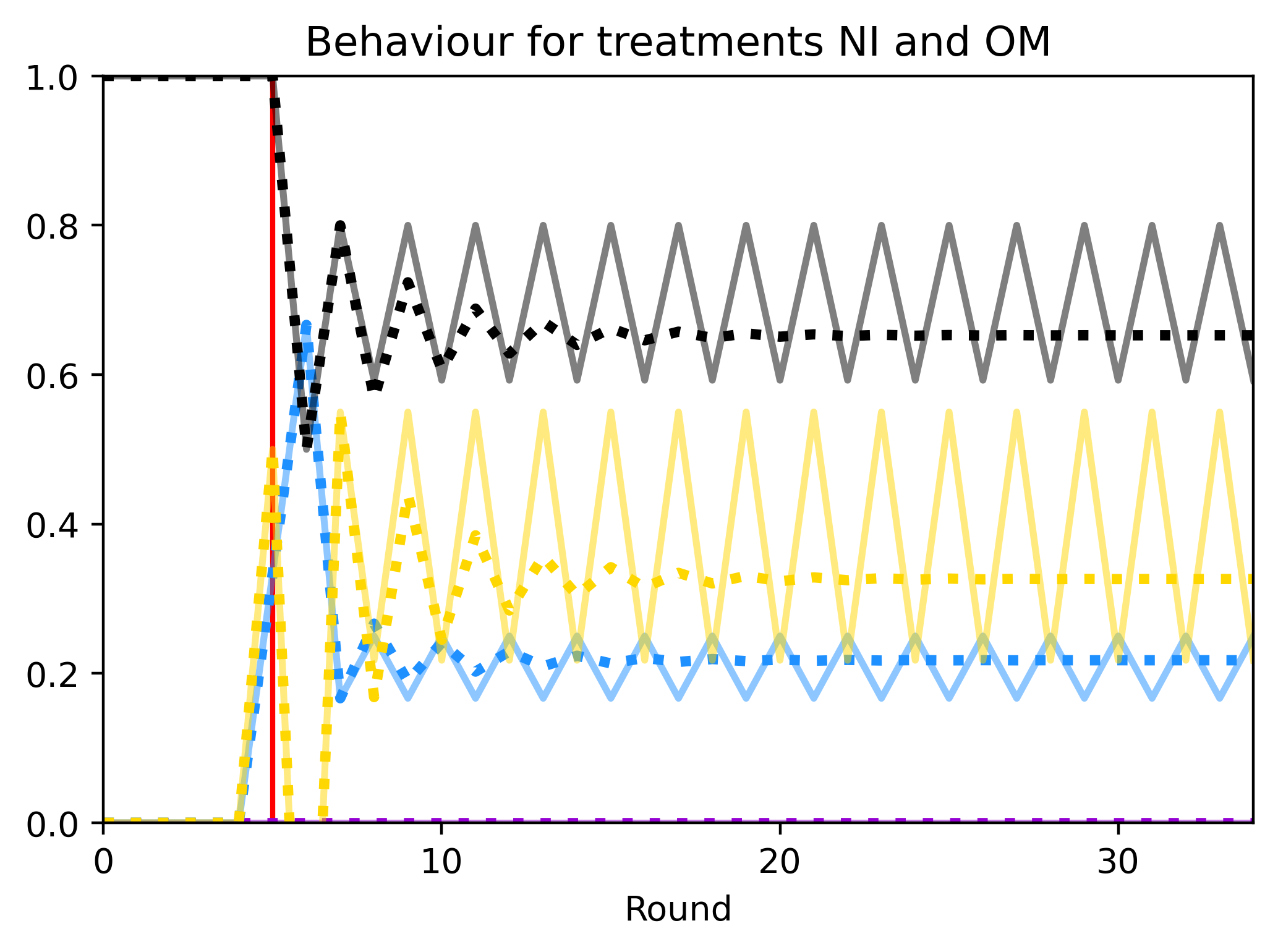}
\end{subfigure}%
\begin{subfigure}{.5\textwidth}
  \includegraphics[width=\linewidth]{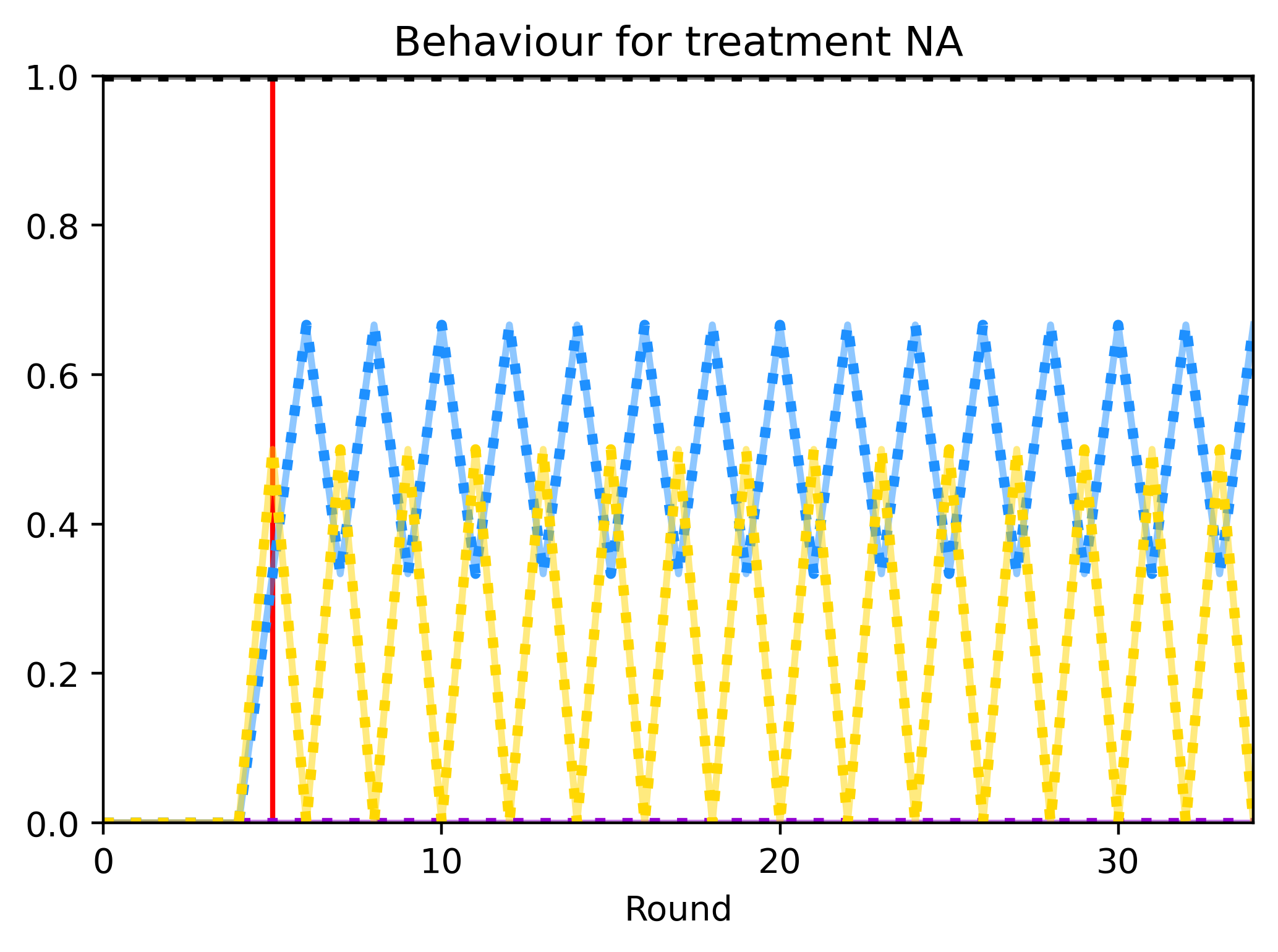}
\end{subfigure}

\hfill
\vspace{5mm}
\centering
\begin{subfigure}{\textwidth}
  \centering
  \includegraphics[width=\linewidth]{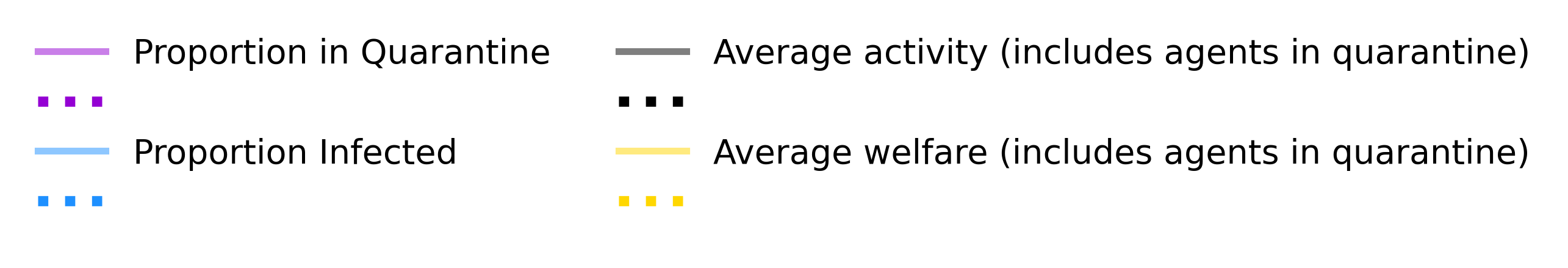}
\end{subfigure}

\caption{Simulated behavior and outcomes. Welfare and activity are scaled by $0.01$ so they share an axis with tracing and quarantine. Solid lines for finite population, dashed lines for infinite population. Red vertical line denotes introduction of COVID-19. $N = 12, I_0 = 4$, other parameters as in Table \ref{tab:parameter values}. }
\label{fig:simulated play}
\end{figure}

\noindent \textbf{Simulations.} Closed form solutions generally do not exist for this model, even when we assume that the number of agents is large. Therefore we show simulated behavior, focusing on the case where agents play symmetric Nash equilibria. Figure \ref{fig:simulated play} shows the simulated behavior in each of the treatments. Solid lines show the simulated play for a group of 12 agents. Dotted lines show simulated play for an infinite population. Average decisions/outcomes are typically the same (or extremely similar) in the 12-agent group and in the infinite population. The exception is infections and welfare in treatment MM. 

These discrepancies between the simulated values for an infinite populate and the 12-agent group are due to rounding issues. With only 12 agents, rounding leads to exactly 1 agent in quarantine in the steady state, rather than $7\%$ of an infinite population (which would translate to 0.8 out of 12 agents). This permits higher activity by agents not in quarantine without a material difference in the level of infection. In turn, this leads to higher welfare in the 12-agent group (compared to the infinite population).
In these simulations, and also in the actual experiment, we calculate the number of infected agents as follows. First, we calculate the expected number of infected agents as in Equation \ref{eq:disease dynamics} (with parameterization set out in Table \ref{tab:parameter values}). We then round the resulting number to the nearest integer and allocate agents to infected status. The probability that an individual agent will become infected is determined as in \Cref{eq:individual infection}. Similarly, when we determine the number of agents that receive an alert from the contact tracing scheme, we calculate the expected number of such agents (using the parameterization in Table \ref{tab:parameter values}) and round it to the nearest integer. Alerts are then sent uniformly at random to agents who are signed up to the contract tracing scheme and are infected in the next round. Notice that the difference between simulated outcomes for infinite populations and fixed-sized group becomes smaller as we increase the number of agents in the group, because the effects of the rounding procedures vanish.

Notice that simulations for treatments OO and MO are identical, since the theory predicts that agents will always participate in the contact tracing scheme (regardless of whether or not they are obliged to) and will never go into quarantine (when it is optional). This is because tracing here provides some useful information without any costs -- so rational and self-interested agents choose to receive that information. Similarly, treatments NA and OM are identical. Self-interested agents do not want to go into quarantine (as this amounts to providing a public good), and therefore they do not participate in the tracing program in treatment OM in order to avoid quarantine. This leaves outcomes the same as treatment NA where the programs are unavailable. In either case, no agent uses them.

\section{Methodology: Data collection} \label{sec:methodology:data_collection}
There are two concurrent data collection phases in this study -- recruitment for the experiment, and the experiment itself. \Cref{sec:methodology_exp_recruitment} outlines the workflow of our recruitment survey and explains its implementation. \Cref{sec:methodology_exp_workflow} describes the workflow of a typical experimental session. \Cref{sec:methodology_exp_implementation} presents implementation details for the experiment. 

\subsection{Recruitment} \label{sec:methodology_exp_recruitment}
We recruited subjects from Amazon Mechanical Turk (MTurk; https://www.mturk.com) using a survey hosted on Qualtrics (https://www.qualtrics.com). The survey was visible to US-residents who have completed at least 500 Human Intelligence Tasks (HITs) on MTurk and have had a ranking of at least 95\%.

During recruitment, we collect subjects' basic demographic information, including: age, gender, race, place of residence, political party affiliation, ideological leaning and self-reported risk-preferences. We also inform subjects about and collect consent for participation in our interactive experiments. Next, subjects pass a qualifying quiz to complete a bonus task. The qualifier is 1 of 3 simple probability questions drawn uniformly at random. The idea of the quiz is to check that subjects display a basic level of mathematical ability. 90.2\% of subjects who completed our recruitment survey passed the quiz.

Subjects are given 3 attempts to answer the quiz correctly, and then, if they are successful, complete the Social Value Orientation (SVO) scale as a bonus task. We use Murphy, Ackermann and Handgraaf version of the task \cite{MAH_2011}. The idea behind the task is that peoples' social preferences vary in a systematic way. On a practical level, the task amounts to deciding how to allocate money between oneself and an anonymous other person, and there is a total of 6 independent decisions to make. The decisions are then used to classify subjects as belonging to one of four categories -- individualistic, prosocial, altruistic, competitive -- with the latter two being relatively rare. Full instructions for the SVO task and screenshots from our implementation are in \Cref{sec:instructions_svo}.

The recruitment survey takes on average 4.9 minutes to complete (st.d. 5.9) and pays \$1. As for the bonus task, we randomly select 2 of every 25 subjects who complete the task, and pay them according to a randomly chosen decision of theirs. Earnings from the bonus task are in the region of \$0.6-4.0. For transparency, a list of MTurk IDs of the winners was published online, but the actual matches and amounts earned were not revealed. 
\subsection{Experiment: Workflow} \label{sec:methodology_exp_workflow}

In order to join an experimental session, subjects need to search for our HIT at the time specified in their invitation, and login to our portal using their MTurk ID. Participation is restricted to the list of invitees only.

Upon login, subjects read the instructions for the main part of the experiment. Instructions are followed by an understanding quiz of 2-3 multiple-choice questions. Subjects must answer all questions correctly to proceed, and they have 3 attempts to do so. Each failed attempt at the quiz is followed by a review section where we show and explain correct answers, and subjects can always go back to the full instructions. The types of questions are the same in all attempts, but the actual questions are randomly varied. Throughout the instructions, subjects are primed to think about COVID-19. Sample instructions and the understanding quiz for one of our treatments are available in \Cref{sec:instructions_main_experiment}.

Once a subject passes the understanding quiz, she joins a waiting room where she waits to be matched with others into a group. To ensure constant attention from a subject, we require that she moves the mouse or presses any key on the keyboard at least every 30 seconds or she gets disqualified. Waiting time is compensated at a rate of \$0.03 for every 10 seconds to a maximum of 12 minutes. Group assignment is randomized, and subjects are not informed of the MTurk IDs of others in their group (and session in general) at any point during or after the experiment. Depending on show-up rates, once a group of 12-15 subjects is formed, they proceed to the main part of the experiment.\footnote{In practice, we never had a group of 15 subjects, with 14 being the maximum.}

The flow of a typical round of the experiment is presented in Figure 1 (see main text). For expository purposes, we focus on the treatment where subjects are asked to make all three decisions -- activity level, contact tracing scheme sign-up and quarantine program participation. 

At the beginning of a round, some subjects are infected (the rest are healthy), and some of whom may be in quarantine.\footnote{The majority of subjects also do not know their health status. Only subjects that received an alert from the contact tracing scheme in round $t-1$ know they are infected in $t$.} Figure 1(a) shows an example where two subjects are infected, of whom one is in quarantine. All members of the group are presented with information on: the number of active participants in their group, the number of infected participants, and the average activity level in their group in the last round. 

Subjects simultaneously choose their activity level (unless they are in quarantine, in which case their activity is set to 0), and whether they want to be signed up to the contact tracing scheme. Activity level is an integer on $[0,100]$ and contact tracing scheme participation is a `Yes/No' decision. Figure 1(b) shows an example, where all subjects choose different activity levels and four of them decide to participate in the contact tracing scheme. Note that subjects in quarantine must still choose whether they want to participate in the contact tracing scheme. In our example, the subject in quarantine decides not to participate in the scheme. Note that we have one treatment (NA) where activity level is fixed for all subjects not in quarantine at 100. Further, in treatments MO and MM, contact tracing scheme participation is mandatory, and so is not a decision.

Next, COVID-19 spreads through the group according to the rules set out in \Cref{sec:model}. For example, Figure 1(c) shows that three subjects get exposed in this round, and so these three subjects will be infected in the next round. Note that, when determining the number of infected subjects, we calculate the expected number of infections and round it to the closest integer. Further, exposure status together with contact tracing scheme participation determine how many subjects will receive an alert and who they will be. In our example, we can see that two of the infected subjects receive an alert. Note that in the actual experiment, alerts are received by one out of three exposed subjects who are signed up to the contact tracing scheme. To achieve this, we calculate the expected number of subjects who will receive an alert and round it to the closest integer.

Those who receive an alert are then asked to decide whether they want to go into quarantine in the next round. This is a 'Yes/No' decision. Going to quarantine amounts to setting activity to zero. The benefit is that subjects in quarantine do not contribute to the spread of COVID-19 in the round. Figure 1(d) shows that in our example, one of the alerted subjects decides to go into quarantine, while another chooses not to. Note that we have two treatments (OM and MM) where quarantine program participation is mandatory and so is not a choice.

The round ends with subjects learning their health status and payoffs from that round. The benefits for the round come from the activity level chosen by the subject, with 1 point earned for each unit of activity. Further, subjects infected with COVID-19 incur a cost of 150 points. Figure 1(e) shows an example of a distribution of points in a group. Subjects are not informed of the earnings of others in their group at any point during or after the experiment.

Based on our example, the next round starts with three subjects infected, of whom one is in quarantine, as in Figure 1(f). The experiment progresses in the same way as described above until it terminates.

Subjects have 20 seconds to make and submit their choices of activity level and contact tracing scheme sign up, 15 seconds for their quarantine program participation decision, and 10 seconds to review their outcomes for the round. If they miss any of the choices in a round, they will earn 0 points in benefits for that round (they will also incur a cost of 150 points if infected). Subjects who miss decisions in three consecutive rounds are disqualified.

The first 5 rounds of the experiment are \emph{without} the presence of COVID-19 or the contact tracing scheme and quarantine program. So in these rounds in all treatments, subjects need to only choose their activity level. At the end of Round 5, COVID-19 is introduced, and 4 subjects are randomly chosen to be infected with COVID-19 during Round 6.\footnote{For groups of 14 and 15, the computer randomly infected five rather than four subjects.}

There are then 30 further rounds of the game as described above. Subsequently, the experiment terminates after any round with probability 50\%. 

Upon completing the main part of the experiment, subjects proceed to the post-experimental survey. Here, we collect demographic information, with some questions being repetitive from the recruitment survey to check for response consistency.\footnote{We repeatedly collected data on sex, age and race. The data is consistent for all three variables for at least 97.5\% of the subjects.} In addition, subjects answer a range of questions relating to their knowledge of, and attitudes towards, a range of COVID-19 related topics including: social distancing; tracing smartphone applications, and tracing programs in general; quarantine rules and behavior; vaccination. Each subject completes 12 primary questions which are on a Likert scale, plus up to 6 clarification questions (the actual number of which depends on the answers to the primary 12 questions). Subjects receive \$2 for completing the survey.

The next part of the experiment is a bonus task which comes in the form of a `Bomb' risk elicitation task \cite{HP_2016}. In this task a subject is presented with 100 boxes arranged on a 10$\times$10 grid. The value of each box is \$0.02 and one of the boxes (chosen uniformly at random) contains a bomb but its exact location is unknown. The subject starts collecting boxes by clicking `Start', at which point the computer starts collecting boxes one after another from the top left corner at a rate of one box per second. She needs to decide when to stop collecting boxes by clicking `Stop'. Then she opens the boxes by clicking `Open'. If the bomb is among collected boxes, the earnings from the bonus task are destroyed and the subject earns 0 from the task. Otherwise she earns the value generated by her collected boxes. Under the assumption of a power utility function -- $u(x)=x^r$, where $r$ is the risk aversion coefficient, -- a risk-averse subject collects fewer that 50 boxes, a risk-loving one -- more than 50 boxes, and a risk-neutral one -- exactly 50 boxes. Instructions for the BRET task and screenshots of the interface are in \Cref{sec:instructions_bret}.

To suppress wealth effects \cite{CGH_2016}, we pay subjects for 10 randomly chosen rounds from the main part of the experiment. Points are converted at a rate of 125 points per \$1. To prevent low hourly earnings, and still ensure that the experiment is incentive-compatible, we include an extra bonus task seen only by subjects whose payment for the experiment would otherwise be below \$5.\footnote{During pilots, we discovered that about 10\% of our subjects received a relatively small payment for the experiment.} The task is a standard 15-item Big-Five personality test \cite{LJLSW_2011}. Compensation for the task is flexible to ensure that the final payment for the experiment is at least \$5. Overall, 9.3\% of subjects completed this bonus task in the experiment, with an average top-up of \$1.45 (st.d. \$1). We do not use the data from this task anywhere in the analysis.

Finally, subjects get to the payment information page where they learn how much they have earned for the experiment. Total earnings consist of: (a) fixed show-up fee of \$1, (b) earnings from the main part of the experiment as explained above, (c) earnings for waiting in the waiting room, and (d) bonus earnings from the bonus task(s). 

\subsection{Experiment: Implementation} \label{sec:methodology_exp_implementation}

We programmed the experiment in o-Tree (v3.3.7; https://www.otree.org) \cite{CSW_2016} and deployed the server on Heroku (https://www.heroku.com). For the BRET task, we use a modified version of the code by Holzmeister and Pfurtscheller \cite{HP_2016}. We ran all 22 sessions of the experiment between 14 June 2021 and 22 July 2021, each with 2-4 independent groups, for a total of 61 groups. 

For a typical session of the experiment, we invite a sample of 200-300 people to fill 40-70 places on a first-come-first-served basis. We send invitations 24-30 hours in advance of a session. The selected sample of invitees is representative of the general US adult population in terms of sex and region of residence. 

Subjects in our final dataset are representative of the adult US population in terms of sex and region of residence (two-sided $\chi^2$, $p = 0.9663$). Note that, when it comes to region of residence, we group US states into 10 Standard Federal Regions as defined by the Office of Management and Budget\footnote{Circular A-105, ``Standard Federal Regions,'' April, 1974.} (see \Cref{fig:distribution_location} for definition).

Treatment assignment to sessions is randomized. We check whether our final dataset is balanced when it comes basic demographics and treatment assignment. We run a chi-squared test on subjects' assignment to treatment (a categorical variable with 6 options) and sex, political party support, and region of residence variables. Results indicate that subjects in all 6 experimental treatments are not significantly different from each other at all conventional significance levels (two-sided $\chi^2$, $p > 0.1$). 

We implement the following rules for exclusion of subjects from participating in the experiment. First, only those subjects who give their consent and answer the qualifying quiz in the recruitment survey correctly are invited to the experiment. Second, we keep track of anonymized IP addresses of subjects who have already completed the experiment and exclude any subjects with duplicate addresses from the list of invitees. Third, subjects must read instructions for the experiment and pass the understanding quiz. 

The experiment takes on average 32.8 minutes to complete (st.d. 5.4 minutes), including an average of 8.5 minutes (st.d. 3 minutes) for instructions and understanding quiz, and 2.5 minutes (st.d. 2.1 minutes) of waiting time. The average per hour pay rate is \$13.54 (st.d. \$4.4).

\section{Dataset} \label{sec:dataset}
The main dataset contains decisions of 731 subjects. Each subject participated in exactly one session. We collected 10 groups for each of the 6 treatments (and an 11th group for treatment MO), for a total of 61 groups. 56 of those started the experiment as groups with 12 subjects, with the remaining 5 groups consisting of 13-14 subjects.\footnote{One group of 13 subjects in each of treatments NA and NI, and one group of 14 in each of treatments NI, OO and MM.} All subjects played for at least 35 rounds, with 52\% subjects playing a further 1-3 rounds. Overall, the main dataset contains 26,180 individual decisions. We also match experimental data with data from our recruitment survey (see \Cref{sec:methodology_exp_recruitment} for details). The full dataset is available from the authors upon request.

In addition to data on subjects' decisions in the experiment, we collect data on a set of variables, which can be broadly categorized as follows: demographic controls, preference controls, political ideology controls and location-based controls. \Cref{tab:data_summmary_stats} presents summary statistics for some of these controls.\\

\begin{table}[t]
\centering
\begin{threeparttable}
\caption{Summary statistics for the main controls}
\label{tab:data_summmary_stats}
\begin{tabular}{llccl}
\toprule
Variable & & 
\multicolumn{1}{c}{$\bar{X}$} & 
\multicolumn{1}{c}{s.d.} & 
Comments \\ \midrule
\multicolumn{5}{l}{\textbf{Demographic controls}} \\
Age & & $42.8$ & $12.1$ & measured in years  \\
Gender & & $0.51$ & $0.02$ & female $= 1$ \\
Race: white only & & $0.78$ & $0.02$ & yes $= 1$ \\
Race: white and other(s) & & $0.05$ & $0.01$ & yes $= 1$ \\
Race: other(s) only & & $0.17$ & $0.01$ & yes $= 1$ \\
Education & & $15.23$ & $2.23$ & measured in years \\
Married & & $0.46$ & $0.02$ & yes $= 1$ \\
Employed & & $0.79$ & $0.02$ & yes $= 1$ \\
Unemployed & & $0.07$ & $0.01$ & yes $= 1$ \\ 
Not in labor force & & $0.14$ & $0.01$ & yes $= 1$ \\
Household earnings & & $50$-$75$ & . & median category ($\$'000$s per year) \\
\multicolumn{5}{l}{\textbf{Preference controls}} \\
Prosocial values & & $0.54$ & $0.02$ & yes $= 1$ \\ 
Individualist values & & $0.45$ & $0.02$ & yes $= 1$ \\ 
Risk score & & $36.65$ & $19.32$ & an integer on $[0,100]$ \\
\multicolumn{5}{l}{\textbf{Political ideology controls}} \\
Republican (lean-Republican) & & $0.29$ & $0.02$ & yes $= 1$ \\ 
Democrat (lean-Democrat) & & $0.63$ & $0.02$ & yes $= 1$ \\ 
Ideology score & & $3.43$ & $1.80$ & an integer on $[1,7]$ \\ 
\bottomrule
\end{tabular}
\begin{tablenotes}
      \item Sample size is 731; $\bar{X}$ -- mean value, or proportion in case of binary variables; s.d. -- standard deviation. 
    \end{tablenotes}
\end{threeparttable}

\end{table}

\noindent \textbf{Demographic controls.} All subjects in our sample are resident in the US, mean age is 42.8 years and 51\% are female. 78\% of subjects identified as white only, 18\% as white and some other category, and 4\% as some other category. The average subject in our sample has 15.2 years of education\footnote{We estimate years of education using subjects' highest qualification reported. We assume that all subjects took the standard number of years to complete each qualification, and undertook no education that did not lead to a qualification.} and 46\% are married. 79\% are employed (either full- or part-time) and the median household earnings in our sample are \$50-75k.\footnote{Household earnings data is categorical.} \\

\begin{figure}[t]
    \centering
    \includegraphics[width=\linewidth]{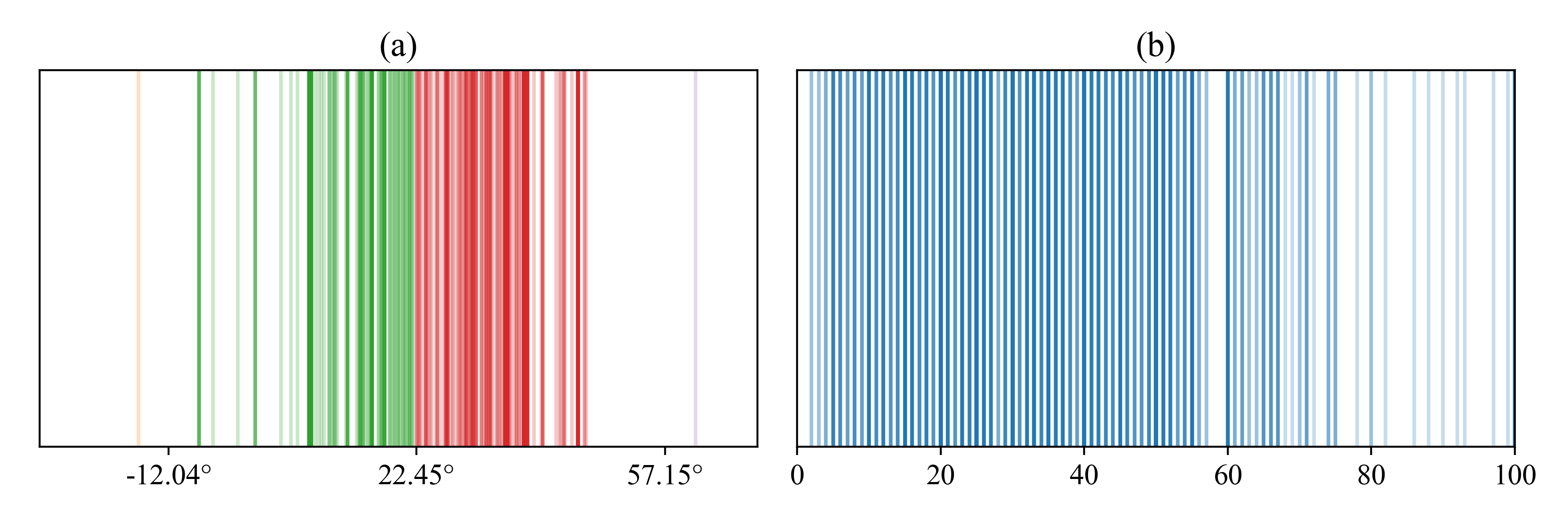}
    \caption{SVO scale (panel a) and BRET (panel b) distributions for the sample. We draw a vertical line on the subplots for each subject whose score in SVO/BRET is of the corresponding value. More intense line color indicates that more subjects are concentrated at that value. For SVO the classification is as follows: angle $\geq$ 57.1$^{\circ}$ -- altruist; $\geq$ 22.45$^{\circ}$ and $<$57.15$^{\circ}$ -- prosocial; $<$22.45$^{\circ}$ and $\geq$ 12.04$^{\circ}$ -- individualist; $<$-12.04$^{\circ}$ -- competitive. Sample size is 731.}
    \label{fig:distribution_svo_bret}
\end{figure}

\noindent \textbf{Preference controls.} As explained in \Cref{sec:methodology_exp_recruitment}, we collect information on subjects' social value orientation (SVO, from recruitment) and risk (BRET, from the bonus task after the experiment) preferences. The distributions of those preferences in our sample are in \Cref{fig:distribution_svo_bret}. 55\% of our subjects are classified as prosocial, with the second largest category -- 45\% -- being individualists. Our sample also has 1 subject with competitive and 1 with altruistic preferences. This is in line with existing research into social values, which shows that very few people exhibit competitive or altruistic preferences \cite{MAH_2011}. When it comes to risk preferences, the mean BRET score in our sample is just under 37 boxes, which translates into moderate risk-aversion. 71\% of our sample can be classified as risk-averse, 18\% are risk-seeking and 11\% are exactly risk-neutral.\\

\begin{figure}[ht]
    \centering
    \includegraphics[width=0.5\linewidth]{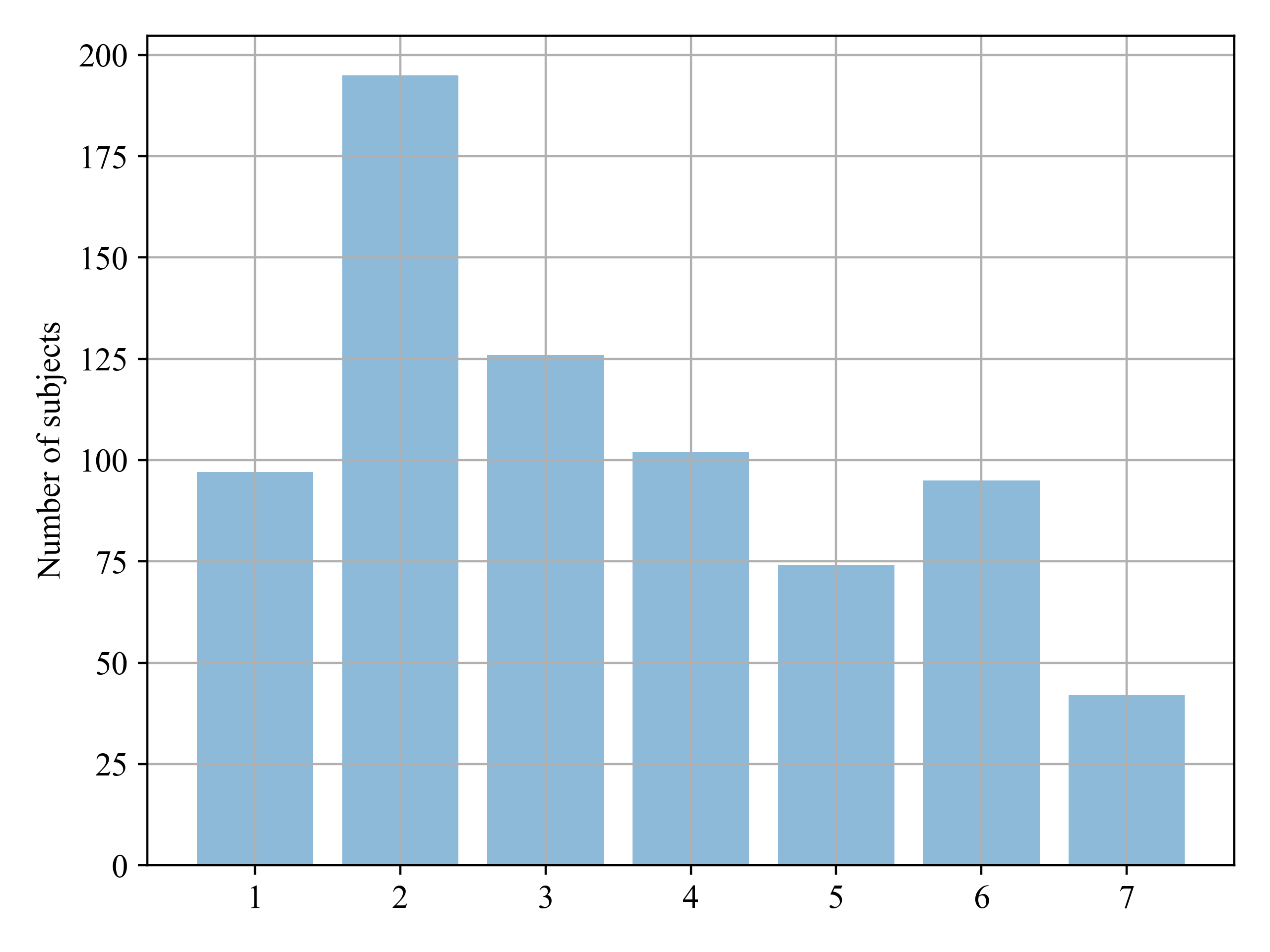}
    \caption{Political ideology distribution for the sample, where 1 stands for strongly liberal and 7 for strongly conservative. Sample size is 731.}
    \label{fig:distribution_ideology}
\end{figure}

\noindent \textbf{Political ideology controls.} As part of the recruitment survey, we collected subjects' party affiliation and political ideology. 29\% of our sample identified as (leaning) Republican and 63\% as (leaning) Democrat. The remaining 8\% indicated no lean. Further, we asked subjects to describe their political views using a 7-point Likert scale where 1 stands for `very liberal' and 7 for `very conservative'. The distribution of the scores for this question in our sample is in \Cref{fig:distribution_ideology}. The mean score is 3.43 (st.d. 1.8). The correlation coefficient for subjects' political ideology and party affiliation is 0.8323 (t-stat $p < 0.0001$). Throughout the analysis, we use the ideology score as our metric of political ideology, as it is a finer measure than political party affiliation.\\

\begin{figure}[ht]
    \centering
    \includegraphics[width=\linewidth]{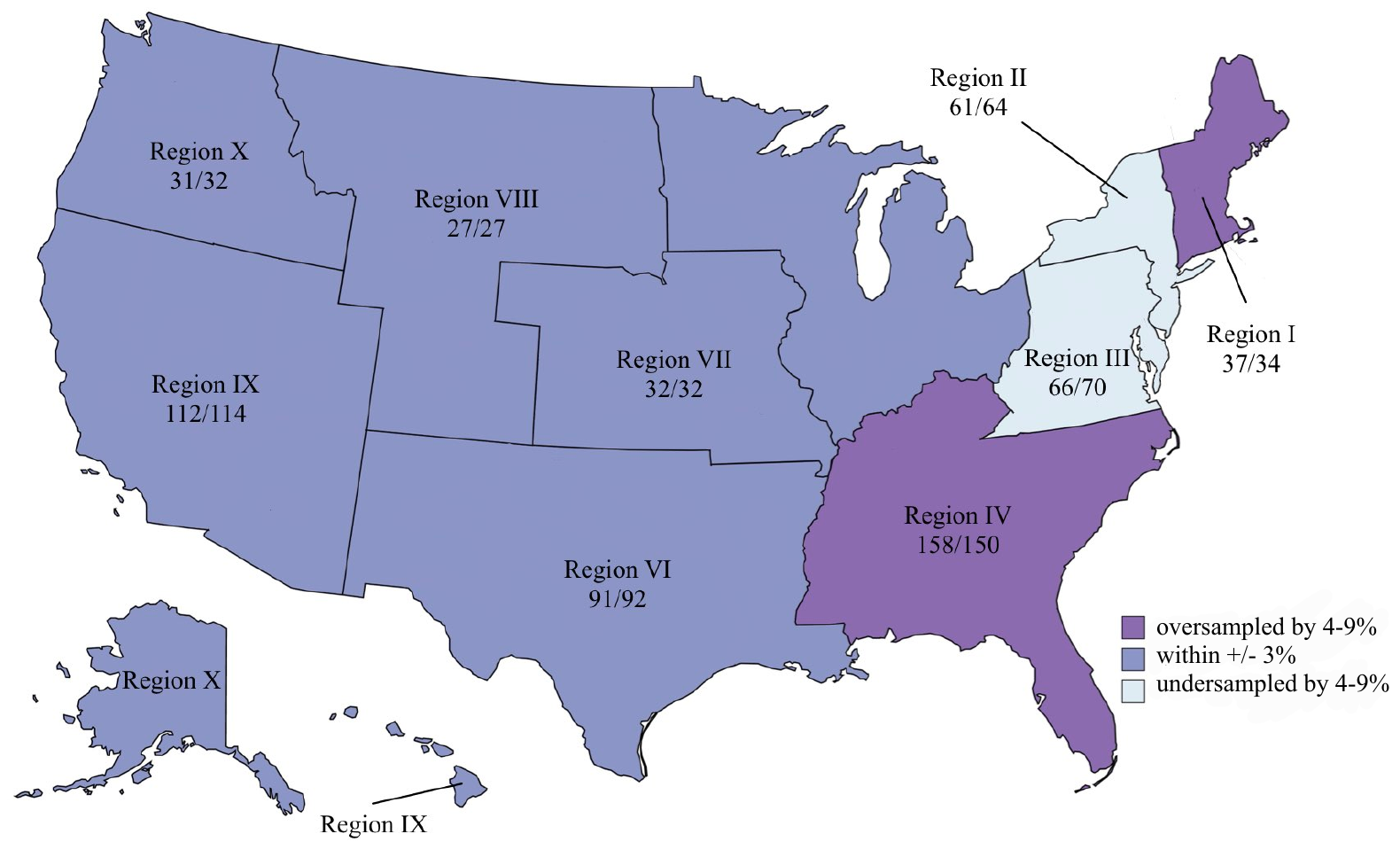}
    \caption{Location distribution, created with www.mapchart.net. X/Y indicate subjects counts, where X indicates the actual count in our sample, and Y -- the target representative count. Sample size is 731. States are divided into 10 Standard Federal Regions (Office of Management and Budget) as follows: I -- CT, ME, MA, NH, RI, VT; II -- NJ, NY; III -- DE, DC, MD, PA, VA, WV; IV -- AL, FL, GA, KY, MS, NC, SC, TN; V -- IL, IN, MI, MN, OH, WI; VI -- AR, LA, NM, OK, TX; VII -- IA, KS, MO, NE; VIII -- CO, MT, ND, SD, UT, WY; IX -- AZ, CA, HI, NV; X -- AK, ID, OR, WA. Note that we exclude US territories.}
    \label{fig:distribution_location}
\end{figure}

\noindent \textbf{Location controls.} We collect information on subjects' state of residence using self-report data and IP-address information. We then convert this information into region of residence. In our analysis, we rely on self-report data, as IP-to-location accuracy is estimated to be only about 50\% when it comes to state identification \cite{KVR_2017}. For 92\% of our subjects, the region of residence from self-report data and IP-address information matches. \Cref{fig:distribution_location} shows the region-level residency distribution of our sample. 

\section{Methodology: Data analysis} \label{sec:methodology:data_analysis}
In this section we explain the tools we use for analysis in \Cref{sec:analysis}. All of our methodology is fairly standard. We also summarize our data exclusion rules, along with the robustness checks we perform.\\

\noindent \textbf{Data exclusion rules.} We exclude some data from our main dataset, using the following rules. First, we exclude data from subjects who drop out due to inactivity midway through the experiment. 7 of the 61 groups experienced a dropout midway through the session and 1 group had 2 dropouts. On average this occurred at round 19. Second, we exclude data from groups where the number of active subjects drops below 10 by the end of the experimental session. We had 1 such group. All of the results in \Cref{sec:analysis} are robust to including this data.

Further, for the analysis in \Cref{sec:analysis}, we use data from rounds 6-34. This is because the game in the initial 5 rounds and rounds 35 onward is different. Rounds 1-5 are excluded because infection only starts in Round 6. Further, from Round 35 onward there is a 50\% chance that the experiment will terminate at the end of that round. While the benefits of activity in round $t$ accrue in round $t$, the costs (in terms of higher likelihood of being infected) only realize in round $t+1$. Therefore, a 50\% termination probability reduces the expected cost of activity, and changes the variance of outcomes of the negative consequences of high activity. 

Finally, it is possible that behavior in a few rounds immediately after the introduction of COVID-19 may be different from that in subsequent rounds due to, for example, learning effects. To ensure that our results are not driven by these differences, we do the following: for parametric analysis, we include a differential time trend in all of our specifications, and for non-parametric analysis, we discard data from Rounds 6-15. Our results are robust to these checks.\\

\noindent \textbf{Parametric analysis.} To look at individual decisions and outcomes, we use standard Ordinary Least Squares (OLS) for decisions and outcomes that are near-continuous\footnote{In our experiment only integer values are allowed, so all variables are in fact discrete. Nevertheless, we treat activity and welfare as approximately continuous. Activity can takes integer values in the range $[0,100]$, and welfare can takes integer value in the range $\{ [-150,-50] , [0,100] \}$.} and a Linear Probability Model (LPM) for binary decisions and outcomes \cite{G_1964, A_1984}. 

Additionally, we use an Instrumental Variables (IV) approach when investigating ideology. This allows us to find a causal impact of ideology on individual decisions. We use survey responses to four questions regarding highly partisan issues that are unrelated to COVID-19 or to medicine, healthcare and science more generally.\footnote{These questions are: (1) \emph{``The U.S. made the right decision in using military force in Iraq''}, (2) \emph{``Thinking about the situation with Israelis and Palestinians these days, do you have a favorable or unfavorable opinion of the Israeli government''}, (3) \emph{``Each year, many global landmarks, including New York's Empire State Building and the Las Vegas strip, take part in Earth Hour. This involves turning their lights off for one hour to show support for nature and the planet. Do you support Earth Hour?''}, (4) \emph{``Do you support allowing gays and lesbians to marry legally?''}. Responses are on a 5-point Likert scale, with low numbers indicating stronger support for the statement, and additional options for ``Don't know'' and ``Refuse to answer''. We collected ideology scores at the recruitment stage, and asked these four questions immediately after the experiment -- typically several days apart. This aims to prevent subjects' answers to one set of questions influencing their answers to the other set (i.e. to mitigate order effects, a la \cite{mcfarland1981effects}.} 
We aggregate these four questions by summing the responses to generate a single index.\footnote{Statements regarding Earth hour and gay marriage enter positively, and the Iraq war and the Israeli government statements enter negatively. Recall that larger numbers indicate \emph{opposition} to the statement.}

For this IV approach to be valid, two conditions must be met. First, instrumental relevance: the instrument must be correlated with ideology. In our data the correlation is $0.7046$ and it is highly significant (t-stat $p < 0.0001$). Second, instrumental validity: the instrument must be uncorrelated with the unobserved error term. This cannot be tested, but amounts to a claim that opinions on the four partisan issues we ask about \emph{only} affect decisions in the experiment via their effect on ideology.

We cluster standard errors at the group level in all regressions. This is because individual decisions depend on behavior and outcomes in the rest of the group and so will be correlated at the group level.\\

\noindent \textbf{Non-parametric analysis.} Since individual decisions are correlated at the group-level, our aggregate level-analysis takes a single group of subjects as an independent unit of observation. For each of the groups, we calculate the values for our key metrics -- individual activity level of subjects not in quarantine, contact tracing scheme sign-up and quarantine participation probabilities, global average activity, shares of infected and quarantined subjects, and per subject welfare, -- averaged over Rounds 6-34. We then run a set of non-parametric tests on those averages. Our analysis is performed on unmatched samples -- i.e. those generated by groups from different treatments. The chosen test is Mann-Whitney U-test (MW) \cite{MW_1947}. 

The reason for using non-parametric test lies in the fact that we have 10-11 independent observations per treatment, meaning that the assumption of data normality is, strictly speaking, not applicable. However, all of our key results are robust to using the parametric analogue (unpaired t-test) of our non-parametric test. 

\section{Results} \label{sec:analysis}
Here, we discuss the main results. \Cref{sec:decisions_outcomes} sets out our main results concerning the effect of treatments on subjects' decisions and outcomes. \Cref{sec:ideology} explores political ideology and seeks to understand its causal impact. \Cref{sec:behav_regularities} documents a set of impacts that features of the environment have on subjects' behavior. \Cref{sec:characteristics} examines associations between individual subject characteristics, and decisions and outcomes. Finally, \Cref{sec:survey_data} shows that subjects' behavior in the experiment is internally consistent with survey data we collect from them.

\subsection{Decisions and outcomes}\label{sec:decisions_outcomes}

\begin{figure}[ht]
    \centering
    \begin{subfigure}{.4\textwidth}
        \includegraphics[width=\linewidth]{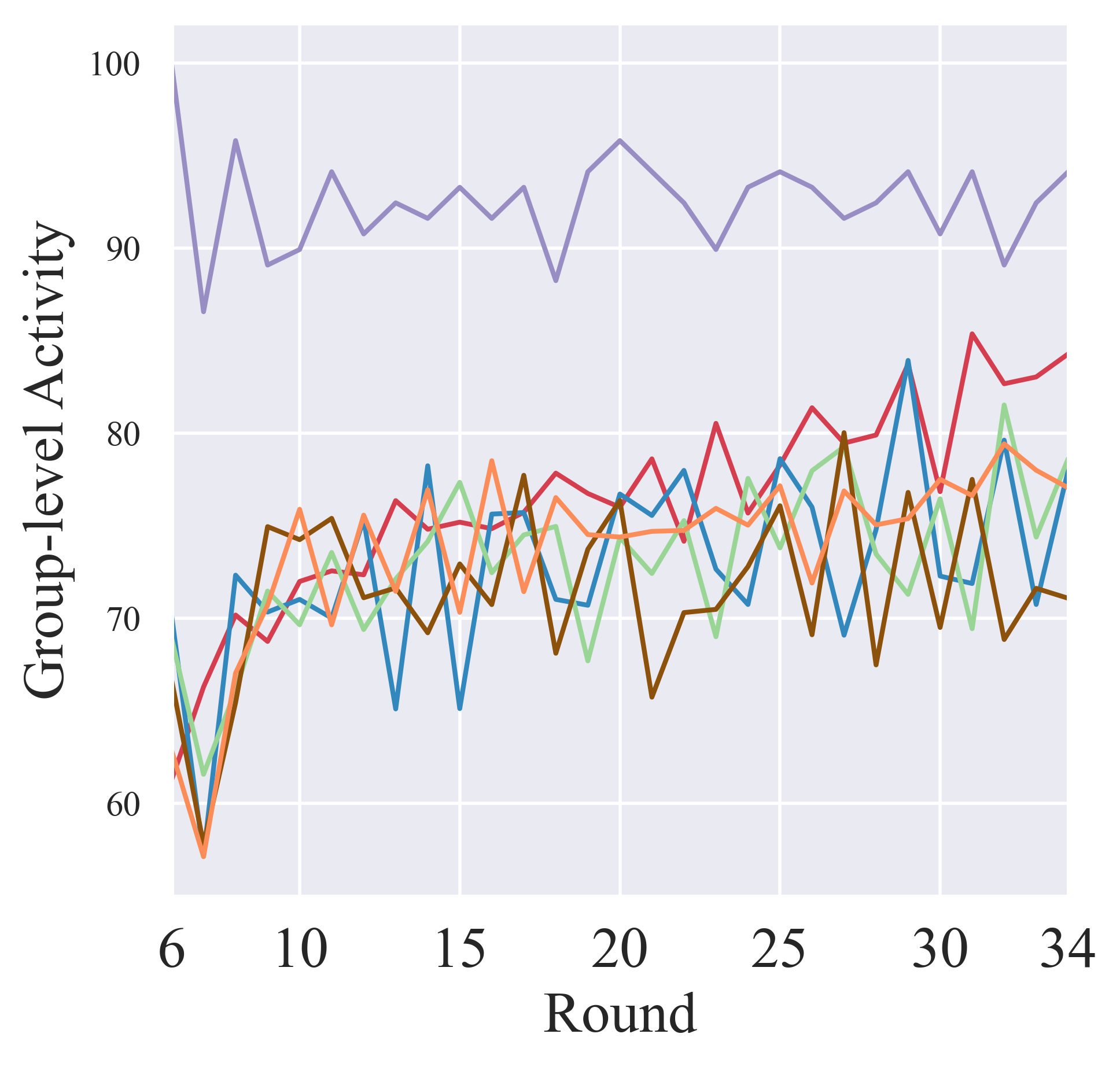}
        \caption{}
        \label{fig:subfig1}
    \end{subfigure} %
    \begin{subfigure}{.4\textwidth}
        \includegraphics[width=\linewidth]{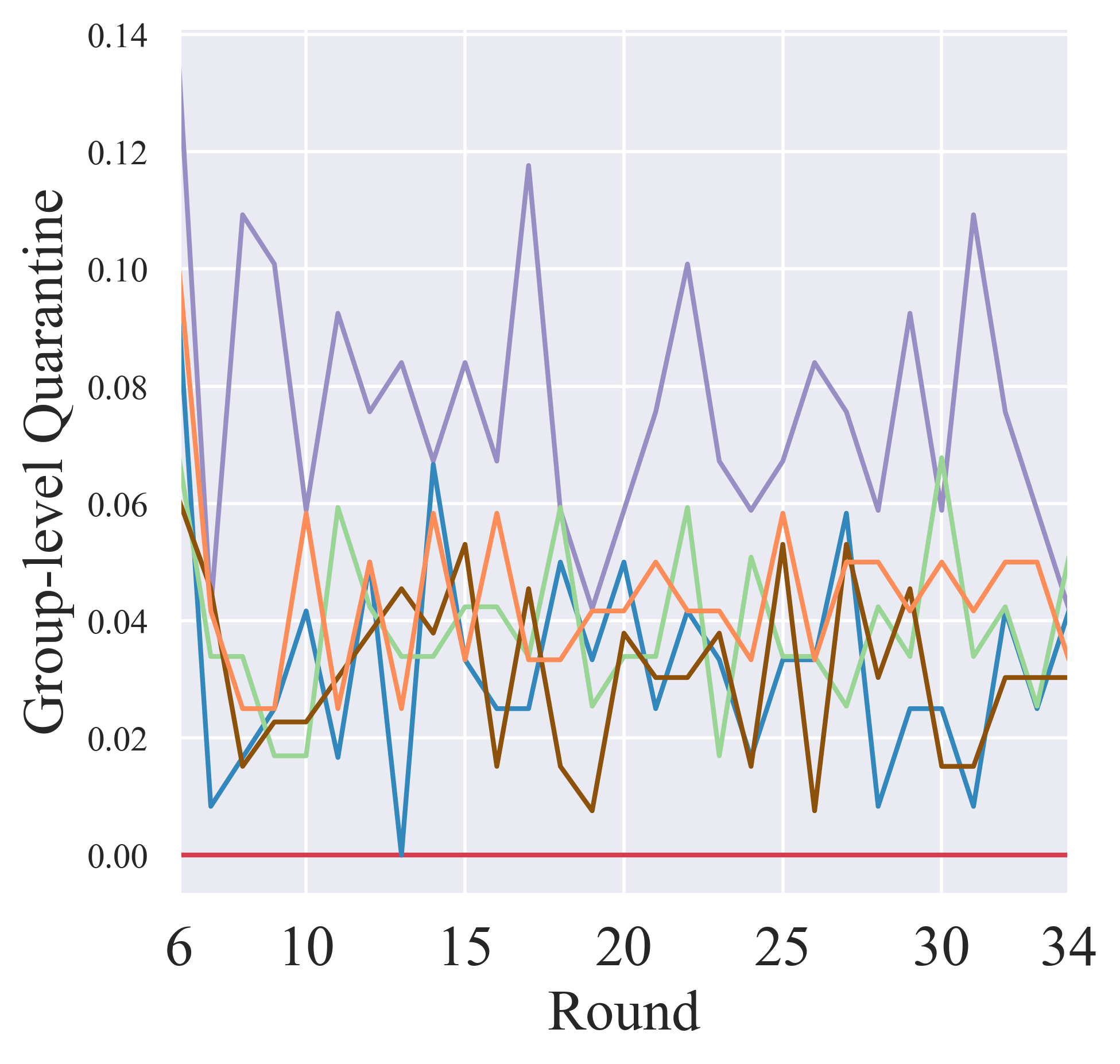}
        \caption{}
        \label{fig:subfig2}
    \end{subfigure}
    \begin{subfigure}{.15\textwidth}
        \includegraphics[width=\linewidth]{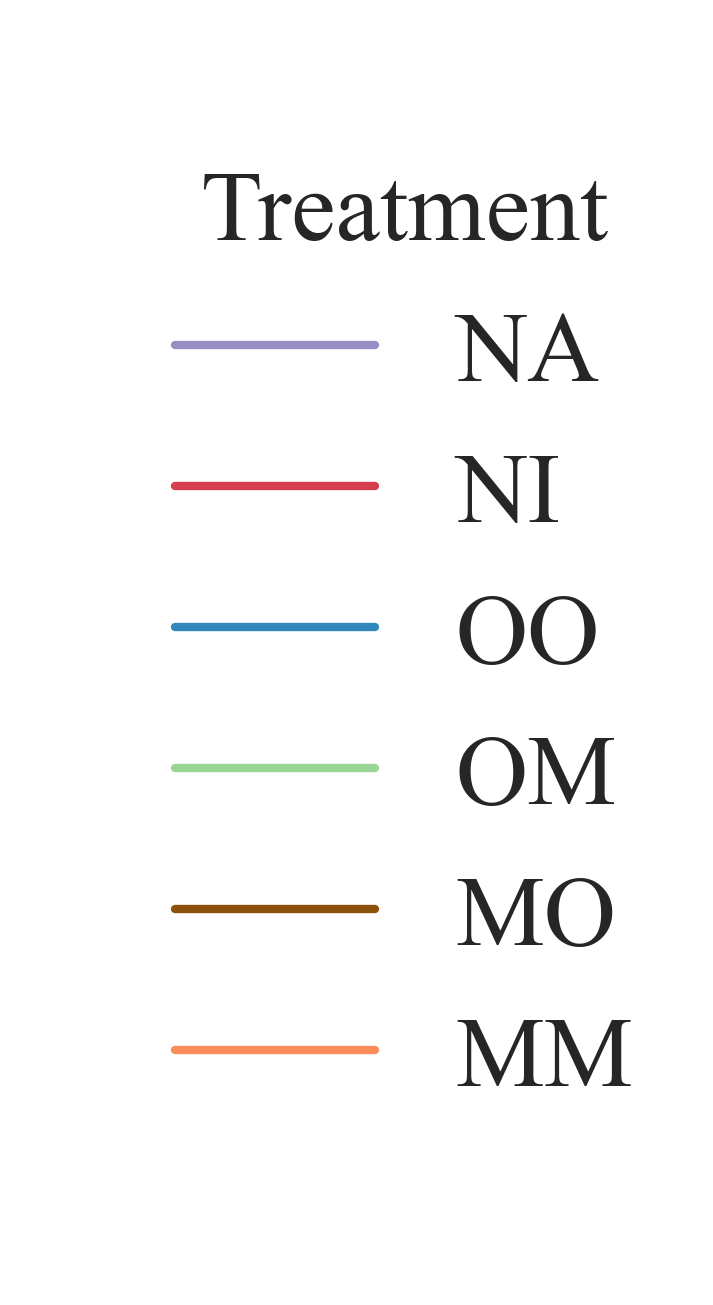}
    \end{subfigure}
    \caption{Treatment-level means of supplementary outcomes over time. COVID-19 is not introduced until Round 6, so Rounds 1-5 are omitted. (a) Average activity in a group. (b) Fraction of subjects in a group in quarantine.}
    \label{fig:extra line graphs}
\end{figure}

Agents make (up to) three decisions in the game: (1) activity, (2) tracing, and (3) quarantine. These decisions, plus fixed features of the environment (the dynamics of the disease, mandatory tracing and/or quarantine) then determine the outcomes. The key outcomes are: (4) the proportion of infected subjects, and (5) welfare (i.e. payoffs). There are also two supplementary outcomes: (6) the average activity level in a group, and (7) the proportion of subjects in quarantine.\footnote{Average activity in a group is affected by agents being in quarantine – as this forces their activity to zero. The number in quarantine depends on the number of alerts received and then the proportion of those receiving an alert who go into quarantine. Therefore these supplementary outcomes depend on all the decisions.} We show line graphs showing treatment-level averages for the decisions and the two main outcomes in the main text (Figure 2). Figure \ref{fig:extra line graphs}, below, shows the equivalent for the supplementary outcomes.

The model predicts rankings of treatments for each of decisions and outcomes (sometimes with ties).\footnote{These rankings are based on the parameterization set out above. For other parameterizations, it is possible that the rankings change -- especially for extreme parameterizations.} Given the stylized setting of the experiment, we focus on the ranking of treatments rather than the point-estimates.\\

\noindent \textbf{How we report results.} For brevity, we report rankings for these variables across the six different treatments when stating results. Pairwise comparisons follow from the rankings. When the variable is fixed by the treatment environment, it is not appropriate to include that treatment in the corresponding parametric analysis. Therefore, when comparing treatments where subjects do make a decision to ones where they do not, we can only use non-parametric analysis. For example, we omit treatments MM, MO and NI from the parametric analysis for tracing because agents do not choose tracing in these treatments. Upper triangles of panels in Table \ref{tab:nonpar_tab} show the p-values for the full set of pairwise comparisons for non-parametric (MW) analysis. Pairwise comparisons for parametric analysis (OLS, IV) are based on the p-values from regressions shown in Table \ref{tab:full_regs} (IV) and Table \ref{tab:ols_regs} (OLS). Throughout this section we report p-values from Table \ref{tab:full_regs}, but the results will not change if we used p-values from Table \ref{tab:ols_regs} instead. Table \ref{tab:full_regs} repeats Table 1 from the main text, but does not omit any variables. In the results in this subsection, strict inequalities (both direct and implied by transitivity) represent statistically significant differences at the $1\%$ level or stronger, \emph{both} for parametric (OLS, IV) and non-parametric (MW) analysis. Similarly, weak inequalities [resp. equalities] (both direct and implied by transitivity) are significant at the $5\%$ level but \emph{not} at the $1\%$ level [resp. are \emph{not} significant at the $5\%$ level], both for parametric (OLS, IV) and non-parametric (MW).\\

\noindent \textbf{Robustness of results.} Our preferred regressions (shown in Table 1 in the main text) uses: (1) an Instrumental Variables approach to identify a causal effect of political ideology, (2) data from all rounds after the introduction of COVID-19 but before the game terminates with some probability (i.e. Rounds 6-34 of the experiment), and (3) treatment-specific time trends to allow for dynamic effects of treatments. The results we show in this section are robust to variations on all of these dimensions. Using OLS (rather than IV), excluding data from Rounds 6-15 (to allow for learning), and excluding a treatment-specific time trend all leave the results substantively unchanged. We also exclude observations from subjects who dropped out of the experiment part-way through, and one group that failed because too many subjects dropped out. Including these observations also has no material impact. For non-parametric analysis, we: (1) exclude data from Rounds 6-15, (2) include data from drop-outs and the failed group, and (3) use a t-test instead of the Mann-Whitney U test. Again, the results are robust to these alterations.\\

\noindent \textbf{A note on theoretical equivalence.} In our stylized model, agents are fully rational and completely self-interested. An obvious implication of this is that they will never provide a pure public good and will under-provide a good which has positive externalities (compared to a socially optimal level).\footnote{See \cite{mas1995microeconomic} for a textbook treatment of public goods and externalities.} Consequently the behavior (and hence outcomes) predicted by the model coincides in two pairs of treatments.

First, treatments OO and MO are \emph{de facto} identical. This is because agents never voluntarily participate in quarantine (i.e. they never provide a pure public good), and always want to participate in the contact tracing scheme. The contact tracing is free, provides some useful information to agents, and entails no obligations to do anything with that information. It allows agents to refine their activity decisions in some circumstances, and so it is optimal to always participate. Since agents make the same ``decisions'' regarding tracing and quarantine, they then face the same situation regarding their activity choice -- and so make the same decisions there too.

Second, treatments OM and NI are also \emph{de facto} identical. This is because in treatment OM agents do not want to participate in quarantine (i.e. they never provide a pure public good), and so choose not to participate in the contact tracing scheme in order to avoid mandatory quarantine. As agents never use either tracing or quarantine in treatment OM, it is equivalent to treatment NI, where neither are available.

For this reason we will have $OO = MO$ and $OM = NI$ in all rankings in the formally stated hypotheses throughout this section. This equivalence does not hold in the results -- as actual behavior departs from our stylized model. Most importantly (and obviously) some subjects in our experiment do in fact provide the public good, which is consistent with the public good game experimental literature (e.g. \citeauthor{ledyard1995public} \cite{ledyard1995public}, \citeauthor{chaudhuri2011sustaining} \cite{chaudhuri2011sustaining}). 

\subsubsection{Decisions}

\noindent \textbf{Theory.} Given our stylized model, tracing and quarantine participation will always have corner solutions. Either all agents participate, or none do. As discussed above, all agents choose to participate in the contact tracing scheme in treatment OO, and all agents are forced to participate in treatments MO and MM.\footnote{Participation in the contact tracing system is indeterminate in treatment NA. This is because agents cannot adjust their activity in response to the information they receive, and will never want to go into quarantine as a result of it. Therefore the information is not helpful -- they cannot do anything with it.} In treatment OM, agents choose \emph{not} to participate, and they are unable to do so in treatment NI. Similarly, no agents go into quarantine after receiving an alert, except in treatments MM and OM, where they are forced to do so.

\begin{theor}[Tracing]\label{h:tracing}
The model predicts the following ranking for participation in the contact tracing scheme:
$1 = MM = MO = OO > OM = 0$
and is indeterminate for treatments NI and NA.
\end{theor}

\begin{theor}[Quarantine]\label{h:quarantine}
The model predicts the following ranking for participation in the quarantine program, conditional on receiving an alert:
$1 = MM = OM > MO = OO = NA = 0$
and is indeterminate for treatment NI.
\end{theor}

Equilibrium activity decisions are generally interior, as agents trade off the direct benefits of additional activity against the cost of increased likelihood of infections. We focus on symmetric Nash equilibria, so all agents (who are not in quarantine) choose the same level of activity in equilibrium. Predicted activity is trivially highest in treatment NA, as it is exogenously fixed at $100$. Next, enforced participation in both tracing and quarantine in treatment MM mitigates the negative impact of activity and is able to keep the number of infections low. So high activity activity is possible in equilibrium. In treatments OM and NI lower average activity is then necessary to keep infections under control as agents do not (or cannot) make use of the tracing and quarantine program. 

Finally, lowest average activity occurs in treatments OO and MO. Again, self-interested agents do not use the quarantine program (as it is a public good). In these treatments they do use the contact tracing scheme -- but it has a perverse effect. When an agent receives an alert from the scheme, she knows for certain she is infected in the next round. Therefore she does as much activity as possible, because she will be infected next period regardless of her actions (and due to the set-up of the model, must also be healthy in the period after). This increases infections and requires \emph{lower} average activity to keep infections under control.

\begin{theor}[Activity]\label{h:activity}
The model predicts the following ranking for individual activity decisions in a steady state: $NA > MM > OM = NI > OO = MO$. 
\end{theor}

\noindent \textbf{Results.} Unsurprisingly, the experimental results regarding participation in tracing and quarantine are less stark than the corner cases the model predicts. About $81\%$ of subjects participate in the contact tracing scheme in treatments OO and NA, even though participation is free. Further, around $60\%$ of subjects participate in treatment OM. Nevertheless, the results support the key qualitative prediction -- that mandatory quarantine discourages participation in the contact tracing scheme. The difference in tracing participation between treatments OO and OM is about $13.7$ percentage points and is significant (\Cref{tab:full_regs} M2 $p = 0.0047$, \Cref{tab:nonpar_tab} MW $p = 0.0493$).

We also find that some subjects do in fact choose to go into quarantine when it is optional. This is line with a very wide body of literature showing that people do provide public goods. However, there is no evidence that the willingness of subjects to go into quarantine after they have received an alert depends on the treatment, as indicated by the lack of significance in quarantine program participation between treatments OO and MO (M3 $p=0.937$, MW $p=0.9438$). However, it is important to note that fewer subjects participate in quarantine program when it is optional compared to mandatory (MW $p=0.0001$).

\begin{table}[ht]
\centering
\caption{Non-parametric analysis of experimental data (independent unit of observation is a group)}
\label{tab:nonpar_tab}
\resizebox{\textwidth}{!}{%
\begin{tabular}{@{}ccccccccrcccccc@{}}
\toprule
NA & NI & OO & OM & MO & MM &  & \multicolumn{1}{l}{\textbf{}} & \multicolumn{1}{l}{\textbf{}} & NA & NI & OO & OM & MO & MM \\
\multicolumn{6}{c}{\textbf{individual activity   (if not in quarantine)}} &  & \multicolumn{1}{l}{} & \multicolumn{1}{l}{} & \multicolumn{6}{c}{\textbf{global average activity}} \\
- & \cellcolor[HTML]{548235}{\color[HTML]{FFFFFF} 0.0001} & \cellcolor[HTML]{548235}{\color[HTML]{FFFFFF} 0.0001} & \cellcolor[HTML]{548235}{\color[HTML]{FFFFFF} 0.0001} & \cellcolor[HTML]{548235}{\color[HTML]{FFFFFF} 0.00001} & \cellcolor[HTML]{548235}{\color[HTML]{FFFFFF} 0.0001} & {\color[HTML]{FFFFFF} } & NA &  & - & \cellcolor[HTML]{548235}{\color[HTML]{FFFFFF} 0.0002} & \cellcolor[HTML]{548235}{\color[HTML]{FFFFFF} 0.0002} & \cellcolor[HTML]{548235}{\color[HTML]{FFFFFF} 0.0002} & \cellcolor[HTML]{548235}{\color[HTML]{FFFFFF} 0.0001} & \cellcolor[HTML]{548235}{\color[HTML]{FFFFFF} 0.0002} \\
\cellcolor[HTML]{FFF2CC}23.35 & - & 0.6232 & 0.5708 & 0.4181 & 0.8501 &  & NI &  & \cellcolor[HTML]{FFF2CC}15.89 & - & 0.2123 & 0.273 & \cellcolor[HTML]{C6E0B4}0.0845 & 0.3847 \\
\cellcolor[HTML]{FFF2CC}25.66 & \cellcolor[HTML]{FFF2CC}2.31 & - & 0.7913 & 0.7513 & 0.7337 &  & OO &  & \cellcolor[HTML]{FFF2CC}20.44 & \cellcolor[HTML]{FFF2CC}4.56 & - & 0.9698 & 0.6985 & 0.9698 \\
\cellcolor[HTML]{FFF2CC}25.16 & \cellcolor[HTML]{FFF2CC}1.81 & \cellcolor[HTML]{FFF2CC}0.50 & - & 0.5974 & 0.9698 &  & OM &  & \cellcolor[HTML]{FFF2CC}20.33 & \cellcolor[HTML]{FFF2CC}4.44 & \cellcolor[HTML]{FFF2CC}-0.12 & - & 0.7513 & 0.9698 \\
\cellcolor[HTML]{FFF2CC}25.28 & \cellcolor[HTML]{FFF2CC}1.93 & \cellcolor[HTML]{FFF2CC}-0.38 & \cellcolor[HTML]{FFF2CC}0.12 & - & 0.4181 &  & MO &  & \cellcolor[HTML]{FFF2CC}20.07 & \cellcolor[HTML]{FFF2CC}4.18 & \cellcolor[HTML]{FFF2CC}-0.37 & \cellcolor[HTML]{FFF2CC}-0.26 & - & 0.5974 \\
\cellcolor[HTML]{FFF2CC}25.21 & \cellcolor[HTML]{FFF2CC}1.86 & \cellcolor[HTML]{FFF2CC}-0.45 & \cellcolor[HTML]{FFF2CC}0.05 & \cellcolor[HTML]{FFF2CC}-0.07 & - &  & MM &  & \cellcolor[HTML]{FFF2CC}20.35 & \cellcolor[HTML]{FFF2CC}4.46 & \cellcolor[HTML]{FFF2CC}-0.09 & \cellcolor[HTML]{FFF2CC}0.02 & \cellcolor[HTML]{FFF2CC}0.28 & - \\
\multicolumn{6}{c}{\textbf{contact tracing   system signup}} &  &  & \multicolumn{1}{l}{} & \multicolumn{6}{c}{\textbf{infected share}} \\
- & \cellcolor[HTML]{548235}{\color[HTML]{FFFFFF} 0.0001} & 0.9097 & \cellcolor[HTML]{C6E0B4}{\color[HTML]{FFFFFF} 0.0538} & \cellcolor[HTML]{548235}{\color[HTML]{FFFFFF} $<$ 0.0001} & \cellcolor[HTML]{548235}{\color[HTML]{FFFFFF} 0.0001} & {\color[HTML]{FFFFFF} } & NA &  & - & \cellcolor[HTML]{548235}0.0091 & \cellcolor[HTML]{548235}{\color[HTML]{FFFFFF} 0.0002} & \cellcolor[HTML]{548235}{\color[HTML]{FFFFFF} 0.0002} & \cellcolor[HTML]{548235}{\color[HTML]{FFFFFF} 0.0001} & \cellcolor[HTML]{548235}{\color[HTML]{FFFFFF} 0.0002} \\
\cellcolor[HTML]{FFF2CC}0.80 & - & \cellcolor[HTML]{548235}{\color[HTML]{FFFFFF} 0.0001} & \cellcolor[HTML]{548235}{\color[HTML]{FFFFFF} 0.0001} & - & - & {\color[HTML]{FFFFFF} } & NI &  & \cellcolor[HTML]{FFF2CC}0.08 & - & \cellcolor[HTML]{A9D08E}{\color[HTML]{FFFFFF} 0.0191} & \cellcolor[HTML]{A9D08E}{\color[HTML]{FFFFFF} 0.0113} & \cellcolor[HTML]{548235}{\color[HTML]{FFFFFF} 0.0025} & \cellcolor[HTML]{548235}{\color[HTML]{FFFFFF} 0.0002} \\
\cellcolor[HTML]{FFF2CC}-0.01 & \cellcolor[HTML]{FFF2CC}-0.81 & - & \cellcolor[HTML]{A9D08E}{\color[HTML]{FFFFFF} 0.0493} & \cellcolor[HTML]{548235}{\color[HTML]{FFFFFF} $<$ 0.0001} & \cellcolor[HTML]{548235}{\color[HTML]{FFFFFF} 0.0001} & {\color[HTML]{FFFFFF} } & OO &  & \cellcolor[HTML]{FFF2CC}0.27 & \cellcolor[HTML]{FFF2CC}0.19 & - & 0.7623 & 0.1808 & \cellcolor[HTML]{548235}{\color[HTML]{FFFFFF} 0.0008} \\
\cellcolor[HTML]{FFF2CC}0.11 & \cellcolor[HTML]{FFF2CC}-0.69 & \cellcolor[HTML]{FFF2CC}0.12 & - & \cellcolor[HTML]{548235}{\color[HTML]{FFFFFF} $<$ 0.0001} & \cellcolor[HTML]{548235}{\color[HTML]{FFFFFF} 0.0001} & {\color[HTML]{FFFFFF} } & OM &  & \cellcolor[HTML]{FFF2CC}0.28 & \cellcolor[HTML]{FFF2CC}0.20 & \cellcolor[HTML]{FFF2CC}0.01 & - & 0.4177 & \cellcolor[HTML]{548235}{\color[HTML]{FFFFFF} 0.0009} \\
\cellcolor[HTML]{FFF2CC}-0.20 & \cellcolor[HTML]{FFF2CC}- & \cellcolor[HTML]{FFF2CC}-0.19 & \cellcolor[HTML]{FFF2CC}-0.31 & - & - &  & MO &  & \cellcolor[HTML]{FFF2CC}0.29 & \cellcolor[HTML]{FFF2CC}0.21 & \cellcolor[HTML]{FFF2CC}0.02 & \cellcolor[HTML]{FFF2CC}0.01 & - & \cellcolor[HTML]{A9D08E}{\color[HTML]{FFFFFF} 0.0289} \\
\cellcolor[HTML]{FFF2CC}-0.20 & \cellcolor[HTML]{FFF2CC}- & \cellcolor[HTML]{FFF2CC}-0.19 & \cellcolor[HTML]{FFF2CC}-0.31 & \cellcolor[HTML]{FFF2CC}- & - &  & MM &  & \cellcolor[HTML]{FFF2CC}0.34 & \cellcolor[HTML]{FFF2CC}0.26 & \cellcolor[HTML]{FFF2CC}0.06 & \cellcolor[HTML]{FFF2CC}0.06 & \cellcolor[HTML]{FFF2CC}0.04 & - \\
\multicolumn{6}{c}{\textbf{quarantine program   participation}} &  &  & \multicolumn{1}{l}{} & \multicolumn{6}{c}{\textbf{quarantined share}} \\
- & \cellcolor[HTML]{548235}{\color[HTML]{FFFFFF} 0.0001} & 0.3447 & \cellcolor[HTML]{548235}{\color[HTML]{FFFFFF} 0.0001} & 0.1927 & \cellcolor[HTML]{548235}{\color[HTML]{FFFFFF} 0.0001} & {\color[HTML]{FFFFFF} } & NA &  & - & \cellcolor[HTML]{548235}{\color[HTML]{FFFFFF} 0.0001} & \cellcolor[HTML]{548235}{\color[HTML]{FFFFFF} 0.0002} & \cellcolor[HTML]{548235}{\color[HTML]{FFFFFF} 0.0003} & \cellcolor[HTML]{548235}{\color[HTML]{FFFFFF} 0.0001} & \cellcolor[HTML]{548235}{\color[HTML]{FFFFFF} 0.0008} \\
\cellcolor[HTML]{FFF2CC}0.63 & - & \cellcolor[HTML]{548235}{\color[HTML]{FFFFFF} 0.0001} & - & \cellcolor[HTML]{548235}{\color[HTML]{FFFFFF} 0.0002} & - & {\color[HTML]{FFFFFF} } & NI &  & \cellcolor[HTML]{FFF2CC}0.0746 & - & \cellcolor[HTML]{548235}{\color[HTML]{FFFFFF} 0.0001} & \cellcolor[HTML]{548235}{\color[HTML]{FFFFFF} 0.0001} & \cellcolor[HTML]{548235}{\color[HTML]{FFFFFF} 0.0002} & \cellcolor[HTML]{548235}{\color[HTML]{FFFFFF} 0.0001} \\
\cellcolor[HTML]{FFF2CC}0.05 & \cellcolor[HTML]{FFF2CC}-0.59 & - & \cellcolor[HTML]{548235}{\color[HTML]{FFFFFF} 0.0001} & 0.9438 & \cellcolor[HTML]{548235}{\color[HTML]{FFFFFF} 0.0001} & {\color[HTML]{FFFFFF} } & OO &  & \cellcolor[HTML]{FFF2CC}0.0459 & \cellcolor[HTML]{FFF2CC}-0.0287 & - & 0.1822 & 0.8861 & \cellcolor[HTML]{A9D08E}{\color[HTML]{FFFFFF} 0.0432} \\
\cellcolor[HTML]{FFF2CC}-0.37 & \cellcolor[HTML]{FFF2CC}- & \cellcolor[HTML]{FFF2CC}-0.41 & - & \cellcolor[HTML]{548235}{\color[HTML]{FFFFFF} $<$ 0.0001} & - &  & OM &  & \cellcolor[HTML]{FFF2CC}0.0372 & \cellcolor[HTML]{FFF2CC}- & \cellcolor[HTML]{FFF2CC}-0.0086 & - & 0.1885 & 0.4018 \\
\cellcolor[HTML]{FFF2CC}0.01 & \cellcolor[HTML]{FFF2CC}-0.62 & \cellcolor[HTML]{FFF2CC}-0.03 & \cellcolor[HTML]{FFF2CC}0.38 & - & \cellcolor[HTML]{548235}{\color[HTML]{FFFFFF} $<$ 0.0001} & {\color[HTML]{FFFFFF} } & MO &  & \cellcolor[HTML]{FFF2CC}0.0459 & \cellcolor[HTML]{FFF2CC}-0.0287 & \cellcolor[HTML]{FFF2CC}0 & \cellcolor[HTML]{FFF2CC}0.0086 & - & \cellcolor[HTML]{C6E0B4}{\color[HTML]{FFFFFF} 0.0748} \\
\cellcolor[HTML]{FFF2CC}-0.37 & \cellcolor[HTML]{FFF2CC}- & \cellcolor[HTML]{FFF2CC}-0.41 & \cellcolor[HTML]{FFF2CC}- & \cellcolor[HTML]{FFF2CC}-0.38 & - &  & MM &  & \cellcolor[HTML]{FFF2CC}0.0372 & \cellcolor[HTML]{FFF2CC}- & \cellcolor[HTML]{FFF2CC}-0.0086 & \cellcolor[HTML]{FFF2CC}0 & \cellcolor[HTML]{FFF2CC}-0.0086 & - \\
\multicolumn{6}{l}{} & \multicolumn{1}{l}{} & \multicolumn{1}{l}{} & \multicolumn{1}{l}{} & \multicolumn{6}{c}{\textbf{individual welfare}} \\
\multicolumn{6}{l}{} & \multicolumn{1}{l}{} & NA & \multicolumn{1}{l}{} & - & 0.1403 & \cellcolor[HTML]{548235}{\color[HTML]{FFFFFF} 0.0022} & \cellcolor[HTML]{548235}{\color[HTML]{FFFFFF} 0.0013} & \cellcolor[HTML]{548235}{\color[HTML]{FFFFFF} 0.0001} & \cellcolor[HTML]{548235}{\color[HTML]{FFFFFF} 0.0002} \\
\multicolumn{6}{l}{} & \multicolumn{1}{l}{} & NI & \multicolumn{1}{l}{} & \cellcolor[HTML]{FFF2CC}4.06 & - & \cellcolor[HTML]{548235}{\color[HTML]{FFFFFF} 0.0036} & \cellcolor[HTML]{548235}{\color[HTML]{FFFFFF} 0.0013} & \cellcolor[HTML]{548235}{\color[HTML]{FFFFFF} 0.0006} & \cellcolor[HTML]{548235}{\color[HTML]{FFFFFF} 0.0002} \\
\multicolumn{6}{l}{} & \multicolumn{1}{l}{} & OO & \multicolumn{1}{l}{} & \cellcolor[HTML]{FFF2CC}-17.15 & \cellcolor[HTML]{FFF2CC}-21.22 & - & 0.3447 & \cellcolor[HTML]{C6E0B4}0.0845 & \cellcolor[HTML]{548235}{\color[HTML]{FFFFFF} 0.0002} \\
\multicolumn{6}{l}{} & \multicolumn{1}{l}{} & OM & \multicolumn{1}{l}{} & \cellcolor[HTML]{FFF2CC}-20.14 & \cellcolor[HTML]{FFF2CC}-24.20 & \cellcolor[HTML]{FFF2CC}-2.99 & - & 0.5495 & \cellcolor[HTML]{548235}{\color[HTML]{FFFFFF} 0.0004} \\
\multicolumn{6}{l}{} & \multicolumn{1}{l}{} & MO & \multicolumn{1}{l}{} & \cellcolor[HTML]{FFF2CC}-23.28 & \cellcolor[HTML]{FFF2CC}-27.34 & \cellcolor[HTML]{FFF2CC}-6.13 & \cellcolor[HTML]{FFF2CC}-3.14 & - & \cellcolor[HTML]{548235}{\color[HTML]{FFFFFF} 0.0004} \\
 \multicolumn{3}{r}{Significance levels:} & \multicolumn{1}{l}{\cellcolor[HTML]{C6E0B4}10\%} & \multicolumn{1}{l}{\cellcolor[HTML]{A9D08E}5\%} & \multicolumn{1}{l}{\cellcolor[HTML]{548235}1\%} & \multicolumn{1}{l}{} & MM & \multicolumn{1}{l}{} & \cellcolor[HTML]{FFF2CC}-29.76 & \cellcolor[HTML]{FFF2CC}-33.83 & \cellcolor[HTML]{FFF2CC}-12.61 & \cellcolor[HTML]{FFF2CC}-9.63 & \cellcolor[HTML]{FFF2CC}-6.49 & - \\
 \midrule
\multicolumn{15}{l}{\multirow{4}{*}{\parbox{21cm}{Notes: H0 for all hypotheses is that there is no statistical difference between column treatment and row treatment wrt individual variables of interest. H1 is always two-sided. All tests are MW. Sample size for all treatments is 10, except for MO with sample size of 11. Lower triangle of each matrix reports [median of column treatment - median of row treatment]. Upper triangle reports corresponding p-values -- those that survive our robustness checks are reported in white.}}}\\
\multicolumn{15}{l}{}\\
\multicolumn{15}{l}{}\\
\multicolumn{15}{l}{}\\
\bottomrule
\end{tabular}%
}
\end{table}

\begin{result}[Tracing]\label{r:tracing}
Both parametric (M2) and non-parametric (MW) analysis finds a ranking for tracing: $1 = MM = MO > OO = NA > OM > 0$.
\end{result}

\begin{result}[Quarantine]\label{r:quarantine}
Both parametric (M3) and non-parametric (MW) analysis finds a ranking for quarantine decisions: $1 = MM = OM > MO = OO = NA > 0$.
\end{result}

Experimentally, subjects' choices of activity do not vary across the treatments. All treatment coefficients in our analysis are statistically insignificant (M1 $p>0.309$, MW $p>0.4181$). Obviously activity is significantly higher in treatment NA (MW $p=0.0001$), but this is trivial.\footnote{There is also a treatment-specific time trend in NI, which we discuss further in \Cref{sec:behav_regularities}.}

\begin{result}\label{r:activity}
Both parametric (M1) and non-parametric (MW) analysis finds a ranking for activity: $1 = NA > MM = OM = NI = OO = OM > 0$.
\end{result}

\begin{table}[ht!]
\caption{Main regression results for individual subject decisions and outcomes, with IV}
\label{tab:full_regs}
\resizebox{\textwidth}{!}{%
\begin{tabular}{@{}lrcrcrcrcrc@{}}
\toprule
\multicolumn{1}{r}{\textbf{Dependent Variable:}} & \multicolumn{2}{c}{Activity} & \multicolumn{2}{c}{Tracing} & \multicolumn{2}{c}{Quarantine} & \multicolumn{2}{c}{Infection} & \multicolumn{2}{c}{Welfare} \\ \midrule
\multicolumn{1}{r}{Model$^{a}$:} & \multicolumn{2}{c}{M1} & \multicolumn{2}{c}{M2} & \multicolumn{2}{c}{M3} & \multicolumn{2}{c}{M4} & \multicolumn{2}{c}{M5} \\
\multicolumn{1}{r}{Data from treatments$^{b}$:} & \multicolumn{2}{c}{all but NA} & \multicolumn{2}{c}{NA, OO, OM} & \multicolumn{2}{c}{NA, OO, MO} & \multicolumn{2}{c}{all} & \multicolumn{2}{c}{all} \\* \midrule
\multicolumn{11}{l}{\textbf{Independent   Variables:}} \\
Treat.NA & -\hspace{6mm} &  & -0.0576*\hspace{4mm} & {\scriptsize (0.0331)} & 0.0398\hspace{6mm} & {\scriptsize (0.0732)} & 0.2280*** & {\scriptsize (0.0178)} & -8.177*** & {\scriptsize (1.886)} \\
Treat.NI & -2.382\hspace{6mm} & {\scriptsize (2.341)} & -\hspace{6mm} &  & -\hspace{6mm} &  & 0.0463\hspace{6mm} & {\scriptsize (0.0365)} & -7.407*\hspace{4mm} & {\scriptsize (3.909)} \\
Treat.OM & 1.986\hspace{6mm} & {\scriptsize (2.041)} & -0.1370*** & {\scriptsize (0.0486)} & -\hspace{6mm} &  & 0.0017\hspace{6mm} & {\scriptsize (0.0218)} & 2.560\hspace{6mm} & {\scriptsize (2.852)} \\
Treat.MO & 0.618\hspace{6mm} & {\scriptsize (2.185)} & -\hspace{6mm} &  & -0.0056\hspace{6mm} & {\scriptsize (0.0710)} & -0.0348\hspace{6mm} & {\scriptsize (0.0221)} & 6.836*** & {\scriptsize (2.636)} \\
Treat.MM & 0.182\hspace{6mm} & {\scriptsize (1.818)} & -\hspace{6mm} &  & -\hspace{6mm} &  & -0.0643*** & {\scriptsize (0.0175)} & 10.11*** & {\scriptsize (2.112)} \\
Round & 0.263*** & {\scriptsize (0.065)} & -0.0006\hspace{6mm} & {\scriptsize (0.0015)} & -0.0100**\hspace{2mm} & {\scriptsize (0.0041)} & 0.0001\hspace{6mm} & {\scriptsize (0.0010)} & 0.281**\hspace{2mm} & {\scriptsize (0.113)} \\
Round \# Treat.NA & -\hspace{6mm} &  & 0.0007\hspace{6mm} & {\scriptsize (0.0016)} & 0.0033\hspace{6mm} & {\scriptsize (0.0045)} & 0.0005\hspace{6mm} & {\scriptsize (0.0010)} & -0.369*** & {\scriptsize (0.120)} \\
Round \# Treat.NI & 0.303**\hspace{2mm} & {\scriptsize (0.128)} & -\hspace{6mm} &  & -\hspace{6mm} &  & 0.0061*** & {\scriptsize (0.0015)} & -0.630*** & {\scriptsize (0.158)} \\
Round \# Treat.OM & 0.041\hspace{6mm} & {\scriptsize (0.081)} & -0.0009\hspace{6mm} & {\scriptsize (0.0022)} & -\hspace{6mm} &  & -0.0011\hspace{6mm} & {\scriptsize (0.0013)} & 0.175\hspace{6mm} & {\scriptsize (0.172)} \\
Round \# Treat.MO & -0.129\hspace{6mm} & {\scriptsize (0.082)} & -\hspace{6mm} &  & 0.0032\hspace{6mm} & {\scriptsize (0.0053)} & -0.0009\hspace{6mm} & {\scriptsize (0.0012)} & -0.016\hspace{6mm} & {\scriptsize (0.166)} \\
Round \# Treat.MM & 0.116\hspace{6mm} & {\scriptsize (0.080)} & -\hspace{6mm} &  & -\hspace{6mm} &  & -0.0021**\hspace{2mm} & {\scriptsize (0.0010)} & 0.403*** & {\scriptsize (0.123)} \\
Ideology & 1.696*** & {\scriptsize (0.415)} & -0.0611*** & {\scriptsize (0.0190)} & 0.0020\hspace{6mm} & {\scriptsize (0.0235)} & 0.0066**\hspace{2mm} & {\scriptsize (0.0029)} & 1.546*** & {\scriptsize (0.490)} \\
Prosocial values & -3.934*** & {\scriptsize (1.127)} & 0.0588*\hspace{4mm} & {\scriptsize (0.0355)} & 0.1580*** & {\scriptsize (0.0480)} & -0.0177*** & {\scriptsize (0.0061)} & -4.071*** & {\scriptsize (1.383)} \\
Risk score & 0.093*** & {\scriptsize (0.028)} & -0.0006\hspace{6mm} & {\scriptsize (0.0009)} & -0.0017\hspace{6mm} & {\scriptsize (0.0014)} & 0.0004**\hspace{2mm} & {\scriptsize (0.0002)} & 0.079**\hspace{2mm} & {\scriptsize (0.036)} \\
Own activity @ t-1 & 0.468*** & {\scriptsize (0.022)} & -\hspace{6mm} &  & -\hspace{6mm} &  & -\hspace{6mm} &  & -\hspace{6mm} &  \\
Infected @ t-1 & 6.646*** & {\scriptsize (1.042)} & -\hspace{6mm} &  & -\hspace{6mm} &  & -\hspace{6mm} &  & -\hspace{6mm} &  \\
Alert @ t-1 & 4.434**\hspace{2mm} & {\scriptsize (2.112)} & -\hspace{6mm} &  & -\hspace{6mm} &  & -\hspace{6mm} &  & -\hspace{6mm} &  \\
Global activity @ t-1 & -0.356*** & {\scriptsize (0.053)} & -\hspace{6mm} &  & -\hspace{6mm} &  & -\hspace{6mm} &  & -\hspace{6mm} &  \\
Number of infected @ t & -1.178*\hspace{4mm} & {\scriptsize (0.705)} & -\hspace{6mm} &  & -\hspace{6mm} &  & -\hspace{6mm} &  & -\hspace{6mm} &  \\
Instructions attempts & -2.729*** & {\scriptsize (0.850)} & -0.0612**\hspace{2mm} & {\scriptsize (0.0296)} & 0.0602\hspace{6mm} & {\scriptsize (0.0421)} & -0.0119**\hspace{2mm} & {\scriptsize (0.0050)} & -1.943*\hspace{4mm} & {\scriptsize (1.024)} \\
Age & -0.057\hspace{6mm} & {\scriptsize (0.046)} & -0.0037**\hspace{2mm} & {\scriptsize (0.0017)} & 0.0029\hspace{6mm} & {\scriptsize (0.0020)} & -0.0003\hspace{6mm} & {\scriptsize (0.0002)} & -0.013\hspace{6mm} & {\scriptsize (0.057)} \\
Sex (female = 1) & -0.557\hspace{6mm} & {\scriptsize (0.973)} & -0.0518\hspace{6mm} & {\scriptsize (0.0389)} & 0.0058\hspace{6mm} & {\scriptsize (0.0444)} & 0.0029\hspace{6mm} & {\scriptsize (0.0073)} & -1.066\hspace{6mm} & {\scriptsize (1.335)} \\
Race = non-white & 0.404\hspace{6mm} & {\scriptsize (1.338)} & 0.0206\hspace{6mm} & {\scriptsize (0.0509)} & 0.0571\hspace{6mm} & {\scriptsize (0.0618)} & -0.0061\hspace{6mm} & {\scriptsize (0.0106)} & 1.166\hspace{6mm} & {\scriptsize (1.846)} \\
Race = white, mixed race & -4.199*\hspace{4mm} & {\scriptsize (2.364)} & 0.0038\hspace{6mm} & {\scriptsize (0.0730)} & -0.153*\hspace{4mm} & {\scriptsize (0.0904)} & -0.0188\hspace{6mm} & {\scriptsize (0.0190)} & -2.815\hspace{6mm} & {\scriptsize (3.173)} \\
Education (years) & -0.048\hspace{6mm} & {\scriptsize (0.250)} & 0.0059\hspace{6mm} & {\scriptsize (0.0092)} & 0.0004\hspace{6mm} & {\scriptsize (0.0102)} & 0.0007\hspace{6mm} & {\scriptsize (0.0014)} & -0.427\hspace{6mm} & {\scriptsize (0.296)} \\
Married & -1.991*\hspace{4mm} & {\scriptsize (1.104)} & 0.0683*\hspace{4mm} & {\scriptsize (0.0409)} & -0.0324\hspace{6mm} & {\scriptsize (0.0434)} & -0.0105\hspace{6mm} & {\scriptsize (0.0093)} & -0.723\hspace{6mm} & {\scriptsize (1.400)} \\
Out of labor force & -2.908*\hspace{4mm} & {\scriptsize (1.708)} & 0.1300*** & {\scriptsize (0.0471)} & 0.1020\hspace{6mm} & {\scriptsize (0.0788)} & -0.0154\hspace{6mm} & {\scriptsize (0.0098)} & -1.325\hspace{6mm} & {\scriptsize (1.802)} \\
Unemployed & 0.033\hspace{6mm} & {\scriptsize (1.939)} & -0.0646\hspace{6mm} & {\scriptsize (0.0611)} & 0.1140\hspace{6mm} & {\scriptsize (0.0830)} & -0.0013\hspace{6mm} & {\scriptsize (0.0142)} & -0.379\hspace{6mm} & {\scriptsize (2.514)} \\
Household income control & Yes &  & Yes &  & Yes &  & Yes &  & Yes &  \\
Federal Region controls & Yes &  & Yes &  & Yes &  & Yes &  & Yes &  \\
Constant & 68.32*** & {\scriptsize (4.995)} & 1.1540*** & {\scriptsize (0.2000)} & 0.4630*\hspace{4mm} & {\scriptsize (0.2520)} & 0.2450*** & {\scriptsize (0.0377)} & 39.11*** & {\scriptsize (6.722)} \\ 
\midrule
Observations: & \multicolumn{2}{c}{17,246} & \multicolumn{2}{c}{10,353} & \multicolumn{2}{c}{859} & \multicolumn{2}{c}{21,199} & \multicolumn{2}{c}{21,199} \\
R-squared: & \multicolumn{2}{c}{0.292} & \multicolumn{2}{c}{0.103} & \multicolumn{2}{c}{0.109} & \multicolumn{2}{c}{0.070} & \multicolumn{2}{c}{0.028} \\
\midrule
\multicolumn{11}{l}{\multirow{6}{*}{\parbox{22cm}{Notes: This table is identical to the regression table in the main text, but does not omit any variables.
Standard errors (reported in parentheses) are clustered at the group level. *** $p<0.01$, ** $p<0.05$, * $p<0.1$. Treatment OO is the baseline in all models.
\textbf{(a)} Activity and Welfare are Ordinary Least-Squares regressions, and the others are Linear Probability Models. All models use a two-stage least squares instrumental variable approach for the ideology variable.
\textbf{(b)} Activity, tracing, and quarantine regressions only use data from treatments where subjects freely made the respective decision. The quarantine regression contains fewer observations because subjects only make this decision after receiving an alert. }}} \\
\multicolumn{11}{l}{} \\
\multicolumn{11}{l}{} \\
\multicolumn{11}{l}{} \\
\multicolumn{11}{l}{} \\
\multicolumn{11}{l}{} \\ \bottomrule
\end{tabular}
}
\end{table}

The key takeaway for policymakers is that the type of tracing scheme and quarantine program used does not appear to affect activity choices. This suggests the economic costs of tracing and quarantine programs -- beyond the direct costs of implementing and running them -- are small. This analysis does not show that tracing and quarantine programs (whether voluntary or mandatory) do not discourage activity in a partial equilibrium setting -- i.e. holding infections constant. Rather, it shows that any reduction in activity caused directly by the programs is fully offset indirectly, through the reduction in infections.

\subsubsection{Outcomes}

\noindent \textbf{Theory.} Predicted infections are lowest in treatment MM because agents are forced to use the quarantine program. Treatments NI and MO have the next lowest number of infections. While agents do not use the quarantine program (which would help control infections), they do not use the contact tracing scheme either. This means they do not ``game the system'' (choose maximal activity when they know they are infected). In contrast, infections are higher in treatments OO and OM precisely because agents \emph{do} ``game the system''. Agents who know they are infected choose maximal activity because they face no personal costs from doing so. This means activity is done \emph{disproportionately} by infected agents, which more than offsets the lower average activity (see \Cref{h:activity}). Finally, infections are highest in treatment NA, as agents do not use quarantine (as it is a public good) and are also unable to reduce infections by limiting their activity.

Welfare in this model is simply the benefits from activity, less the costs of infections. With the exception of treatment NA, the theoretical rankings for activity and infections are the inverse of one another. Therefore, the overall welfare ranking is unambiguous -- it is the same as the ranking for activity. For treatment NA, the theory (trivially) predicts very high activity, and also predicts very high infections. Self-interested agents are unwilling to provide a public good (quarantine) and are unable to protect themselves by reducing their activity. This results in low welfare as the cost of infections outweighs the benefits of high activity.

\begin{theor}\label{h:infections}
The model predicts the following ranking for infections: $NA > OO = MO > NI = OM > MM$.
\end{theor}

\begin{theor}\label{h:welfare}
The model predicts the following ranking for welfare: $MM > OM = NI > OO = MO > NA$.
\end{theor}

\noindent \textbf{Results.} As noted above, activity decisions do not differ systematically across treatments. Therefore the experimental welfare ranking is simply the inverse of the experimental infections ranking.\footnote{With the exception of treatment NA, which has higher activity because it is fixed by the environment.} 
In line with the theory, treatment MM has the lowest level of infections (M4 $p=0.0467$). Fully mandatory tracing and quarantine is successful at keeping infections low. It is also successful at reducing infections over time -- there is a negative time trend in infections for treatment MM (M2 $p=0.041$). Making only one of the tracing and quarantine programs mandatory appears to be ineffective at reducing infections, relative to a fully optional set-up. 

Infections in treatments MO and OM are no lower than in treatment OO (M4 $p>0.115$, MW $p>0.181$), and are higher than treatment MM (t-test $p<0.047$, MW $p<0.0009$ ). There does not appear to be a time trend in infections for either of these treatments. For welfare, these two treatments are not statistically different from one another (t-test $p=0.152$).\footnote{However, treatment MO has statistically higher welfare than OO (M5 $p=0.009$), while treatment OM does not (M5 $p=0.369$). This is due to differences in the time dynamics of participation in optional tracing versus optional quarantine. We discuss this further in \Cref{sec:behav_regularities}.}

Infections are then higher in treatment NI. In the parametric analysis this shows up in a significant time trend in infections (M4 $p<0.0001$), rather than in an immediate effect. Finally, infections are highest in treatment NA, where subjects cannot choose their economic activity. This suggests that voluntary reductions in activity may be more effective at reducing infections than voluntary tracing and quarantine. In terms of welfare, treatment NA does no worse than NI (t-test $p=0.8222$). While infections are higher in NA, this is offset by higher economic activity.

\begin{result}
Both parametric (LPM) and non-parametric (MV) analysis finds a ranking for infections: $NA > NI > OO = MO = OM > MM$.
\end{result}

\begin{result}
Both parametric (OLS, IV) and non-parametric (MW) analysis finds a ranking for welfare: $MM > OO = MO = OM > NI = NA$.
\end{result}

\subsubsection{Supplementary outcomes}

As shown in the left-hand panel of \Cref{fig:extra line graphs}, the results for activity do not change when we look at the group-level averages instead of individual decisions. This is because (1) the fraction of subjects in quarantine is fairly small, and (2) it is similar across all treatments except treatment NA. This means that whatever effect accounting for quarantine does have on overall activity levels applies symmetrically to treatments. While treatment NA does have more agents in quarantine, this is clearly not sufficient to offset the high activity that comes with activity being fixed. Both of these features are shown in the right-hand panel of \Cref{fig:extra line graphs}. In turn, treatment NA has more subjects in quarantine because it has similar participation in the tracing and quarantine programs (see sections immediately above), but greater number of infections. This means more alerts and so more subjects in quarantine.

\begin{result}
Non-parametric (MW) analysis finds the following ranking for group-level activity: $NA > MM = OO = MO = OM = NI$.
\end{result}

\begin{result}
Non-parametric (MW) analysis finds the following ranking for the fraction of agents in quarantine: $NA > NI > MM = OO = MO = OM$.
\end{result}

Non-parametric results are shown in \Cref{tab:nonpar_tab}. We omit parametric analysis here for brevity, as it provides no additional insight.

\begin{table}
\caption{Main regression results for individual subject decisions and outcomes, without IV}
\label{tab:ols_regs}
\resizebox{\textwidth}{!}{%
\begin{tabular}{@{}lcccccccccc@{}}
\toprule
\multicolumn{1}{r}{\textbf{Dependent Variable:}} & \multicolumn{2}{c}{Activity} & \multicolumn{2}{c}{Tracing} & \multicolumn{2}{c}{Quarantine} & \multicolumn{2}{c}{Infection} & \multicolumn{2}{c}{Welfare} \\ \midrule
\multicolumn{1}{r}{Model$^{a}$:} & \multicolumn{2}{c}{M1'} & \multicolumn{2}{c}{M2'} & \multicolumn{2}{c}{M3'} & \multicolumn{2}{c}{M4'} & \multicolumn{2}{c}{M5'} \\
\multicolumn{1}{r}{Data from treatments$^{b}$:} & \multicolumn{2}{c}{all but NA} & \multicolumn{2}{c}{NA, OO, OM} & \multicolumn{2}{c}{NA, OO, MO} & \multicolumn{2}{c}{all} & \multicolumn{2}{c}{all} \\* \midrule
\multicolumn{11}{l}{\textbf{Independent   Variables$^{c}$:}} \\
Ideology & 0.974*** & {\scriptsize (0.271)} & -0.0396*** & {\scriptsize (0.0121)} & 0.0133\hspace{6mm} & {\scriptsize (0.0146)} & 0.0039**\hspace{2mm} & {\scriptsize (0.0017)} & 0.942*** & {\scriptsize (0.340)} \\
Treat.NA & -\hspace{6mm} &  & -0.0434\hspace{6mm} & {\scriptsize (0.0345)} & 0.0462\hspace{6mm} & {\scriptsize (0.0738)} & 0.2260*** & {\scriptsize (0.0181)} & -8.638*** & {\scriptsize (1.860)} \\
Treat.NI & -2.590\hspace{6mm} & {\scriptsize (2.364)} & -\hspace{6mm} &  & -\hspace{6mm} &  & 0.0456\hspace{6mm} & {\scriptsize (0.0369)} & -7.568*\hspace{4mm} & {\scriptsize (3.918)} \\
Treat.OM & 1.614\hspace{6mm} & {\scriptsize (2.064)} & -0.1290**\hspace{2mm} & {\scriptsize (0.0489)} & -\hspace{6mm} &  & 0.0004\hspace{6mm} & {\scriptsize (0.0222)} & 2.259\hspace{6mm} & {\scriptsize (2.819)} \\
Treat.MO & 0.246\hspace{6mm} & {\scriptsize (2.141)} & -\hspace{6mm} &  & 0.0010\hspace{6mm} & {\scriptsize (0.0711)} & -0.0362\hspace{6mm} & {\scriptsize (0.0224)} & 6.515**\hspace{2mm} & {\scriptsize (2.668)} \\
Treat.MM & -0.130\hspace{6mm} & {\scriptsize (1.797)} & -\hspace{6mm} &  & -\hspace{6mm} &  & -0.0654*** & {\scriptsize (0.0178)} & 9.859*** & {\scriptsize (2.083)} \\
Other Controls & Yes &  & Yes &  & Yes &  & Yes &  & Yes &  \\
Constant & 71.470*** & {\scriptsize (4.910)} & 1.0210*** & {\scriptsize (0.1850)} & 0.3610\hspace{6mm} & {\scriptsize (0.2480)} & 0.2570*** & {\scriptsize (0.0345)} & 41.74*** & {\scriptsize (6.790)} \\
\midrule
Observations: & \multicolumn{2}{c}{17,246} & \multicolumn{2}{c}{10,353} & \multicolumn{2}{c}{859} & \multicolumn{2}{c}{21,199} & \multicolumn{2}{c}{21,199} \\
R-squared: & \multicolumn{2}{c}{0.293} & \multicolumn{2}{c}{0.110} & \multicolumn{2}{c}{0.110} & \multicolumn{2}{c}{0.070} & \multicolumn{2}{c}{0.028} \\
 \midrule
\multicolumn{11}{l}{\multirow{5}{*}{\parbox{22cm}{Notes: This table is identical to Table \ref{tab:full_regs}, but does not use instrumental variable approach.
Standard errors (reported in parentheses) are clustered at the group level. *** $p<0.01$, ** $p<0.05$, * $p<0.1$. Treatment OO is the baseline in all models.
\textbf{(a)} Activity and Welfare are Ordinary Least-Squares regressions, and the others are Linear Probability Models.
\textbf{(b)} As in Table \ref{tab:full_regs}. 
\textbf{(c)} Regressions include exactly the same set of controls as the corresponding model in Table \ref{tab:full_regs}. Many are omitted here for brevity. Point estimates and significance levels are not materially different for these variables compared to Table \ref{tab:full_regs}.}}} \\
\multicolumn{11}{l}{} \\
\multicolumn{11}{l}{} \\
\multicolumn{11}{l}{} \\
\multicolumn{11}{l}{} \\ \bottomrule
\end{tabular}
}
\end{table}

\subsection{Ideology}\label{sec:ideology}

\Cref{tab:ols_regs} shows that political ideology has an important association with subjects' decisions in the experiment, and therefore the outcomes they experience. Political conservatism is associated with higher economic activity (\Cref{tab:ols_regs}, M1' $p<0.0008$) and lower rates of participation in the contact tracing scheme (M2' $p = 0.0027$). Consequently, they are more likely to become infected (M4' $p = 0.0271$), but also receive higher welfare on average (M5' $p=0.0075$).

However, subjects were not randomly allocated to their political ideology -- this is clearly not feasible. Therefore, the association between ideology and decisions (and outcomes) may be endogenous. Political ideology could be correlated with some omitted variable that is the true cause of the observed association. We use an Instrumental Variables approach to address this possible endogeneity. We discuss this approach, and the instrument we use, in \Cref{sec:methodology:data_analysis}. To reiterate, the critical assumption underlying this approach is that subjects' opinions regarding the four partisan issues we ask about \emph{only} affect decisions in the game via their effect on ideology. They must have no independent effect on decisions separate from ideology. 

Using the IV approach, we find larger effects of ideology on economic activity and participation in the contact tracing scheme, and also probability of infection and welfare. The magnitude of the point estimates increases by a factor of approximately $1.5-1.8$ (compare Tables \ref{tab:ols_regs} and \ref{tab:full_regs}). The point estimates are statistically different from zero with or without the IV approach. With the IV approach, we can now attach a causal interpretation to the estimates. Political conservatism causes subjects to choose higher economic activity and makes them less likely to participate in the contact tracing scheme.

The point estimates are also economically important. A 4-point increase in our ideology measure (from `moderate liberal' to `moderate conservative', on our 7-point scale) induces a $6.7$ point increase in economic activity (on a scale of $0-100$, where the overall mean is $76.6$) and a $24.4$ percentage point decrease in the probability of signing up to the contact tracing scheme (where the overall mean is $71.6$ percent). In contrast, there is no evidence that ideology affects the probability a subject chooses to go into quarantine \emph{conditional on} receiving an alert from the contact tracing scheme. While the point estimate is close to zero, the standard error is relatively large. This is partly because there are not many observations in our dataset, so the effect is likely to be imprecisely estimated. 

These decisions then feed through into infection probability and overall welfare. A 4-point increase in our ideology measure induces a $2.6$ percentage point increase in the probability of infection (where the overall mean is $26.4$ percent), and a $6.2$ point increase in welfare (where the overall mean is $36.9$).

We can interpret the higher activity and lower probability of signing up to contact tracing as partial free-riding by political conservatives.\footnote{The free-riding is only partial as political conservatives do experience some costs to their actions -- they are more likely to become infected.} 
These benefits to political conservatives are at the individual level. Implicitly, the regression analysis only considers the effect of changing a single individual's ideology on her own decisions and outcomes. It treats the ideological make-up of the rest of the group as given. These benefits may be due to the mix of political ideology present in each group. That is, the benefits political conservatives obtain by their partial free-riding may be \emph{at the expense of political liberals}. 

It is not possible to directly test this hypothesis using the experimental data -- an individual's decisions (and outcomes) depend on the decisions made by the other members of their group. Therefore, if group composition changes, we would expect behavior to change accordingly. Simply taking the observed decisions of subjects of a given political ideology would miss these behavioral changes. We carry out a set of simulations in \Cref{sec:simulations} to better understand the impact of ideology at the \emph{group} level. That is, the impact of ideology when the group is more homogeneous with respect to ideology. \\

\noindent \textbf{A comment on causality.} While our IV approach allows us to identify the causal impact of political ideology, it is important to remember that ideology only affects \emph{decisions} directly. The effect on infections and welfare is indirect -- ideology causes decisions, which \emph{in turn} cause outcomes. In many respects, this point is obvious. Nevertheless, to provide a sense check we add subjects' decisions to the main regressions for infections and welfare.

As expected, the ideology variable is no longer significant (\Cref{tab:control_decisions}). It does in fact only operate through subjects' decisions. Unsurprisingly, controlling for subjects' decisions also removes the statistical significance of other individual characteristics (the next subsection shows that certain other characteristics are associated with decisions and outcomes).

\begin{table}[th!]
\centering
\caption{Regression results for main outcomes, controlling for individual decisions, with IV}
\label{tab:control_decisions}
\begin{tabular}{@{}lcccc@{}}
\toprule
\multicolumn{1}{r}{\textbf{Dependent   Variable:}} & \multicolumn{2}{c}{Infection} & \multicolumn{2}{c}{Welfare} \\ \midrule
\multicolumn{1}{r}{Model:} & \multicolumn{2}{c}{M4$^{\dagger}$} & \multicolumn{2}{c}{M5$^{\dagger}$} \\
\multicolumn{1}{r}{Data from treatments:} & \multicolumn{2}{c}{all} & \multicolumn{2}{c}{all} \\ \midrule
\multicolumn{5}{l}{\textbf{Independent   Variables:}} \\ \midrule
Treat.NA & 0.1420*** & (2.387) & -21.35*** & (2.387) \\
Treat.NI & 0.0563*\hspace{4mm} & (4.768) & -8.450*\hspace{4mm} & (4.768) \\
Treat.OM & -0.0072\hspace{6mm} & (2.761) & 1.074\hspace{6mm} & (2.761) \\
Treat.MO & -0.0332*\hspace{4mm} & (2.599) & 4.985*\hspace{4mm} & (2.599) \\
Treat.MM & -0.0636*** & (2.076) & 9.536*** & (2.076) \\
Round & -0.000002\hspace{2mm} & (0.145) & 0.0004\hspace{4mm} & (0.145) \\
Round \# Treat.NA & 0.0008\hspace{6mm} & (0.151) & -0.120\hspace{6mm} & (0.151) \\
Round \# Treat.NI & 0.0053*** & (0.206) & -0.787*** & (0.206) \\
Round \# Treat.OM & -0.0014\hspace{6mm} & (0.182) & 0.212\hspace{6mm} & (0.182) \\
Round \# Treat.MO & -0.0006\hspace{6mm} & (0.183) & 0.090\hspace{6mm} & (0.183) \\
Round \# Treat.MM & -0.0026*** & (0.147) & 0.397*** & (0.147) \\
Ideology & 0.000\hspace{6mm} & (0.366) & -0.053\hspace{6mm} & (0.366) \\
Prosocial values & -0.0063\hspace{6mm} & (0.825) & 0.948\hspace{6mm} & (0.825) \\
Risk score & 0.0001\hspace{6mm} & (0.019) & -0.008\hspace{6mm} & (0.019) \\
Own activity @ t & -0.0005*** & (0.024) & 1.075*** & (0.024) \\
Own activity @ t-1 & 0.0030*** & (0.028) & -0.445*** & (0.028) \\
Tracing participation @ t & -0.0207**\hspace{2mm} & (1.384) & 3.110**\hspace{2mm} & (1.384) \\
In quarantine @ t & 0.6860*** & (4.536) & -102.9*** & (4.536) \\
Constant & 0.0559\hspace{6mm} & (5.606) & -8.384\hspace{6mm} & (5.606) \\ \midrule
Observations: & \multicolumn{2}{c}{21,199} & \multicolumn{2}{c}{21,199} \\
R-squared: & \multicolumn{2}{c}{0.208} & \multicolumn{2}{c}{0.379} \\ \midrule
\multicolumn{5}{l}{\multirow{7}{*}{\parbox{13cm}{Notes: Regression specifications are identical to models M4 and M5 in Table \ref{tab:full_regs}, but add subjects' own decisions.
Standard errors (reported in parentheses) are clustered at the group level. *** $p<0.01$, ** $p<0.05$, * $p<0.1$. Treatment OO is the baseline in all models. Infections uses a Linear Probability Model, and Welfare uses an Ordinary Least-Squares regression. Both models use a two-stage least squares instrumental variable approach for the ideology variable. }}}                                   \\
\multicolumn{5}{l}{} \\
\multicolumn{5}{l}{} \\
\multicolumn{5}{l}{} \\
\multicolumn{5}{l}{} \\
\multicolumn{5}{l}{} \\
\multicolumn{5}{l}{} \\ \bottomrule
\end{tabular}%
\end{table}

\subsection{Other behavioral regularities}\label{sec:behav_regularities}

In this subsection we examine a set of responses to the environment. First, we show that some subjects use information from the contact tracing scheme to ``game the system''. That is, they choose \emph{higher} economic activity after receiving an alert. Second, we show that mandatory quarantine does dissuade subjects from participating in the contact tracing scheme. These are both in line with the predictions of our stylized model. Next, we explore some time dynamics. We show that activity is increasing at a \emph{faster rate} in treatment NI than in other treatments.\\

\noindent \textbf{Gaming the system.} 
For the totally self-interested agents in our stylized model, the key benefit of the contact tracing scheme is that they can use it to ``game the system''. When an agent receives an alert in round $t$, she know that she will be infected in round $t+1$. As a consequence of the SIS framework (which does not allow consecutive periods of infection), she also knows she will be healthy in round $t+2$. Therefore her activity choice at $t+1$ cannot affect her health at any point in time.\footnote{Except very indirectly, through its impact on the overall number of infections, which may in turn influence the probability she becomes infected in the future. This channel does not exist in the large population setting, but can be present in small groups.} 
Therefore it is optimal to choose activity of 100. Unsurprisingly, our experimental finding is not this stark. Nevertheless, we do find evidence that for subjects who do not go into quarantine, activity choices are higher following an alert (\Cref{tab:full_regs}, M1 $p=0.0358$). At least some subjects do ``game the system'' to an extent. \\

\begin{figure}[ht]
    \centering
    \begin{subfigure}{.5\textwidth}
        \includegraphics[width=\linewidth]{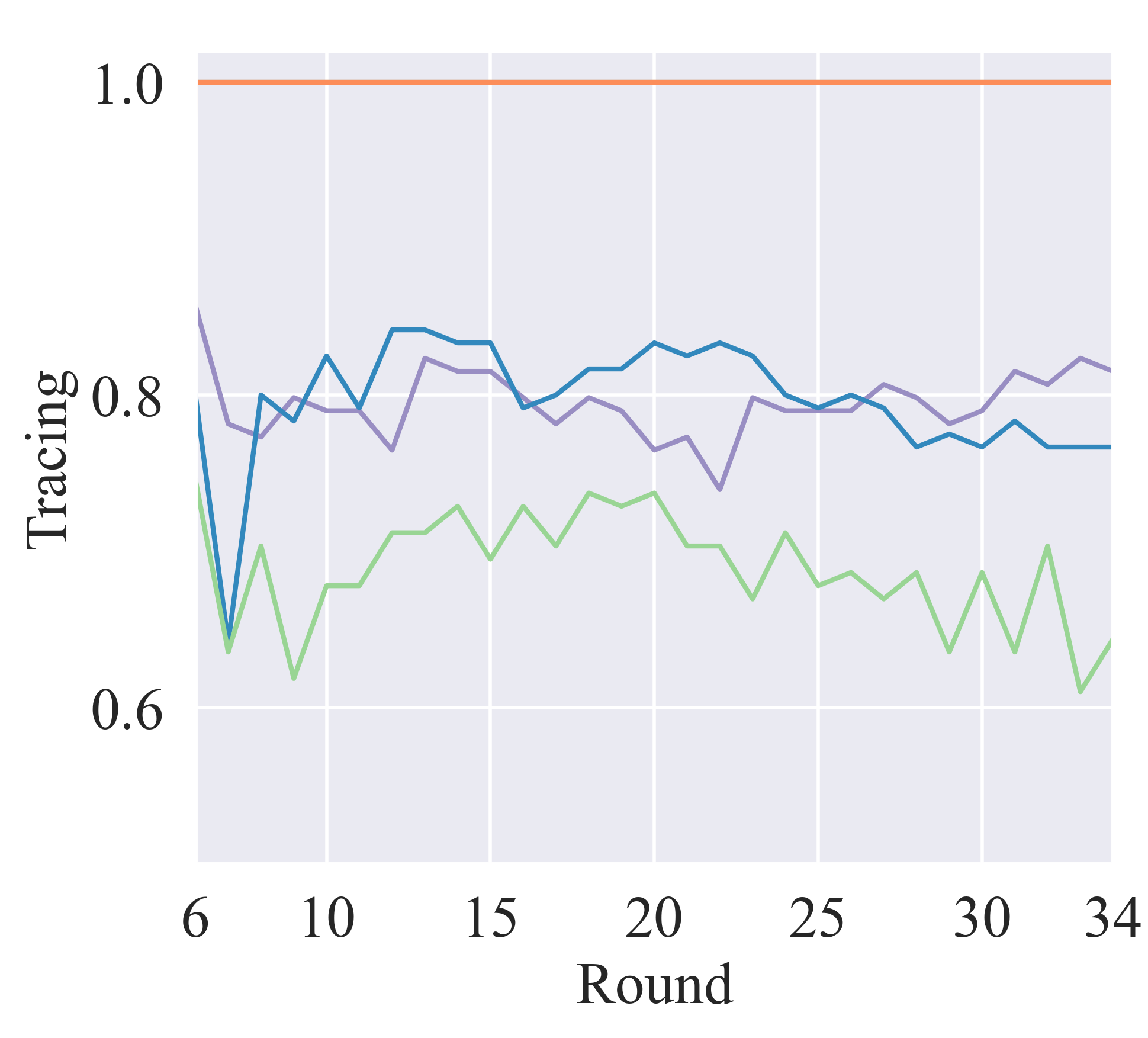}
    \end{subfigure}
    \begin{subfigure}{.15\textwidth}
        \includegraphics[width=\linewidth]{figures/other/legend.png}
    \end{subfigure}
    \caption{Treatment-level means of proportion of subjects participating in the contact tracing scheme over time. COVID-19 is not introduced until Round 6, so Rounds 1-5 are omitted.}
    \label{fig:tracing signup}
\end{figure}

\noindent \textbf{Quarantine discourages tracing.} Our model predicts that when quarantine is optional \emph{everyone} participates in the contact tracing scheme, while when quarantine is mandatory, \emph{nobody} does (unless they are forced to do so). Again, our experimental results are less stark. Nevertheless, there is a clear effect -- shown in \Cref{fig:tracing signup}. Subjects participate in tracing $81\%$ of the time in treatment OO, but only $69\%$ of the time in treatment OM. This difference is highly significant (M2 $p=0.0047$, MW $p=0.0493$).

\noindent \textbf{Time dynamics.} In our main parametric analysis we include a linear time trend and a full interaction with the treatment controls. This allows us to examine time trends in behavior and outcomes, and whether they are peculiar to certain treatments. The most obvious reason for the presence of a time trend is learning and convergence to a steady state. 

For example, our main regressions in \Cref{tab:full_regs} show that the rate of participation in quarantine is falling over time (M3 $p=0.0157$) and that activity is rising over time (M1 $p<0.0001$). Both of these effects are driven by the first 10 rounds after COVID-19 is introduced. \Cref{tab:trend_regs}, which repeats the analysis from \Cref{tab:full_regs} but drops data from these first 10 rounds, shows this. Looking at the corresponding regressions for activity and for quarantine, we can see that the round control is no longer significant. Therefore they appear to be learning and/or convergence effects.

In contrast, the time trend in activity specific to treatment NI remains. When subjects do not have access to the tracing and quarantine programs, they appear to increase their activity choices over time. Model M1'' of \Cref{tab:trend_regs} shows that it is statistically and economically significant. It is not immediately clear what is driving this behavior. A possible interpretation is a ``fatigue'' effect, where subjects give up on trying to moderate their economic activity. 

\begin{table}[ht!]
\centering
\caption{Main regression results for individual subject decisions, excluding data from first 10 rounds, with IV}
\label{tab:trend_regs}
\begin{tabular}{@{}lcccccc@{}}
\toprule
\multicolumn{1}{r}{\textbf{Dependent Variable:}} & \multicolumn{2}{c}{Activity} & \multicolumn{2}{c}{Tracing} & \multicolumn{2}{c}{Quarantine} \\
\midrule
\multicolumn{1}{r}{Model:} & \multicolumn{2}{c}{M1''} & \multicolumn{2}{c}{M2''} & \multicolumn{2}{c}{M3''} \\
\multicolumn{1}{r}{Data from treatments:} & \multicolumn{2}{c}{all but NA} & \multicolumn{2}{c}{NA, OO, OM} & \multicolumn{2}{c}{NA, OO, MO} \\
\midrule
\multicolumn{7}{l}{\textbf{Independent   Variables:}} \\
Treat.NA & -\hspace{6mm} &  & -0.1470*** & (0.0476) & 0.0261\hspace{6mm} & (0.1750) \\
Treat.NI & -6.164\hspace{6mm} & (4.211) & -\hspace{6mm} &  & -\hspace{6mm} &  \\
Treat.OM & -0.622\hspace{6mm} & (3.179) & -0.1150*\hspace{4mm} & (0.0656) & -\hspace{6mm} &  \\
Treat.MO & -1.109\hspace{6mm} & (3.036) & -\hspace{6mm} &  & -0.1120\hspace{6mm} & (0.1680) \\
Treat.MM & -0.136\hspace{6mm} & (3.198) & -\hspace{6mm} &  & -\hspace{6mm} &  \\
Round & 0.030\hspace{6mm} & (0.096) & -0.0032**\hspace{2mm} & (0.0016) & -0.0082\hspace{6mm} & (0.0081) \\
Round \# Treat.NA & -\hspace{6mm} &  & 0.0051**\hspace{2mm} & (0.0020) & 0.0038\hspace{6mm} & (0.0096) \\
Round \# Treat.NI & 0.415**\hspace{2mm} & (0.180) & -\hspace{6mm} &  & -\hspace{6mm} &  \\
Round \# Treat.OM & 0.172\hspace{6mm} & (0.142) & -0.0019\hspace{6mm} & (0.0025) & -\hspace{6mm} &  \\
Round \# Treat.MO & -0.044\hspace{6mm} & (0.124) & -\hspace{6mm} &  & 0.0066\hspace{6mm} & (0.0093) \\
Round \# Treat.MM & 0.168\hspace{6mm} & (0.115) & -\hspace{6mm} &  & -\hspace{6mm} &  \\
Ideology & 1.665*** & (0.406) & -0.0626*** & (0.0195) & 0.0015\hspace{6mm} & (0.0270) \\
Prosovial values & -4.117*** & (1.078) & 0.0574\hspace{6mm} & (0.0372) & 0.1540*** & (0.0481) \\
Risk score & 0.090*** & (0.027) & -0.0008\hspace{6mm} & (0.0010) & -0.0019\hspace{6mm} & (0.0015) \\
Own activity @ t-1 & 0.471*** & (0.023) & -\hspace{6mm} &  & -\hspace{6mm} &  \\
Infected @ t-1 & 9.672*** & (1.052) & -\hspace{6mm} &  & -\hspace{6mm} &  \\
Alert @ t-1 & 6.751*** & (1.762) & -\hspace{6mm} &  & -\hspace{6mm} &  \\
Global activity @ t-1 & -0.400*** & (0.070) & -\hspace{6mm} &  & -\hspace{6mm} &  \\
Number of infected @ t & -0.733\hspace{6mm} & (0.819) & -\hspace{6mm} &  & -\hspace{6mm} &  \\
Other controls & Yes &   & Yes &  & Yes &  \\
Constant & 74.98*** & (6.190) & 1.2370*** & (0.2100) & 0.5300*\hspace{4mm} & (0.3000) \\
\midrule
Observations: & \multicolumn{2}{c}{11,295} & \multicolumn{2}{c}{6,783} & \multicolumn{2}{c}{574} \\
R-squared: & \multicolumn{2}{c}{0.304} & \multicolumn{2}{c}{0.111} & \multicolumn{2}{c}{0.119} \\
\midrule
\multicolumn{7}{l}{\multirow{7}{*}{\parbox{16cm}{Notes: Regressions in this table are identical to the those in Table \ref{tab:full_regs}, but exclude data from the first 10 rounds after the introduction of COVID-19. These are Rounds 6-15 of the experiment.
Standard errors (reported in parentheses) are clustered at the group level. *** $p<0.01$, ** $p<0.05$, * $p<0.1$. Treatment OO is the baseline in all models.
\textbf{(a)} Activity uses an OLS regression, and Tracing and Quarantine use an LPM model. All models use a two-stage least squares instrumental variable approach for the ideology variable.
\textbf{(b)} Regressions only use data from treatments where subjects freely made the respective decision. }}} \\
\multicolumn{7}{l}{} \\
\multicolumn{7}{l}{} \\
\multicolumn{7}{l}{} \\
\multicolumn{7}{l}{} \\
\multicolumn{7}{l}{} \\
\multicolumn{7}{l}{} \\ \bottomrule
\end{tabular}%
\end{table}

\subsection{Other individual characteristics}\label{sec:characteristics}

Our preferred regressions contain a battery of demographic and preference controls, in addition to all of the variables discussed earlier in this section. \Cref{tab:full_regs} (which repeats Table 1 from the main text, but showing all variables) shows that most demographic controls have no statistically significant association with either subject's decisions or outcomes.

The only controls showing significant associations with decisions and outcomes are social values, risk preferences, and understanding of the experimental instructions. Understanding of the experimental instructions is proxied by the number of attempts subjects required to complete a short quiz at the end of the instructions.\footnote{Potential subjects who failed this quiz 3 times were not allowed to participate in the experiment. So all subjects in our dataset failed this quiz at most twice.}\\

\noindent \textbf{Social values.} Subjects with ``pro-social'' values\footnote{We group subjects into two categories: `pro-social' and `other'. All but 2 subjects in the `other' category are individualistic.} do lower activity ($3.9$ points, M1 $p=0.0005$), are marginally more likely to sign up to the contact tracing program (approx $5.9$ pct. points, M2 $p=0.098$), and are significantly more likely to go into quarantine conditional on receiving an alert ($15.8$ pct. points, M3 $p=0.001$).

As a consequence, they are less likely to be infected ($1.8$ pct. points, M4 $p=0.004$) and have lower overall welfare ($4.1$ points, M5 $p=0.003$). For all decisions and outcomes, except quarantine, these effects are similar to a $1-3$ point decrease in our ideology measure. That is, along most lines, having ``pro-social'' values has a similar impact to that of a more liberal political ideology. Quarantine decisions, however, stand in stark contrast. Pro-social values are associated with a very large increase in the probability of going into quarantine, whereas there is no evidence that political ideology has an impact on quarantine decisions. An important simplification of our set-up is that if a subject receives an alert, then she is definitely infected in the next round. This may lead to different decision-making compared to a scenario where false positives are likely.\\

\noindent \textbf{Risk preferences.} The effect of risk preferences is in line with basic economic intuition. Economic activity yields known benefits for certain, but also entails a risk of infection (and the associated costs). Unsurprisingly, subjects who are more willing to take risk do more economic activity in the experiment.\footnote{Note that our measure of risk preferences is coded \emph{positively} -- higher numbers indicate greater willingness to take risk, \emph{not} greater aversion to risk.} Further, risk preferences are not associated with decisions to participate in the contract tracing scheme or the quarantine program. This is intuitive -- as neither of these decisions entail taking any risk (the tracing scheme provides better information, and the quarantine involves a deterministic outcome when a subject has definitely been exposed to the disease). Higher activity then feeds into a higher probability of being infected and higher overall welfare. It is not a given that welfare increases -- in principle the effect could go either way. However, in this experiment the benefits from higher activity associated with greater willingness to take risks more that outweigh the corresponding costs from increased infections.\\

\noindent \textbf{Understanding of the experiment.} Subjects who took more attempts to answer the short quiz that followed the experimental instructions choose lower activity on average (M1 $p=0.001$) and are less likely to participate in the contact tracing scheme (M2 $p=0.04$). While they are no more or less likely to choose to go into quarantine conditional on receiving an alert (M3 $p=0.153$) this is again imprecisely estimated, likely due to the small number of observations for this decision. These effects are not small -- taking on additional attempt to pass the quiz has a similar sized effect on decisions to moving one point to the right on our ideology scale.

\subsection{Consistency with survey data}\label{sec:survey_data}

As part of our post-experimental survey, we collect data on subjects' attitudes towards, and support for, a range of interventions -- including social distancing, contact tracing and quarantine. 

\begin{table}[th!]
\centering
\caption{Regression results for individual subject decisions, with policy support indices}
\label{tab:regs_survey}
\resizebox{\textwidth}{!}{%
\begin{tabular}{lcccccc}
\toprule
\multicolumn{1}{r}{\textbf{Dependent  Variable:}} & \multicolumn{2}{c}{Activity} & \multicolumn{2}{c}{Tracing} & \multicolumn{2}{c}{Quarantine} \\
\midrule
\multicolumn{1}{r}{Model:} & \multicolumn{2}{c}{M1$^{*}$} & \multicolumn{2}{c}{M2$^{*}$} & \multicolumn{2}{c}{M3$^{*}$} \\
\multicolumn{1}{r}{Data from treatments:} & \multicolumn{2}{c}{all but NA} & \multicolumn{2}{c}{NA, OO, OM} & \multicolumn{2}{c}{NA, OO, MO} \\
\midrule
\multicolumn{7}{l}{\textbf{Independent   Variables:}} \\
Treat.NA & -\hspace{8mm} &  & -0.034\hspace{8mm} & (0.032) & 0.104\hspace{6mm} & (0.069) \\
Treat.NI & 2.242\hspace{10mm} & (2.832) & -\hspace{6mm} &  & -\hspace{6mm} &  \\
Treat.OM & 1.079\hspace{10mm} & (2.258) & -0.118***\hspace{2mm} & (0.038) & -\hspace{6mm} &  \\
Treat.MO & -0.492\hspace{10mm} & (2.451) & -\hspace{6mm} &  & 0.017\hspace{6mm} & (0.072) \\
Treat.MM & 2.071\hspace{8mm} & (2.567) & -\hspace{6mm} &  & -\hspace{6mm} &  \\
Social distancing support index & -2.072***\hspace{2mm} & (0.475) & 0.031**\hspace{2mm} & (0.011) & 0.028\hspace{6mm} & (0.024) \\
Contact tracing support index & -0.330\hspace{8mm} & (0.272) & 0.017*** & (0.006) & -0.012\hspace{8mm} & (0.009) \\
Quarantine support index & -0.412\hspace{8mm} & (0.415) & 0.006\hspace{6mm} & (0.009) & 0.019\hspace{6mm} & (0.014) \\
Constant & 91.97*** & (2.648) & 0.463*** & (0.071) & 0.308**\hspace{2mm} & (0.113) \\
\midrule
Observations: & \multicolumn{2}{c}{17,246} & \multicolumn{2}{c}{10,353} & \multicolumn{2}{c}{859} \\
R-squared: & \multicolumn{2}{c}{0.0398} & \multicolumn{2}{c}{0.0982} & \multicolumn{2}{c}{0.0316} \\ \midrule
\multicolumn{7}{l}{\multirow{2}{*}{\parbox{17cm}{Notes: Standard errors (reported in parentheses) are clustered at the group level. *** $p<0.01$, ** $p<0.05$, * $p<0.1$. Treatment OO is the baseline in all models.}}} \\
\multicolumn{7}{l}{} \\ \bottomrule
\end{tabular}%
}
\end{table}

We use this information to create indices of support for each of these three intervention mechanisms.\footnote{We have two questions for social distancing attitudes: (1) \emph{Reduction of social activity creates a strain on the economy which is not justified. (*)} (2) \emph{In light of the COVID-19 outbreak, I have reduced my social activity (e.g. by not going to restaurants and avoiding large gatherings of people).} For tracing, there are three questions: (1) \emph{In light of the COVID-19 outbreak, I would be comfortable sharing information with a public health official on places I have recently visited.} (2) \emph{In light of the COVID-19 outbreak, I would be comfortable sharing information with a public health official on the names of people I might have been in contact with recently.} (3) \emph{In light of the COVID-19 outbreak, I would be comfortable sharing information with a public health official on location data from my cell phone.}. Finally. for quarantine there are also three questions: (1) \emph{I would quarantine myself for at least 14 days if a public health official has told me I have been in contact with somebody who tested positively for COVID-19.} (2) \emph{Everyone who has tested positive for COVID-19 should be required to quarantine themselves. Those who do not comply should be ticketed and fined.} (3) \emph{Those who have tested positive for COVID-19 should be required to stay at a central isolation location (like a dorm or a field hospital), where medical staff would monitor their health. They should not be allowed to go home until they are cleared from isolation by the appropriate public health authorities.} Responses to all questions are on a 5-point Likert scale, with higher numbers indicating stronger support for the statement, and additional options for ``Don't know'' and ``Refuse to answer''. We then add the values together for each one of social distancing, tracing and quarantine, to obtain three separate indices of support for the corresponding measure. Note that answer to \emph{(*)} enters the index for social distancing support negatively.} Higher values of these indices indicate stronger self-reported support for social distancing, contact tracing and quarantine measures. 

We check whether self-reported attitudes to these COVID-19 related interventions are consistent with subjects' decisions in the experiment. To do this, we regress activity, tracing and quarantine decisions on the set of treatment dummies, plus the three support indices as defined above. The output is in \Cref{tab:regs_survey}.

The results are consistent. We find that subjects who support social distancing do \emph{less} activity in the experiment (M1$^{*}$ $p=0.0001$). The other two indices are not significant in the regression for activity decisions. When it comes to tracing, subjects who support social distancing and contact tracing do \emph{more} tracing (M2$^{*}$ $p=0.0113$ and $p=0.0049$ respectively). Finally, none of the three indices appears to be significantly associated with quarantine decisions (M3$^{*}$ $p>0.1663$), but this may be due to the low number of observations for these decisions.

Overall, these results show that subjects' behavior in the experiment and self-reported support for COVID-19 related interventions are internally consistent. As our experiment is highly stylized, this analysis provides some external validity for our results in this section.

\section{Simulations} \label{sec:simulations}
We carry out a set of simulations to better understand the effects of ideology. \Cref{subsec:sims method} explains how we carry out these simulations. \Cref{subsec:sims results} reports the results.

\subsection{Methodology}\label{subsec:sims method}
First, we form a group a 12 agents. The group is our basic unit of analysis for the simulations -- as individual agents' decisions depend on what others in the group do. \\

\noindent \textbf{Endowments.} We endow each simulated agent with: (1) a risk aversion score (drawn from a normal distribution with location and scale parameters 36.65 and 19.32 respectively, truncated at 2 and 100), (2) whether she is prosocial (random uniform draw wp 54.7\%), and (3) an instructions understanding score (random draw s.t. the agent passes the quiz at the first attempt with probability 72.8\%, and needs a further 1 or 2 attempts with probability 21.3\% and 5.1\% respectively). The distributions in (1)-(3) are calibrated from our main (experimental) dataset. We do not account for other demographics in our simulations as our analysis finds that they are not a significant (and/or robust) determinants of decisions.

We also endow each agent with an ideology score -- an integer on [1,7]. 1 stands for ‘strongly liberal’ and 7 for ‘strongly conservative’. As we are trying to understand the effects of ideology at the \emph{group level} we endow all agents in a given group with similar ideologies. Agents in a group are either: all liberal (ideology score of 1 or 2), all moderate (ideology score of 3, 4, or 5), or all conservative (ideology score of 6 or 7). The proportions assigned to each score are drawn from the distribution in our main experimental dataset.\\

\noindent \textbf{Decisions.} We introduce the random COVID-19 infections, which happen after Round 5 in the experiment, in Round 1 ,and then simulate for 29 rounds. Where a treatment allows agents to make a decision regarding activity, simulated activity choices are the predicted values from a regression analysis. We use a reduced version of specification M1 from \Cref{tab:full_regs}, omitting demographic characteristics and any other variables that are not significant at 5\%.\footnote{This helps us better match the moments of the experimental dataset. Using coefficients from M1(\Cref{tab:full_regs}) directly does not change the qualitative predictions of these simulations, but lowers the accuracy of the fit of simulated to experimental data.} 

Similarly, we use simulated tracing system sign-up and quarantine participation decisions predicted values from modified versions of M2 and M3 (\Cref{tab:full_regs}), respectively. The rules for excluding/including variables are the same as those for activity. Note that ideology score features in modeling tracing system sign-up but not quarantine participation -- as dictated by the parametric analysis.\\

\noindent \textbf{Runs.} We simulate behavior and outcomes for each group under each of the six different treatments (NA, NI, OO, OM, MO, and MM). We can therefore compare groups that are identical except for the treatment. This helps isolate the effect of ideology \emph{and} how it might interact with the treatment.

To allow more direct comparability to the experimental dataset, we simulate 10 groups with each ideology under each of the six treatments, for a total of 180 group-level observations (one observation is the full set of decisions and outcomes for 12 agents over the 29 rounds). This process described above constitutes 1 `run' of the simulation. In everything that follows, results and analysis are based on 1000 `runs'.

\begin{table}[ht]
\centering
\caption{Moments of the actual dataset and calibrated dataset from simulations}
\resizebox{\textwidth}{!}{%
\label{tab:calibration_to_acta}
\begin{tabular}{@{}rllllllllllllll@{}}
\toprule
 & \multicolumn{2}{c}{\multirow{2}{*}{\begin{tabular}[c]{@{}c@{}}individual\\ activity\end{tabular}}} & \multicolumn{2}{c}{\multirow{2}{*}{\begin{tabular}[c]{@{}c@{}}tracing\\ signup\end{tabular}}} & \multicolumn{2}{c}{\multirow{2}{*}{\begin{tabular}[c]{@{}c@{}}quarantine\\ participation\end{tabular}}} & \multicolumn{2}{c}{\multirow{2}{*}{\begin{tabular}[c]{@{}c@{}}global\\ activity\end{tabular}}} & \multicolumn{2}{c}{\multirow{2}{*}{\begin{tabular}[c]{@{}c@{}}individual\\ welfare\end{tabular}}} & \multicolumn{2}{c}{\multirow{2}{*}{infected}} & \multicolumn{2}{c}{\multirow{2}{*}{quarantined}} \\
\multicolumn{1}{l}{} & \multicolumn{2}{c}{} & \multicolumn{2}{c}{} & \multicolumn{2}{c}{} & \multicolumn{2}{c}{} & \multicolumn{2}{c}{} & \multicolumn{2}{c}{} & \multicolumn{2}{c}{} \\ \midrule
 & \multicolumn{14}{l}{\textbf{data from the actual experiment}} \\ \midrule

NA & 100 & (0) & 0.80 & (0.40) & 0.63 & (0.48) & 92 & (26) & 23 & (87) & 0.47 & (0.50) & 0.08 & (0.26) \\
NI & 76 & (30) & \multicolumn{1}{r}{-} &  & \multicolumn{1}{r}{-} &  & 76 & (30) & 22 & (77) & 0.37 & (0.48) & \multicolumn{1}{r}{-} &  \\
OO & 75 & (29) & 0.80 & (0.4) & 0.52 & (0.50) & 73 & (32) & 38 & (73) & 0.23 & (0.42) & 0.03 & (0.17) \\
OM & 76 & (28) & 0.69 & (0.46) & 1.00 & (0) & 73 & (31) & 41 & (71) & 0.21 & (0.41) & 0.04 & (0.19) \\
MO & 74 & (31) & 1.00 & (0) & 0.55 & (0.50) & 71 & (33) & 44 & (68) & 0.18 & (0.38) & 0.03 & (0.17) \\
MM & 77 & (29) & 1.00 & (0) & 1.00 & (0) & 74 & (32) & 53 & (66) & 0.13 & (0.34) & 0.04 & (0.20) \\ \midrule
 & \multicolumn{14}{l}{\textbf{data from simulations}} \\ \midrule
NA & 100 & (0) & 0.77 & (0.42) & 0.61 & (0.49) & 93 & (26) & 23 & (86) & 0.46 & (0.500) & 0.07 & (0.26) \\
NI & 76 & (9) & \multicolumn{1}{r}{-} &  & \multicolumn{1}{r}{-} &  & 76 & (9) & 20 & (73) & 0.38 & (0.48) & \multicolumn{1}{r}{-} &  \\
OO & 75 & (10) & 0.83 & (0.38) & 0.50 & (0.50) & 73 & (16) & 40 & (67) & 0.22 & (0.41) & 0.03 & (0.16) \\
OM & 77 & (11) & 0.69 & (0.46) & 1.00 & (0) & 74 & (18) & 44 & (68) & 0.20 & (0.40) & 0.04 & (0.20) \\
MO & 73 & (9) & 1.00 & (0) & 0.52 & (0.50) & 72 & (14) & 46 & (61) & 0.17 & (0.38) & 0.02 & (0.15) \\
MM & 77 & (11) & 1.00 & (0) & 1.00 & (0) & 74 & (18) & 53 & (61) & 0.14 & (0.35) & 0.04 & (0.19) \\ \midrule
\multicolumn{15}{l}{\multirow{2}{*}{Notes: we report mean values of the variables of interest and standard deviations (in parentheses).}} \\
\multicolumn{15}{l}{} \\ \bottomrule
\end{tabular}%
}
\end{table}

\noindent \textbf{Consistency.} To check that our simulations generate similar data (at least in terms of moments), we run simulations as described above except that the ideology score for each subject is a random draw on $[1,7]$ from a distribution of our experimental dataset. \Cref{tab:calibration_to_acta} shows means of the variables of interest (both decisions and outcomes) from the actual experiment (upper panel), and from the data generated through simulations (lower panel). The means are matched very closely. Simulated standard deviations also match the experimental dataset, except for activity and welfare. Simulations substantially underestimate variability of activity, and this feeds through to the variability of welfare.\\

\noindent \textbf{Statistical analysis.} Our analysis is based on group-level means (over the 29 rounds we simulated) of each of our key metrics: global activity, proportion signed up to tracing, share of infected and quarantined subjects in the group, and per subject welfare. We run two sets of non-parametric tests (using a two-sided Mann-Whitney test on unmatched samples):
\begin{enumerate}
    \item Within each treatment, compare median outcomes for liberals and for conservatives. This allows us to examine whether liberals or conservatives do ``better'' in each treatment. 
    \item A difference-in-differences test. First, take the difference across treatments. This calculates the effect of changing treatment environments for an ideologically homogeneous group. Second, compare these differences across ideologies. This allows us to explore whether the gain associated with moving from one treatment to another is larger for a particular ideology.
\end{enumerate}

For each of these sets of tests, we report the proportion of times the test rejects the null hypothesis of no difference at the 5\% level. This is the Monte Carlo estimate of power for our tests (see \cite{Z_2014} for details). We consider power of 80\% and above as sufficiently strong evidence of statistical difference \cite{C_1992, VM_2007}.

\subsection{Results}\label{subsec:sims results}

\begin{figure}[ht!]
\centering
\begin{subfigure}{.5\textwidth}
  \includegraphics[width=\linewidth]{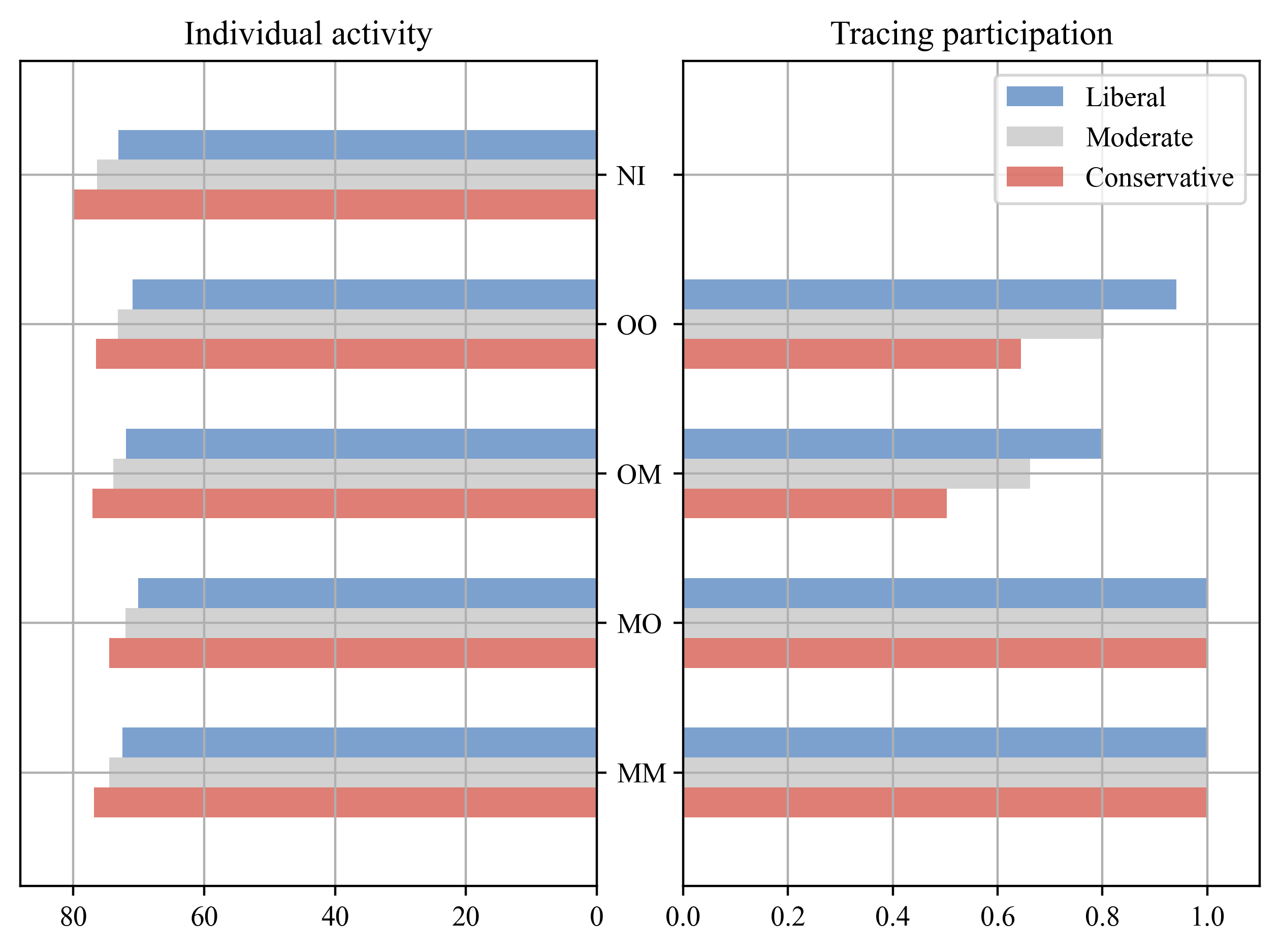}
  \caption{}
  \label{fig:sub1}
\end{subfigure}%
\begin{subfigure}{.5\textwidth}
  \hspace{0cm}
  \includegraphics[width=\linewidth]{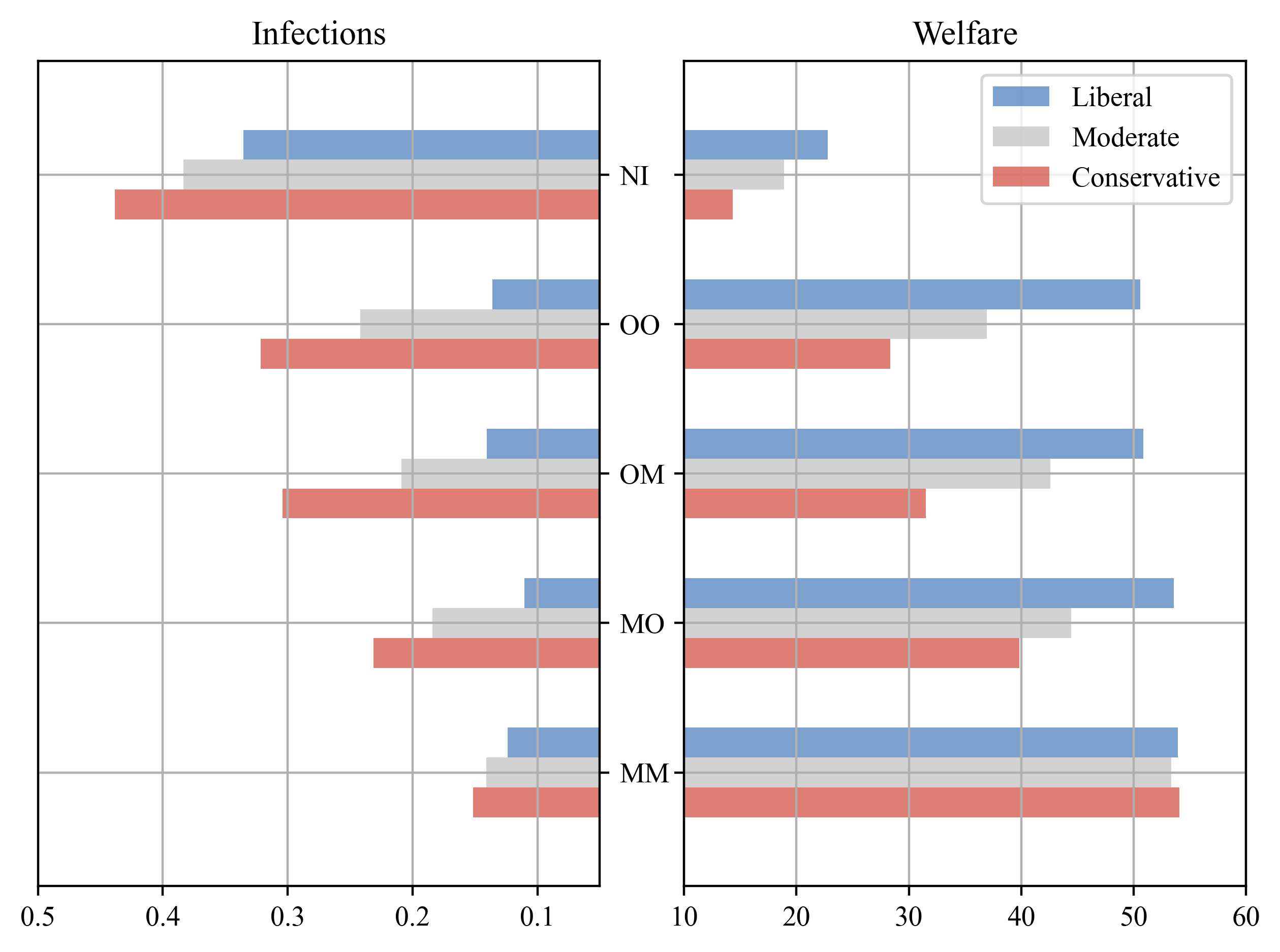}
  \caption{}
  \label{fig:sub2}
\end{subfigure}
\caption{Average simulated decisions (panel a) and outcomes (panel b) for three ideologically homogeneous groups.}
\label{fig:simulations_output}
\end{figure}

\Cref{fig:simulations_output} shows averages of some of the variables of interest for each of the three ideologies. We can see that decisions and outcomes depend on both treatment and ideology. Liberals do less activity than moderates, who in turn do less than conservatives (left part of panel (a)). Conversely, liberals participate in contact tracing system more than moderates, who in turn participate more than conservatives (right part of panel (b)). As a result, liberals have lowest levels of infection and highest individual welfare, while conservatives have highest infections and lowest welfare. 

Notice that the differences in outcomes practically vanish under the fully mandatory program (treatment MM). Ideology score is not a significant determinant of quarantine decisions and so is omitted from \Cref{fig:simulations_output}.

Non-parametric tests on the data from simulations confirm the observations above (see \Cref{tab:nonpar_simulations}(a)). Our analysis focuses on comparing the simulated outcomes of liberal and conservative groups against each other. We do not compare them against the moderate group.

Activity of agents not in quarantine is higher in conservative groups relative to liberal ones for all treatments except NA (where it is fixed at 100) (MW power $=1$). Contact tracing system participation is lower in conservative groups whenever it is optional (MW power $=1$). Consequently, the share of infected subjects is higher in conservative groups for all treatments (MW power $=1$). This leads to average individual welfare in conservative groups being lower than that in liberal groups in most treatments (MW power $=1$). The exceptions are treatments NA where the power is just below our cut-off rule (MW power $=0.76$), and MM where very low power indicates extremely weak evidence against the null hypothesis (MW power $=0.09$).

Further, \Cref{tab:nonpar_simulations}(b) shows that the gain (in terms of welfare) from going from no intervention (treatment NI) to either of treatments OO or OM is higher for liberals than conservatives (MW power $=1$). In contrast, the gain from moving from treatment NI to treatment MM is significantly higher for conservatives (MW power $=1$). Finally, both conservatives and liberals experience similar benefits from moving to treatment MO (from treatment NI) (MW power $=0.69$). 

When it comes to a move from a fully optional program (treatment OO) to a partially (or fully) mandatory we observe the following (\Cref{tab:nonpar_simulations}(c)):
\begin{enumerate}
    \item there is no evidence that one ideology benefits more from the introduction of mandatory quarantine (treatment OM) (MW power $=0.23$);
    \item conservatives benefit more than liberals from introduction of mandatory tracing (treatment MO) (MW power $=0.88$);
    \item conservatives benefit more than liberals from introduction of a fully mandatory program (treatment MM) (MW power $=0.1$).
\end{enumerate}

\begin{table}[ht!]
\centering
\caption{Non-parametric analysis of data from simulations for liberal and conservative groups (independent unit of observation is a group)}
\label{tab:nonpar_simulations}
\begin{tabular}{@{}lrlrlrlrl@{}}
\toprule
 & \multicolumn{2}{c}{\multirow{2}{*}{\begin{tabular}[c]{@{}c@{}}individual\\ activity\end{tabular}}} & \multicolumn{2}{c}{\multirow{2}{*}{\begin{tabular}[c]{@{}c@{}}tracing\\ participation\end{tabular}}} & \multicolumn{2}{c}{\multirow{2}{*}{\begin{tabular}[c]{@{}c@{}}infected\\ share\\ \end{tabular}}} & \multicolumn{2}{c}{\multirow{2}{*}{\begin{tabular}[c]{@{}c@{}}individual\\ welfare\end{tabular}}} \\
 & & & & & & & & \\
 \midrule
\multicolumn{9}{l}{\textbf{panel (a)}} \\
 \midrule
NA\hspace{6cm} & - &  & -0.3 & 1 & 0.03 & 1 & -1.43 & 0.76 \\
NI\hspace{6cm} & 7 & 1 & - &  & 0.11 & 1 & -8.82 & 1 \\
OO\hspace{6cm} & 8 & 1 & -0.3 & 1 & 0.19 & 1 & -23.36 & 1 \\
OM\hspace{6cm} & 7 & 1 & -0.3 & 1 & 0.17 & 1 & -20.14 & 1 \\
MO\hspace{6cm} & 7 & 1 & - &  & 0.12 & 1 & -13.72 & 1 \\
MM\hspace{6cm} & 6 & 1 & - &  & 0.03 & 1 & 0.12 & 0.09 \\
 \midrule
\multicolumn{9}{l}{\textbf{panel (b)}} \\
 \midrule
OO\hspace{6cm} & - &  & - &  & 0.09 & 1 & -14.64 & 1 \\
OM\hspace{6cm} & - &  & - &  & 0.06 & 1 & -11.35 & 1 \\
MO\hspace{6cm} & - &  & - &  & 0.02 & 0.21 & -5.06 & 0.69 \\
MM\hspace{6cm} & - &  & - &  & -0.08 & 1 & 8.89 & 1 \\
\midrule
\multicolumn{9}{l}{\textbf{panel (c)}} \\
\midrule
OM\hspace{6cm} & - &  & - &  & -0.03 & 0.26 & 3.31 & 0.23 \\
MO\hspace{6cm} & - &  & - &  & -0.07 & 0.91 & 9.59 & 0.88 \\
MM\hspace{6cm} & - &  & - &  & -0.16 & 1 & 23.45 & 1 \\
 \midrule
\multicolumn{9}{l}{\multirow{10}{*}{\parbox{17cm}{Notes: H0 for all hypotheses is that there is no statistical difference between different ideological groups for the specific variables of interest (columns). H1 is always two-sided. All tests are MW. We generate 10 groups and subject them to all treatments, and repeat this exercise 1,000 times. We report: \textbf{(a)} the average difference in medians of variables between liberals and conservatives for all six treatments; \textbf{(b)} the average (for 1,000 simulations) difference in median (for 10 groups) gain from going from no program (treatment NI) to a program between liberals and conservatives; \textbf{(c)} the average (for 1,000 simulations) difference in median (for 10 groups) gain from going from a fully optional program (treatment OO) to another program between liberals and conservatives. For all three panels we also report the estimate of power -- i.e. the proportion of times (out of 1,000) that H0 is false at 5\% significance level.}}} \\
\multicolumn{9}{l}{} \\
\multicolumn{9}{l}{} \\
\multicolumn{9}{l}{} \\
\multicolumn{9}{l}{} \\
\multicolumn{9}{l}{} \\
\multicolumn{9}{l}{} \\
\multicolumn{9}{l}{} \\
\multicolumn{9}{l}{} \\
\multicolumn{9}{l}{} \\ \bottomrule
\end{tabular}
\end{table}


\clearpage

\bibliographystyle{IEEEtranN}
\bibliography{si_bib.bib} 

\clearpage
\section{Experimental instructions} \label{sec:instructions}
\Cref{sec:instructions_main_experiment} contains instructions, understanding quiz and screenshots from the interface for the experiment. Next, \Cref{sec:instructions_svo,,sec:instructions_bret} present instructions and interfaces for the social (SVO) and risk (BRET) preferences elicitation tasks.

\subsection{Main experiment}
\label{sec:instructions_main_experiment}

This section contains sample instructions and understanding quiz for treatment OO of our experiment, which has the greatest number of choices that subjects need to make. Instructions and quizzes for other treatments are available on request. The section concludes with screenshots from the interface of the experiment.

\textbf{Welcome to the experiment!}

You will earn a fixed fee of \textbf{\$1} for completing the whole experiment. You can also earn points based on your decisions during the experiment, which will be converted to \$ at the end of the session. There may also be bonus tasks at the end, giving you a chance to earn more.

It will take around \textbf{30-40 minutes} to do the whole experiment, and most people will earn \textbf{\$6-10}. You can earn more or less than this, depending on the choices you and other people make.

During the experiment you will interact with a group of other real people recruited through MTurk. Everybody will remain anonymous -- nobody will learn your identity and you won't learn anyone else's identity at any point during or after the experiment.

The experiment is interactive, so please pay attention throughout the whole experiment. Not doing so will slow down the session, ruin the experience for others and may even result in your disqualification.

The instructions on the next 8 pages explain the rules of the experiment. Please read them carefully. Everyone will read the same instructions. Once you finish reading the instructions, you will need to complete a short quiz to check you understand the game. If you fail the quiz too many times, you will be disqualified and you will not earn the fixed fee of \$1.

To continue to the instructions, click 'Next'.

\textbf{Page 1: Set-up}

Before the experiment starts, you will be allocated to a group of 12-15 participants. The participants in your group do not change during the experiment.

The experiment has at least 35 rounds. In each round after Round 35, the computer flips a virtual coin. If it is Heads, there will be another round. If it is Tails, the experiment ends.

During a round you can earn points from activity. In the real world, activity involves going out of the house and meeting other people. Examples of activity include going to work in-person, going to a bar or restaurant, going out shopping, or meeting friends.

In a round, you choose your activity level, which can be any whole number between 0 and 100. You can think of 100 as the amount of time you typically spent out of the house before the COVID-19 pandemic started in early 2020.

To choose an activity level you can either drag the slider or type a number into the box to the right of the slider, as shown in the interface below. The two methods do the same thing.

\centerline{\includegraphics[scale=.6]{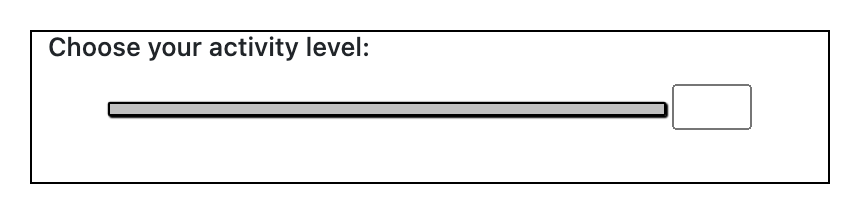}}

\textbf{You earn one point for each unit of activity you choose to do.} For example, if you choose an activity level of 85, you will earn 85 points.

\textbf{For Rounds 1-5, this is all that happens.} Doing activity has benefits, but has no costs. You can think of this as the world before the start of the COVID-19 pandemic.

To continue to the next page, click 'Next'. To go to the previous page, click 'Back'.

\textbf{Page 2: COVID-19}

At the beginning of \textbf{Round 6}, the COVID-19 pandemic starts. The computer randomly chooses 4-5 participants to be infected with COVID-19 (commonly known as coronavirus). Every participant faces the same probability of being infected, but you will not know whether \textbf{you} are infected with COVID-19 until the end of Round 6. The interface below shows the information available to you.

\centerline{\includegraphics[scale=.6]{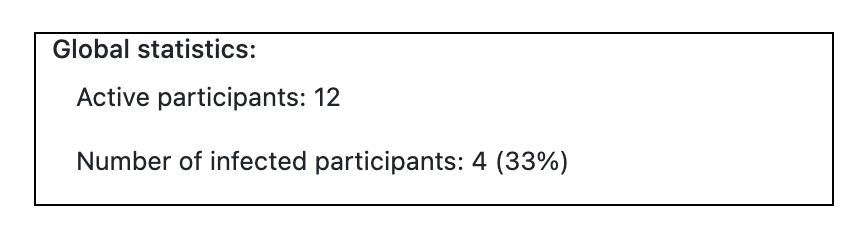}}

\textbf{If you are infected with COVID-19 in a round, you pay a cost of 150 points.} You can think of this cost as the inconvenience of mild symptoms of COVID-19, which are experienced by most healthy adults. The main symptoms of COVID-19 are shortness of breath, a high fever and a new, continuous cough. Most patients experience mild symptoms and recover in 1-2 weeks, but cases can progress to pneumonia and organ failure in the most vulnerable individuals.

\textbf{Infection lasts for one round, then you recover.} This means you pay a cost of 150 points only in the round you are infected. Infection confers no immunity so you can be reinfected in any of the following rounds.

To continue to the next page, click 'Next'. To go to the previous page, click 'Back'.

\textbf{Page 3: Spreading}

After everyone in your group has chosen their activity level, COVID-19 spreads from infected to healthy participants. \textbf{The number of participants who become exposed to COVID-19 depends on the number of infected participants, the size of the group and the average activity level in the group.} The average activity level is calculated by adding up activity levels of all participants and dividing by the size of the group.

The number of people who get exposed to COVID-19 in a round is:

$$\text{exposed} = 3 \times \frac{\text{infected}}{\text{group size}} \times (\text{group size} - \text{infected}) \times \frac{(\text{average activity level})^2}{10,000}$$

rounded to the nearest whole number.

For example, suppose that there are 12 participants in the group (including you), and 4 of them are infected. And suppose that the average activity level is 80. Using the formula above, you can see that 5 participants become exposed to COVID-19:

$$\text{exposed} = 3 \times \frac{4}{12} \times (12 - 4) \times \frac{(80)^2}{10,000} = 5.12 \approx 5$$

Not everyone has the same chance of becoming exposed. \textbf{If you are already infected, then you cannot be exposed. If you are currently healthy, then the chance you are exposed depends on the total number of participants who get exposed, the number of currently infected and your own activity level.} Consider the example above, where 5 participants in a group of 12 are exposed, 4 are infected, and the average activity level is 80. Assuming you are not infected in a round, your probability of becoming exposed in this round is calculated as:

$$\text{probability you become exposed} = \frac{\text{exposed}}{\text{group size}} \times \frac{\text{your activity level}}{\text{average activity level}} \times 100\%$$

$$ = \frac{5}{12-4} \times \frac{\text{your activity level}}{80} \times 100\% = \text{your activity level} \times 0.781\%$$

On the graph below, you can move the mouse along the green line to see how the probability of you becoming exposed varies with your activity level in this example:

\centerline{\includegraphics[scale=.6]{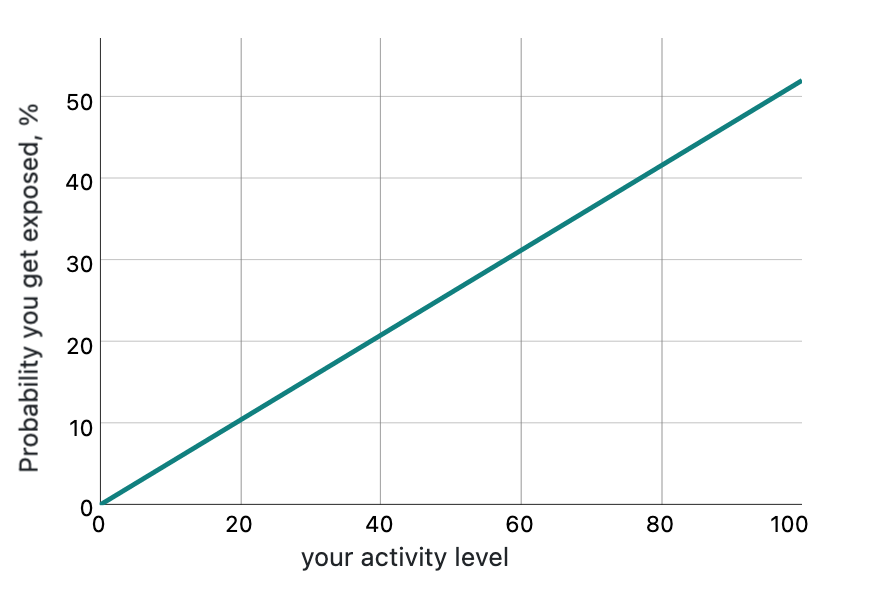}}

To continue to the next page, click 'Next'. To go to the previous page, click 'Back'.

\textbf{Page 4: Tracing}

One way to reduce the spread of COVID-19 is to implement a contact tracing scheme. You can think of this system as an application installed on your smartphone which sends you notifications if you have been in contact with someone who has COVID-19. Applications like this are currently available in most of the USA.

\textbf{From Round 6 onwards}, once the COVID-19 pandemic has started, there is a contact tracing system. \textbf{You need to choose whether you want to sign up to it.} The decision to sign up to contact tracing is in addition to choosing your activity level. \textbf{Signing up does not cost you any points.} The interface below shows how you can sign up by selecting 'Yes' or opt out by selecting 'No'.

\centerline{\includegraphics[scale=.6]{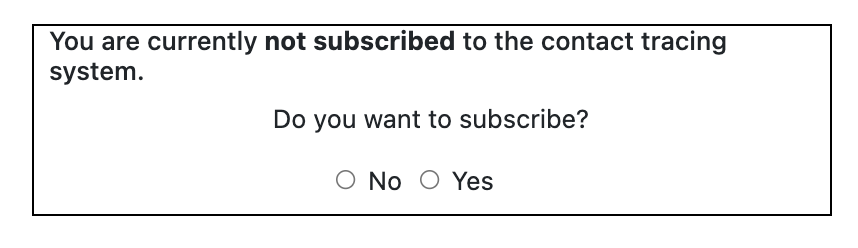}}

\textbf{In each round after Round 6 you can change your decision} about whether to be signed up to the contact tracing scheme. The interface below shows what the contact tracing part of the interface looks like in a round, when you were subscribed to the scheme in the previous round:

\centerline{\includegraphics[scale=.6]{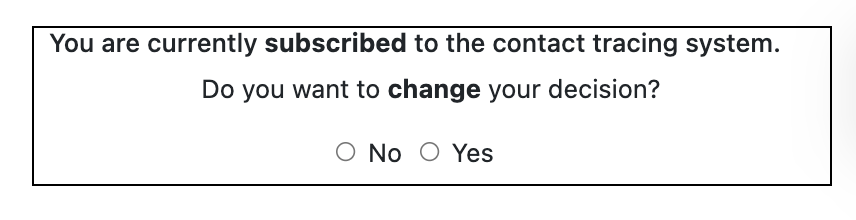}}

In this example, you can unsubscribe from the tracing scheme by clicking 'Yes'. To remain signed up, click 'No'.

The benefit of the contact tracing system is that it can send you an alert if you become exposed to COVID-19.

The contact tracing system is not perfect, so it does not notify everyone who has been exposed to COVID-19. If you are signed up to the contact tracing system and become exposed to COVID-19, there is a one in three chance it sends you an alert. \textbf{If you get an alert, then you have definitely been exposed to COVID-19.}

To continue to the next page, click 'Next'. To go to the previous page, click 'Back'.

\textbf{Page 5: Quarantine}

\textbf{If you receive an alert, you need to choose whether to go into quarantine.} The interface below shows how you can make your choice about quarantine by selecting either 'Yes' or 'No':

\centerline{\includegraphics[scale=.6]{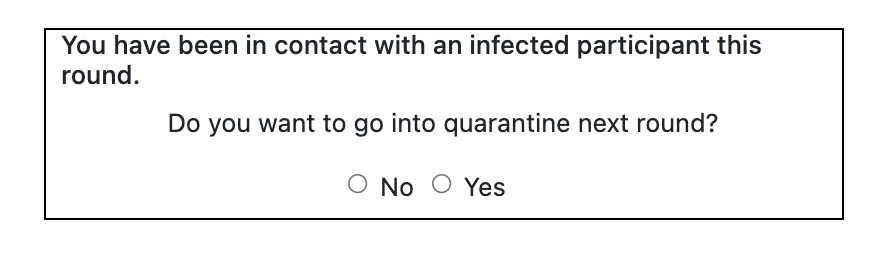}}

Selecting 'Yes' means you go into quarantine in the next round. \textbf{Your activity in the next round will be set to zero}, so you cannot earn any points in that round. Others will benefit, as you will not infect anyone else. Selecting 'No' does not affect your ability to choose your activity level in the next round.

To continue to the next page, click 'Next'. To go to the previous page, click 'Back'.

\textbf{Page 6: Payoffs from one round}

At the end of each round, you earn points from your activity level and may incur a cost from being infected.

You earn \textbf{one point for each unit of activity} you choose to do. If you are infected, you incur a cost of \textbf{150 points}.

You need to choose both your activity level and whether you want to participate in the contact tracing scheme. After you make your choices, you have to click 'Submit' to confirm. For the quarantine decision, just choose 'Yes' or 'No', and then click ‘Submit'.

\textbf{If you fail to make and submit any of your choices in a round, you will receive a penalty of 100 points for that round.} For any choice you have not made, a default option will be implemented in order for the experiment to continue. The table below shows the default option for each of the choices.

\centerline{\includegraphics[scale=.6]{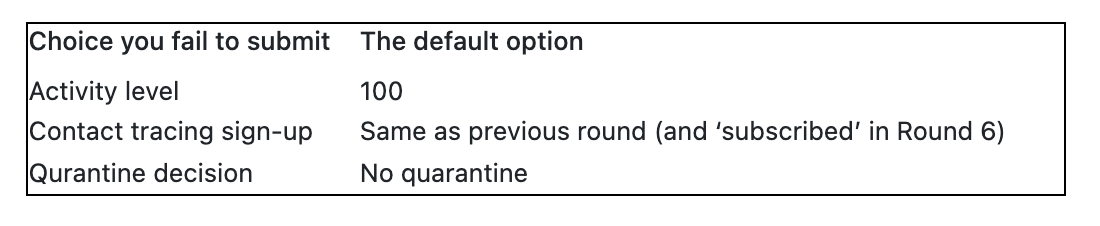}}

A separate box on the interface shows all the information about your earnings.

\centerline{\includegraphics[scale=.6]{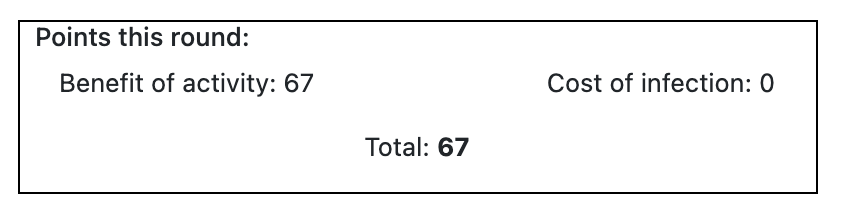}}

\textbf{If you fail to submit any of your choices in 3 consecutive rounds you will be disqualified from the experiment.} In this case, you will not earn any compensation for your participation in the experiment.

Throughout the experiment there will be a timer to the right of the interface which shows how much time you have left to submit your choices.

Finally, throughout the whole experiment, you can use a small tool designed to calculate the number of points that you earn in a round depending on your activity level and health status. To use the tool, you need to click a button located near the timer. Once you click the button, the following window pops up:

\centerline{\includegraphics[scale=.6]{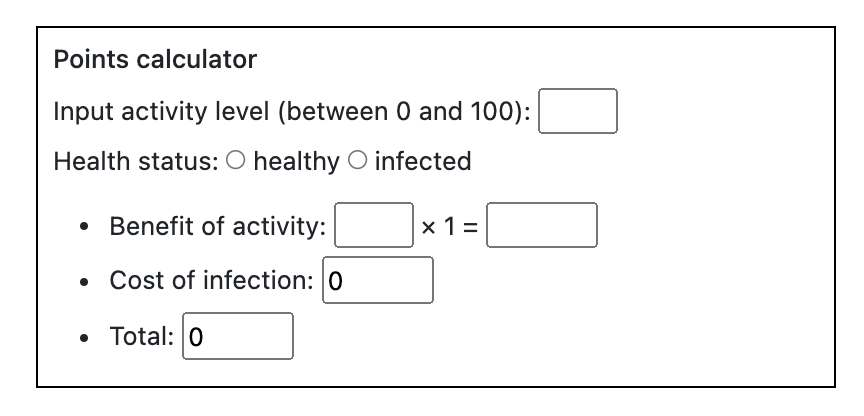}}

Feel free to check how the calculator works using the interface above.

To continue to the next page, click 'Next'. To go to the previous page, click 'Back'.

\textbf{Page 7: Dynamics (moving between rounds)}

Up to this point, we have explained to you how a single round of the experiment works.

From Round 2 onwards, you learn the average activity level from the previous round, and how many participants are infected in the current round. Remember that nobody can be infected until Round 6.

The overall number of infected participants in a round is equal to the number of participants who got exposed to COVID-19 in the previous round. \textbf{You will not know whether you are infected unless you received an alert from the contact tracing system in the previous round.}

Note that \textbf{you do not know how many participants (if any) are in quarantine in a round.} However, any participant who is in quarantine during the round will not contribute to the spread of COVID-19.

Suppose that there are 12 participants in your group (including you), 5 participants got exposed to COVID-19 in Round 10, and 2 participants decided to go into quarantine. Also suppose that the average activity level in Round 11 was 70. Then the number of participants who got exposed to COVID-19 in Round 11 is:

$$\text{exposed} = 3 \times \frac{\text{infected} - \text{quarantined}}{\text{group size}} \times (\text{group size} - \text{infected}) \times \frac{(\text{average activity level})^2}{10,000}$$

$$= 3 \times \frac{5 - 2}{12} \times (12 - 5) \times \frac{(70)^2}{10,000} = 2.57 \approx 3$$

To continue to the next page, click 'Next'. To go to the previous page, click 'Back'.

\textbf{Page 8: Payment}

When the experiment ends, the computer randomly picks \textbf{12} rounds to determine your payment for the experiment. The points are converted to \$ at a rate of \textbf{125 points per \$1}.

Suppose that you earn 695 in the 12 randomly drawn rounds. Then your payment for the experiment is \$5.56, plus the fixed fee and earnings from the bonus tasks.

\textbf{You will not receive any information regarding your payment until the end of the experiment.} At the end of the experiment, you will be able to see a detailed breakdown of your payment (including any bonuses) and a full history of your choices in the experiment.

To continue to the short Quiz, click 'Next'. To go to the previous page, click 'Back'.

\textbf{Quiz}

Before you can start the experiment, you must complete a short Quiz. You need to answer all questions correctly before you can continue.

You have 3 attempts in total. If you \textbf{fail all 3 attempts}, you will be disqualified from the experiment.

1. Some participants will be randomly infected with COVID-19 at some point during the experiment. In which round will this occur?

2. In total, how many points would you get in Round $T$ if your activity level was $X$ and you were \textit{[not]} infected in this round? [$X \in \{10,20,...,100\}$]

3. If you received an alert in Round $L$ and decided to \textit{[not]} go into quarantine, would this decision affect your choice of activity in the next round?

To submit your answers, press the 'Submit' button below. To go back to the instructions, press the 'Back' button.

\textbf{The experiment: Interface}

\begin{figure}[ht]
\begin{minipage}{\linewidth}
\centering
\subfloat[Activity level and tracing sign-up decisions]{\label{}
\includegraphics[width=0.95\textwidth]{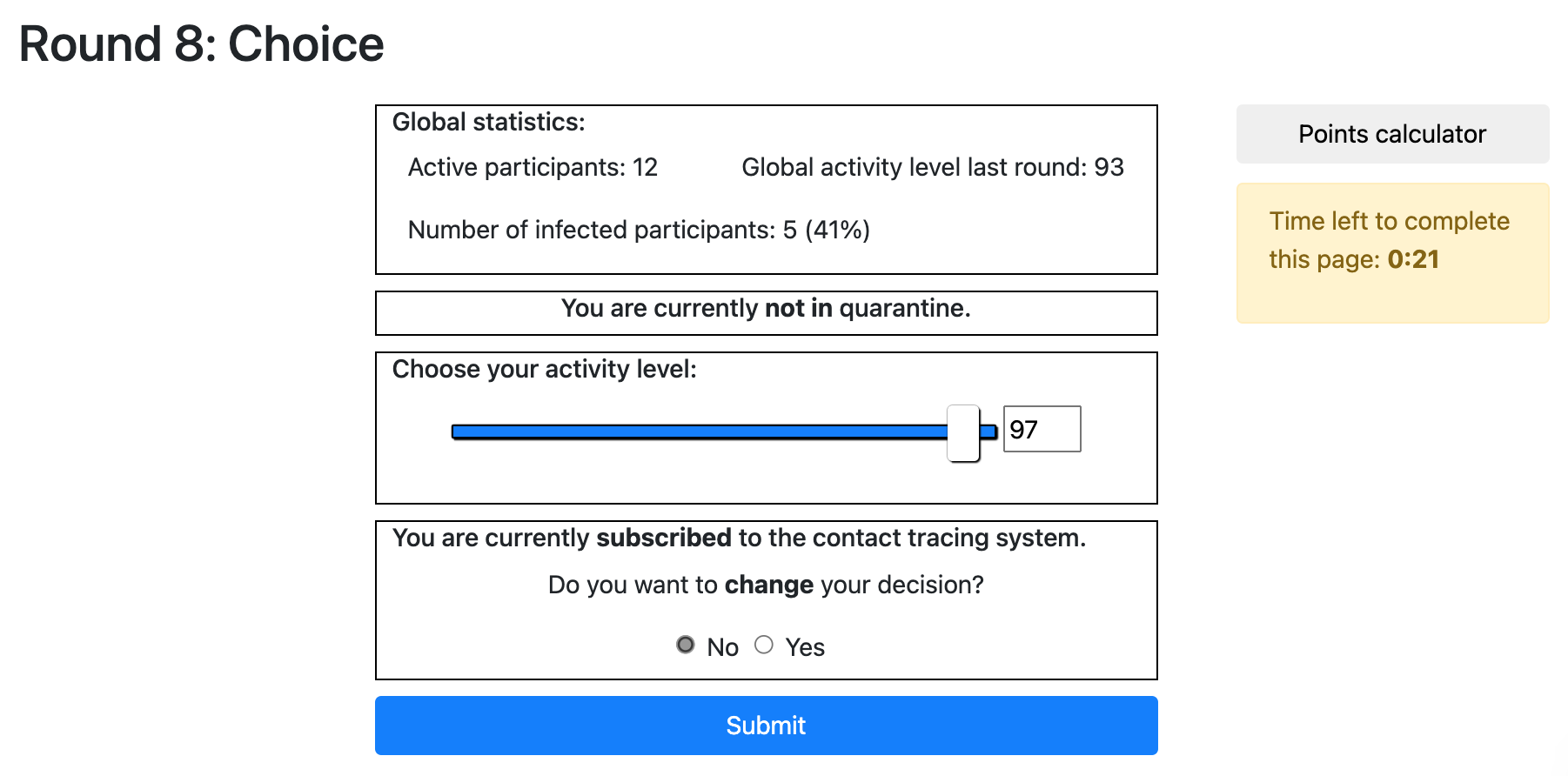}
}
\end{minipage}
\begin{minipage}{\linewidth}
\centering
\subfloat[Alert]{\label{}
\includegraphics[width=0.95\textwidth]{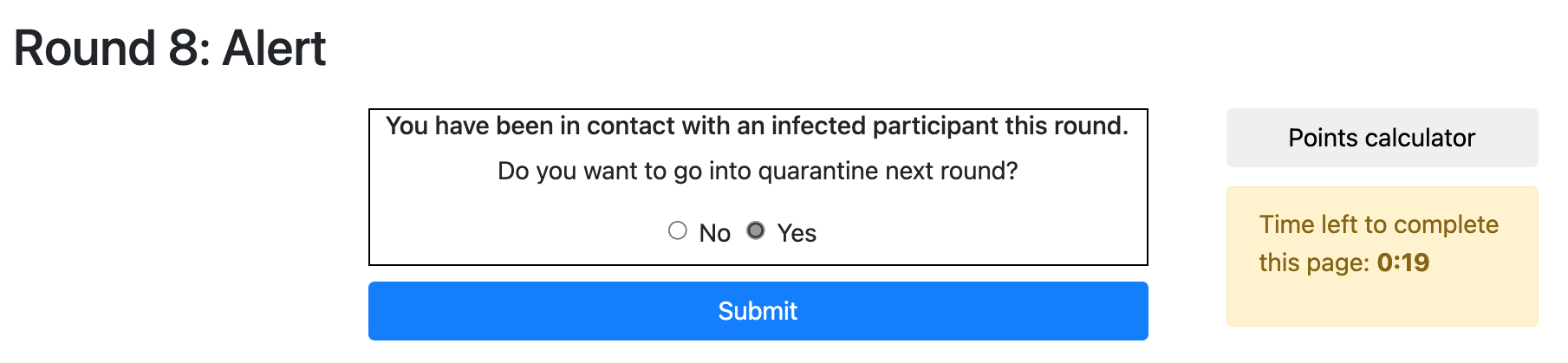}
}
\end{minipage}
\begin{minipage}{\linewidth}
\centering
\subfloat[Results]{\label{}
\includegraphics[width=0.95\textwidth]{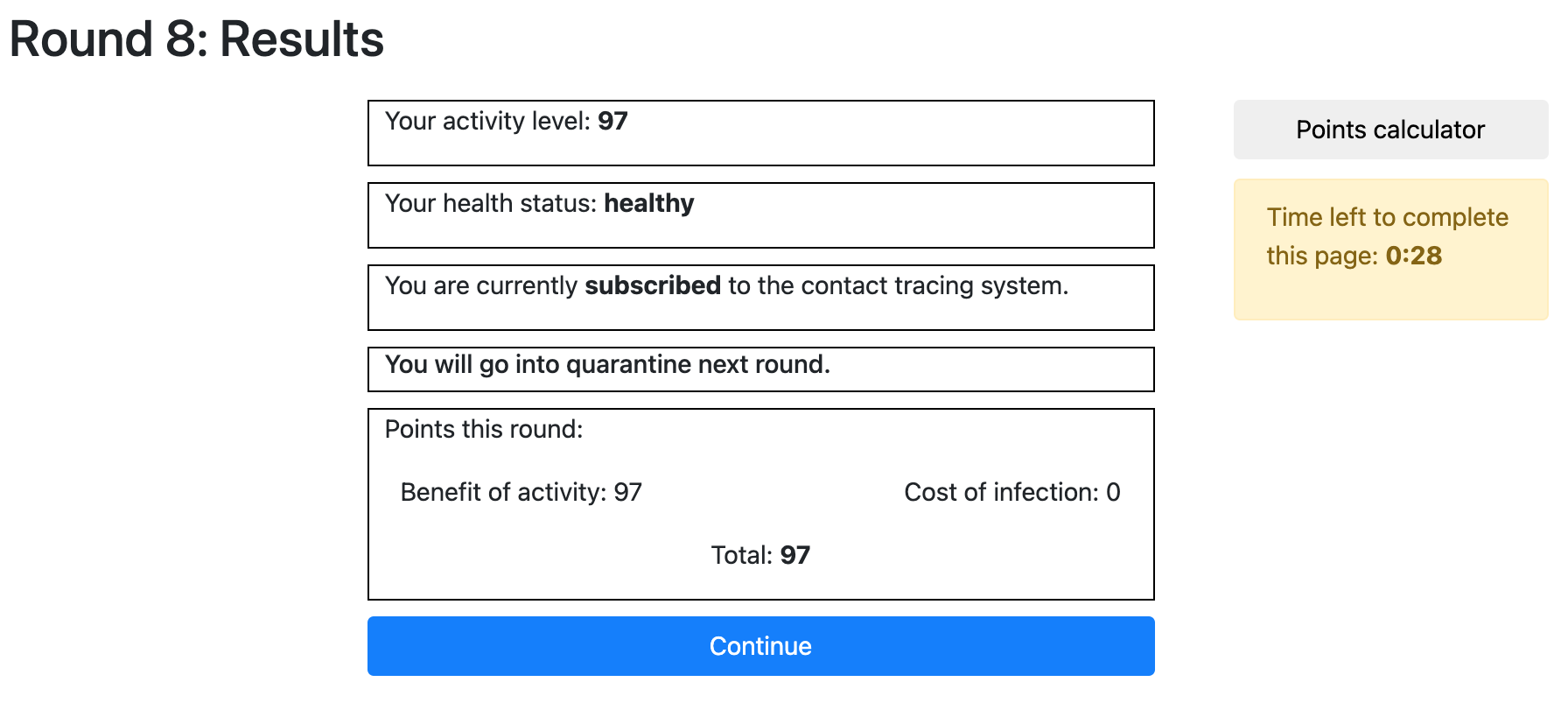}
}
\end{minipage}
\caption{Round 8 of the experiment (treatment OO): interface}
\label{fig:experiment_interface}
\end{figure}

\Cref{fig:experiment_interface} shows the decisions and results interfaces from the main experiment. We focus on Round 8 of treatment OO which has the maximum number (i.e. 3) of decisions to be made by subjects. In this example, the subject chooses and activity level of 97 and to be signed up to tracing (\Cref{fig:experiment_interface}(a)). She then receives an alert from the contact tracing system and decides to go into quarantine in Round 9 (\Cref{fig:experiment_interface}(b)). She earns 97 points for Round 8 (\Cref{fig:experiment_interface}(c)).

\subsection{Social value orientation (SVO) slider measure}
\label{sec:instructions_svo}

This section presents instructions and interface from the Social value Orientation (SVO) task, completed by subjects as part of the recruitment survey. 

\textbf{SVO: Instructions}

You have answered the qualifying questions correctly and are now in the Bonus Task.

In this Bonus Task you will make a series of decisions about allocating money between you and another anonymous Turker (called, Other). All of your decisions will be completely confidential.

There are a total of 6 decisions. They are independent of each other. For each decision, you need to decide how you want to share some money between yourself and Other. All values are in cents. Once you have made your decision, click the button below your choice. Your choices affect the amount of money you receive, and the amount of money the Other receives.

There are no right or wrong answers, this is all about personal preferences.

Every time 25 Turkers complete the task, we will randomly pick two of them and pay them for the Bonus Task as follows. We will pick one of the 6 decisions, and randomly implement the decision of one of the two chosen Turkers.

For example, suppose we randomly chose Turkers A and B out of those 25 Turkers, and that we further randomly chose to implement decision 3 of Turker B. Suppose that in decision 3 Turker B allocated 150 cents to themselves, and 140 cents to Other. Then Turkers A and B will be paid 140 and 150 cents respectively.

\textbf{SVO: Interface}

\begin{figure}[ht]
    \centering
    \includegraphics[width=0.9\textwidth]{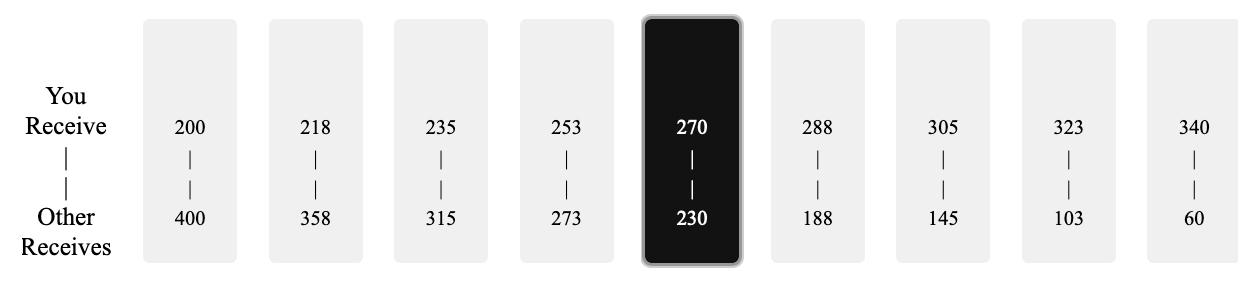}
    \caption{Social Value Orientation (SVO) Slider Measure: decision interface}
    \label{fig:svo_interface}
\end{figure}

\Cref{fig:svo_interface} shows the interface for the SVO task for one of the six decisions. In this example, 

presents part of the interface of the SVO task. We focus on one of the six decisions. In this example, a participant decided to allocate 270 cents to self and 230 cents to Other.

\subsection{Bomb risk elicitation task (BRET)}
\label{sec:instructions_bret}

This section includes instructions from and screenshots of the interface of the Bomb Risk Elicitation Task (BRET), which subjects complete after the main experiment and the post-experimental survey.

\textbf{BRET: Instructions}

Thank you very much for taking part in the experiment!

You are now in the Bonus Task in which you have an opportunity to earn an extra payment.

On the next page, you will see 100 boxes. As soon as you start the task by pressing the \textbf{'Start'} button, one box is collected per second starting from the top left corner. Once collected, the box is marked by a tick symbol.

\textbf{For each box collected, you earn \$0.02.}

One of the 100 boxes contains a \textbf{bomb} that destroys all of your earnings. \textbf{You do not know where the bomb is located.} You only know that the bomb can be in any box with equal probability.

Your task is to choose when to stop collecting the boxes and open those you have collected. You stop collecting boxes by pressing \textbf{'Stop'} at any time. After that you open the boxes you have collected by pressing the \textbf{'Open'} button. Note that once you press 'Stop' you cannot restart collecting boxes.

A dollar or a bomb symbol will be shown on each of the boxes you have collected.

If the bomb symbol does not appear, that means that you have \textbf{not collected the box with the bomb}. In this case, you \textbf{earn the amount accumulated by the boxes you have collected}.

If the bomb symbol appears, that means that you \textbf{have collected the box with the bomb}. In this case, you \textbf{earn zero} for the Bonus Task.

To proceed to the Bonus Task, press the button below.

\textbf{BRET: Interface}

\begin{figure}[ht]
\begin{minipage}{.5\linewidth}
\centering
\subfloat[Decision]{\label{fig:bret_choice}
\includegraphics[scale=0.6]{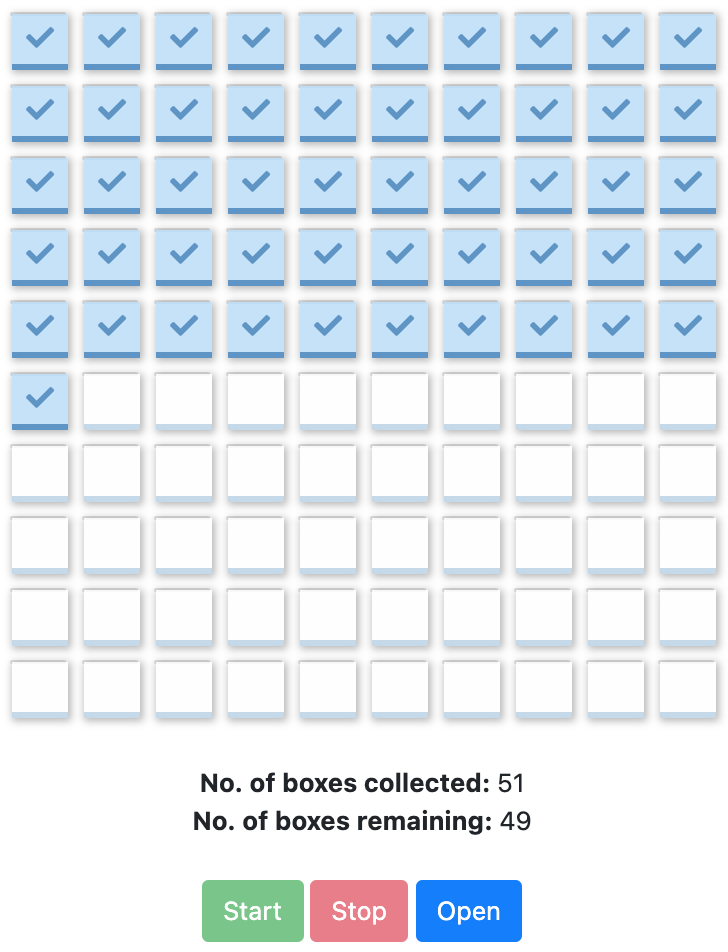}
}
\end{minipage}
\begin{minipage}{.5\linewidth}
\centering
\subfloat[Results]{\label{fig:bret_result}
\includegraphics[scale=0.6]{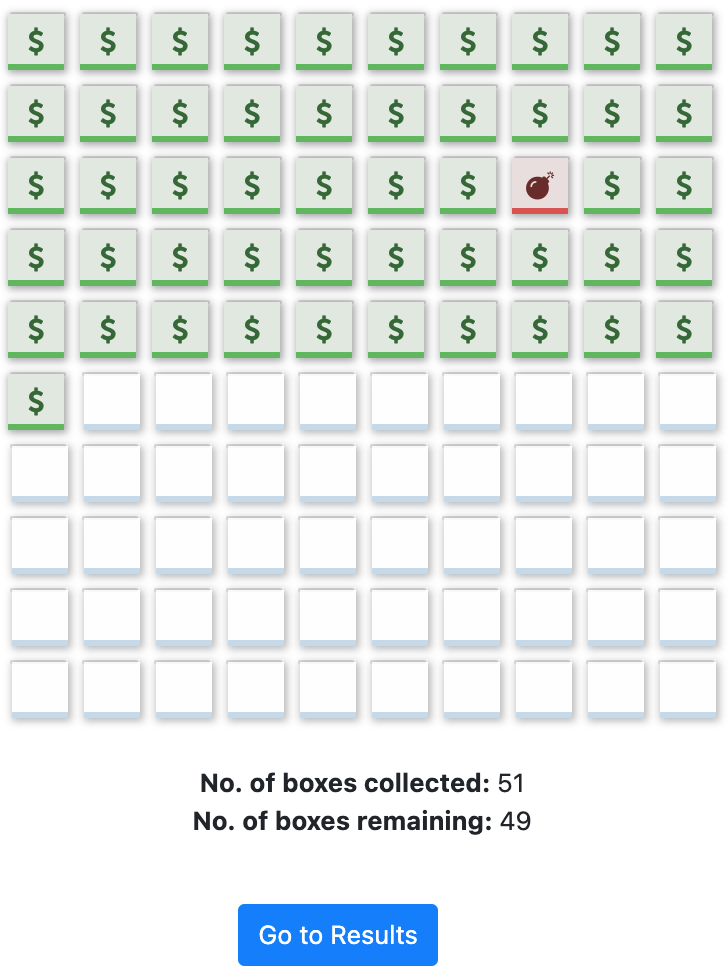}
}
\end{minipage}
\caption{Bomb Risk Elicitation Task (BRET): interface}
\label{fig:bret_interface}
\end{figure}

\Cref{fig:bret_interface} presents the interface of the BRET. \Cref{fig:bret_choice} is an example of the interface after a subject has clicked 'Stop', having opened 51 boxes. \Cref{fig:bret_result} shows the interface after she clicked 'Open'. We can see that the bomb is in 28th box -- i.e. among the 51 collected boxes. Consequently, in this example the subject earns zero for the BRET.

